\documentclass[12pt]{article}

  \usepackage{graphicx}
  \usepackage{epsfig}
  \usepackage{amssymb}
  \usepackage{subfigure}
    \usepackage{feynmf}%%%{feynmp}
\evensidemargin - 1.truecm
\begin{document}

\newcommand{\be}{\begin{equation}}
\newcommand{\ee}{\end{equation}}
\newcommand{\nn}{\nonumber}
\newcommand{\bea}{\begin{eqnarray}}
\newcommand{\eea}{\end{eqnarray}}
\newcommand{\bfig}{\begin{figure}}
\newcommand{\efig}{\end{figure}}
\newcommand{\bc}{\begin{center}}
\newcommand{\ec}{\end{center}}

\begin{fmffile}{NONab}

%%%%%%%%%%%%%%%%%%%%%%%%%%%%%%%%%%%%%%%%%%%%%%%%%%%%%%%%%%%%%%%%%%%%%%%%

\begin{titlepage}
\nopagebreak
{\flushright{
        \begin{minipage}{5cm}
        Freiburg-THEP 03/19\\
        TTP03-34\\
        CERN-TH/2003-274\\
        {\tt hep-ph/0311145}\\
        \end{minipage}        }

}
\renewcommand{\thefootnote}{\fnsymbol{footnote}}
%\vspace*{-1.5cm}                          
\vskip 0.5cm
\begin{center}
\boldmath
{\Large\bf Master Integrals for the 2-loop QCD virtual\\[3mm]
corrections to the Forward-Backward  Asymmetry}\unboldmath
\vskip 1.cm
{\large  R.~Bonciani 
\footnote{This work was supported  by the European Union under
contract HPRN-CT-2000-00149}
\footnote{Email: Roberto.Bonciani@physik.uni-freiburg.de} ,}
\vskip .2cm
{\it Fakult\"at f\"ur Mathematik und Physik, 
Albert-Ludwigs-Universit\"at
Freiburg, \\ D-79104 Freiburg, Germany} 
\vskip .2cm
{\large P.~Mastrolia\footnote{Email: Pierpaolo.Mastrolia@bo.infn.it}},
\vskip .2cm
{\it Dipartimento di Fisica dell'Universit\`a di Bologna, 
I-40126 Bologna, Italy} 
\vskip .1cm
{\it and Institut f\"ur Theoretische Teilchenphysik,
Universit\"at Karlsruhe, \\ D-76128 Karlsruhe, Germany}
\vskip .2cm
{\large E.~Remiddi\footnote{Ettore.Remiddi@bo.infn.it}}
\vskip .2cm
{\it Theory Division, CERN, CH-1211 Geneva 23, Switzerland} 
\vskip .1cm
{\it Dipartimento di Fisica dell'Universit\`a di Bologna
and INFN Sezione di Bologna, I-40126 Bologna, Italy}
\end{center}
\vskip 1.2cm

\begin{abstract}
We present the Master Integrals needed for the calculation of the 
two-loop QCD corrections to the forward-backward asymmetry of a 
quark-antiquark pair produced in electron-positron annihilation events.
The abelian diagrams entering in the evaluation of the vector form 
factors were calculated in a previous paper. We consider here the 
non-abelian diagrams and the diagrams entering in the computation of 
the axial form factors, for arbitrary space-like momentum transfer $Q^2$ 
and finite heavy quark mass $m$. Both the UV and IR divergences are 
regularized in the continuous $D$-dimensional scheme. The Master 
Integrals are Laurent-expanded around $D=4$ and evaluated by the 
differential equation method; the coefficients of the 
expansions are expressed as 1-dimensional harmonic polylogarithms of 
maximum weight 4.

\end{abstract}
\vfill
\end{titlepage}

\section{Introduction \label{Intro}}

The measurement of $A_{FB}^{q \bar{q}}$, the forward-backward asymmetry
of the production of quark-antiquark pairs in the processes
$e^+ e^- \to q \bar{q}$, is a stringent test of the Standard Model,
as it provides a precise determination of the effective weak mixing 
angle $\sin^2 \theta_W^{eff}$ \cite{Yellow}. 
In particular, the most precise determination of $\sin^2 \theta_W^{eff}$
comes from the measurement of the forward-backward asymmetry in heavy
flavour production (c and b quarks) on the $Z^0$ peak 
\cite{EWWorkingGroup}. At the next generation of Linear 
Colliders it will be also possible the measurement of 
$A_{FB}^{q \bar{q}}$ for the top quark \cite{TOPexp}. 

The precision reached by the actual and aimed at future measurements 
requires,
from the theoretical counterpart, the control of the second-order 
perturbative corrections. As concerning QCD, the order 
${\mathcal O}(\alpha_S^2)$ corrections for massless quarks were calculated
numerically in \cite{Altarelli} and analytically in \cite{vanNeerven}.
For the b quarks the order ${\mathcal O}(\alpha_S^2)$ corrections were 
calculated numerically in \cite{Catani}, retaining terms that do not vanish in 
the small-mass limit (constants and logarithmically-enhanced terms), but 
neglecting both terms containing linear mass corrections, like $m_b/Q$,
and terms in which such a ratio is enhanced by a power of the logarithm 
$\log(Q/m_b)$. In order to take into 
account also this kind of terms, a full analytic calculation in which 
the mass of the heavy quark is kept systematically different from zero 
is required. 

In this paper we explore the possibility of such analytic computation, 
limiting, for the moment, the analysis to the ${\mathcal O}(\alpha_S^2)$
virtual corrections. 

The Feynman diagrams necessary for the calculation are shown in Fig. 
\ref{fig1}. We indicate with a double line the heavy quark and with 
a simple line the light quark of the diagram. The dashed line, carrying
momentum $Q$, stands for the vector or axial current.

All the scalar integrals entering the calculation of the diagrams 
(a)--(e) were evaluated analytically in 
\cite{RoPieRem1}. Among them, those corresponding to diagrams (a),
(c), (d), and (e) have been tested numerically  by the TOPSIDE collaboration
\cite{Passa}. The vector form factors, corresponding to 
the coupling of the fermion line with a photon of momentum $Q$, have 
been already calculated in \cite{RoPieRem2}, in the framework of QED.
In this work we present the analytical results for the scalar integrals 
entering the calculation of the diagrams of Fig. \ref{fig1} (f)--(l), 
for arbitrary space-like momentum transfer $Q^2$ (the continuation to 
time-like values can be carried out with the usual replacement 
$Q^2 = - (s+i\epsilon) $). We keep the mass $m$ of the 
heavier quark in the diagram as finite and we consider the lighter 
flavours as massless. The explicit values of the form factors 
will be given elsewhere. 

All the amplitudes are regularized within the continuous 
$D$-dimensional regularization 
scheme \cite{DimReg} in which both the IR and UV divergences, 
parametrized by the same parameter $D$, show up as poles in $(D-4)$. 

%%%%%%%%%%%%%%%%%%%% 2-loop Vertex ABELIAN %%%%%%%%%%%%%%%%%%%%%%%%%%%%%%%
\bfig
\bc
\subfigure[]{
\begin{fmfgraph*}(30,30)
\fmfleft{i1,i2}
\fmfright{o}
\fmf{double}{i1,v1}
\fmf{double}{i2,v2}
\fmf{dashes}{v5,o}
\fmflabel{$p_{2}$}{i1}
\fmflabel{$p_{1}$}{i2}
\fmflabel{$Q$}{o}
\fmf{double,tension=.4}{v2,v3}
\fmf{double,tension=.3}{v3,v5}
\fmf{double,tension=.4}{v1,v4}
\fmf{double,tension=.3}{v4,v5}
\fmf{gluon,tension=0}{v2,v1}
\fmf{gluon,tension=0}{v4,v3}
\end{fmfgraph*} }
%
%%%%%%%%%%%%%%%%%%%%%%%
%
%
\hspace{8mm}
\subfigure[]{
\begin{fmfgraph*}(30,30)
\fmfleft{i1,i2}
\fmfright{o}
\fmfforce{0.3w,0.6h}{v10}
\fmfforce{0.3w,0.4h}{v11}
\fmf{double}{i1,v1}
\fmf{double}{i2,v2}
\fmf{dashes}{v5,o}
\fmflabel{$p_{2}$}{i1}
\fmflabel{$p_{1}$}{i2}
\fmflabel{$Q$}{o}
\fmf{double,tension=.3}{v2,v3}
\fmf{double,tension=.3}{v3,v5}
\fmf{double,tension=.3}{v1,v4}
\fmf{double,tension=.3}{v4,v5}
\fmf{gluon,tension=0}{v2,v4}
\fmf{gluon,tension=0}{v1,v3}
\end{fmfgraph*} }
%
%%%%%%%%%%%%%%%%%%%%%%%
%
\hspace{8mm}
\subfigure[]{
\begin{fmfgraph*}(30,30)
\fmfleft{i1,i2}
\fmfright{o}
\fmfforce{0.2w,0.93h}{v2}
\fmfforce{0.2w,0.07h}{v1}
\fmfforce{0.2w,0.5h}{v3}
\fmfforce{0.8w,0.5h}{v5}
\fmfforce{0.2w,0.3h}{v10}
\fmf{double}{i1,v1}
\fmf{double}{i2,v2}
\fmf{dashes}{v5,o}
\fmflabel{$p_{2}$}{i1}
\fmflabel{$p_{1}$}{i2}
\fmflabel{$Q$}{o}
\fmf{double,tension=0}{v2,v5}
\fmf{double,tension=0}{v3,v4}
\fmf{gluon,tension=.4}{v1,v4}
\fmf{double,tension=.4}{v4,v5}
\fmf{double,tension=0}{v1,v3}
\fmf{gluon,tension=0}{v2,v3}
\end{fmfgraph*} } \\
%
%%%%%%%%%%%%%%%%%%%%%%%
%
\subfigure[]{
\begin{fmfgraph*}(30,30)
\fmfleft{i1,i2}
\fmfright{o}
\fmfforce{0.2w,0.93h}{v2}
\fmfforce{0.2w,0.07h}{v1}
\fmfforce{0.8w,0.5h}{v5}
\fmfforce{0.2w,0.4h}{v9}
\fmfforce{0.5w,0.45h}{v10}
\fmfforce{0.2w,0.5h}{v11}
\fmf{double}{i1,v1}
\fmf{double}{i2,v2}
\fmf{dashes}{v5,o}
\fmflabel{$p_{2}$}{i1}
\fmflabel{$p_{1}$}{i2}
\fmflabel{$Q$}{o}
\fmf{double}{v2,v3}
\fmf{gluon,tension=.25,left}{v3,v4}
\fmf{double,tension=.25}{v3,v4}
\fmf{double}{v4,v5}
\fmf{double}{v1,v5}
\fmf{gluon}{v1,v2}
\end{fmfgraph*} }
%
%%%%%%%%%%%%%%%%%%%%%%%
%
\hspace{8mm}
\subfigure[]{
\begin{fmfgraph*}(30,30)
\fmfleft{i1,i2}
\fmfright{o}
\fmfforce{0.2w,0.93h}{v2}
\fmfforce{0.2w,0.07h}{v1}
\fmfforce{0.2w,0.3h}{v3}
\fmfforce{0.2w,0.7h}{v4}
\fmfforce{0.8w,0.5h}{v5}
\fmf{double}{i1,v1}
\fmf{double}{i2,v2}
\fmf{dashes}{v5,o}
\fmflabel{$p_{2}$}{i1}
\fmflabel{$p_{1}$}{i2}
\fmflabel{$Q$}{o}
\fmf{double}{v2,v5}
\fmf{gluon}{v1,v3}
\fmf{gluon}{v2,v4}
\fmf{double}{v1,v5}
\fmf{double,right}{v4,v3}
\fmf{double,right}{v3,v4}
\end{fmfgraph*} }
%
%%%%%%%%%%%%%%%%%%%%%%%
%
\hspace{8mm}
\subfigure[]{
\begin{fmfgraph*}(30,30)
\fmfleft{i1,i2}
\fmfright{o}
\fmfforce{0.2w,0.93h}{v2}
\fmfforce{0.2w,0.07h}{v1}
\fmfforce{0.2w,0.3h}{v3}
\fmfforce{0.2w,0.7h}{v4}
\fmfforce{0.8w,0.5h}{v5}
\fmf{double}{i1,v1}
\fmf{double}{i2,v2}
\fmf{dashes}{v5,o}
\fmflabel{$p_{2}$}{i1}
\fmflabel{$p_{1}$}{i2}
\fmflabel{$Q$}{o}
\fmf{double}{v2,v5}
\fmf{gluon}{v1,v3}
\fmf{gluon}{v2,v4}
\fmf{double}{v1,v5}
\fmf{plain,right}{v4,v3}
\fmf{plain,right}{v3,v4}
\end{fmfgraph*} } \\
%
%%%%%%%%%%%%%%%%%%%%%%%
%
\subfigure[]{
\begin{fmfgraph*}(30,30)
\fmfleft{i1,i2}
\fmfright{o}
\fmfforce{0.2w,0.93h}{v2}
\fmfforce{0.2w,0.07h}{v1}
\fmfforce{0.2w,0.3h}{v3}
\fmfforce{0.2w,0.7h}{v4}
\fmfforce{0.8w,0.5h}{v5}
\fmf{double}{i1,v1}
\fmf{double}{i2,v2}
\fmf{dashes}{v5,o}
\fmflabel{$p_{2}$}{i1}
\fmflabel{$p_{1}$}{i2}
\fmflabel{$Q$}{o}
\fmf{double}{v2,v5}
\fmf{gluon}{v1,v3}
\fmf{gluon}{v2,v4}
\fmf{double}{v1,v5}
\fmf{gluon,left}{v3,v4}
\fmf{gluon,left}{v4,v3}
\end{fmfgraph*} } 
%
%%%%%%%%%%%%%%%%%%%%
%
\hspace{8mm}
\subfigure[]{
\begin{fmfgraph*}(30,30)
\fmfleft{i1,i2}
\fmfright{o}
\fmfforce{0.2w,0.93h}{v2}
\fmfforce{0.2w,0.07h}{v1}
\fmfforce{0.2w,0.5h}{v3}
\fmfforce{0.8w,0.5h}{v5}
\fmf{double}{i1,v1}
\fmf{double}{i2,v2}
\fmf{dashes}{v5,o}
\fmflabel{$p_{2}$}{i1}
\fmflabel{$p_{1}$}{i2}
\fmflabel{$Q$}{o}
\fmf{double,tension=0}{v1,v5}
\fmf{gluon,tension=0}{v4,v3}
\fmf{double,tension=.4}{v2,v4}
\fmf{double,tension=.4}{v4,v5}
\fmf{gluon,tension=0}{v1,v3}
\fmf{gluon,tension=0}{v2,v3}
\end{fmfgraph*}}
%
%%%%%%%%%%%%%%%%%%%%
%
\hspace{8mm}
\subfigure[]{
\begin{fmfgraph*}(30,30)
\fmfleft{i1,i2}
\fmfright{o}
\fmf{double}{i1,v1}
\fmf{double}{i2,v2}
\fmf{dashes}{v5,o}
\fmflabel{$p_{2}$}{i1}
\fmflabel{$p_{1}$}{i2}
\fmflabel{$Q$}{o}
\fmf{gluon,tension=.3}{v2,v3}
\fmf{double,tension=.3}{v3,v5}
\fmf{gluon,tension=.3}{v1,v4}
\fmf{double,tension=.3}{v4,v5}
\fmf{double,tension=0}{v2,v1}
\fmf{double,tension=0}{v4,v3}
\end{fmfgraph*} } \\
%
%%%%%%%%%%%%%%%%%%%%
%
\subfigure[]{
\begin{fmfgraph*}(30,30)
\fmfleft{i1,i2}
\fmfright{o}
\fmf{double}{i1,v1}
\fmf{double}{i2,v2}
\fmf{dashes}{v5,o}
\fmflabel{$p_{2}$}{i1}
\fmflabel{$p_{1}$}{i2}
\fmflabel{$Q$}{o}
\fmf{gluon,tension=.3}{v2,v3}
\fmf{plain,tension=.3}{v3,v5}
\fmf{gluon,tension=.3}{v1,v4}
\fmf{plain,tension=.3}{v4,v5}
\fmf{double,tension=0}{v2,v1}
\fmf{plain,tension=0}{v4,v3}
\end{fmfgraph*} }
%
%
%%%%%%%%%%%%%%%%%%%%%%%
%
\hspace{8mm}
\subfigure[]{
\begin{fmfgraph*}(30,30)
\fmfleft{i1,i2}
\fmfright{o}
\fmf{plain}{i1,v1}
\fmf{plain}{i2,v2}
\fmf{dashes}{v5,o}
\fmflabel{$p_{2}$}{i1}
\fmflabel{$p_{1}$}{i2}
\fmflabel{$Q$}{o}
\fmf{gluon,tension=.3}{v2,v3}
\fmf{double,tension=.3}{v3,v5}
\fmf{gluon,tension=.3}{v1,v4}
\fmf{double,tension=.3}{v4,v5}
\fmf{plain,tension=0}{v2,v1}
\fmf{double,tension=0}{v4,v3}
\end{fmfgraph*} }
%
%%%%%%%%%%%%%%%%%%%%
%
\hspace{8mm}
\subfigure[]{
\begin{fmfgraph*}(30,30)
\fmfleft{i1,i2}
\fmfright{o}
\fmfforce{0.2w,0.93h}{v2}
\fmfforce{0.2w,0.07h}{v1}
\fmfforce{0.2w,0.3h}{v3}
\fmfforce{0.2w,0.7h}{v4}
\fmfforce{0.8w,0.5h}{v5}
\fmf{plain}{i1,v1}
\fmf{plain}{i2,v2}
\fmf{dashes}{v5,o}
\fmflabel{$p_{2}$}{i1}
\fmflabel{$p_{1}$}{i2}
\fmflabel{$Q$}{o}
\fmf{plain}{v2,v5}
\fmf{gluon}{v1,v3}
\fmf{gluon}{v2,v4}
\fmf{plain}{v1,v5}
\fmf{double,left}{v3,v4}
\fmf{double,left}{v4,v3}
\end{fmfgraph*} } 
%
%%%%%%%%%%%%%%%%%%%%
%
\vspace*{5mm}
\caption{\label{fig1} 
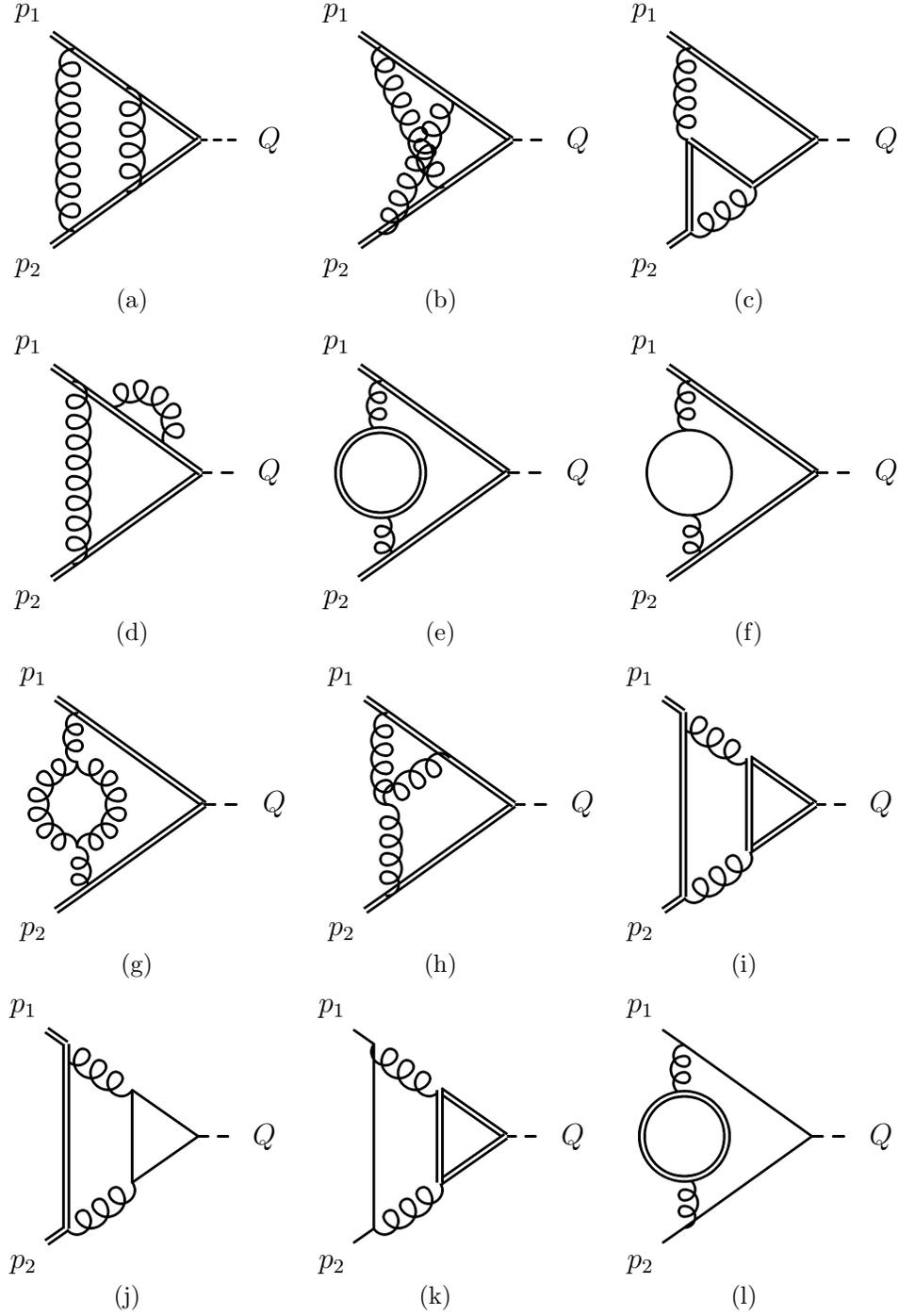
The 2-loop vertex diagrams involved in the calculation
of $A_{FB}$ at order ${\mathcal O}(\alpha_{S}^{2})$. The curly lines 
are massless gluons; the double straight lines, quarks of mass $m$;
the single straight lines, massless quarks. All the 
external fermion lines are on the mass-shell: $p_1^2 = p_2^2 = -m^{2}$, 
the double lines; $p_1^2 = p_2^2 = 0$, the single lines. The
dashed line on the r.h.s. carries  momentum 
$Q=p_{1}+p_{2}$, with the metrical convention $Q^{2}>0$ when 
$Q$ is space-like. } 
\ec
\efig
%%%%%%%%%%%%%%%%%%%%%%%%%%%%%%%%%%%%%%%%%%%%%%%%%%%%%%%%%%%%%%%%%%%%%%%%

We will follow closely the approach already used in 
\cite{RoPieRem1,RoPieRem2}. By systematic use of integration by parts 
identities (IBPs) \cite{Chet}, Lorentz invariance identities (LI) 
\cite{Rem3}, and general symmetry relations, any scalar integral 
entering the game is expressed as a linear combination of a relatively 
small number of independent scalar integrals, the so called Master 
Integrals (MIs). In the present work we are left with 35 MIs. 17 of 
them were already calculated in \cite{RoPieRem1}. 
We present here the analytical evaluation of the remaining 18, obtained
by means of the differential equations method \cite{Kot,Rem1,Rem2} or, 
when all the propagators are massless, {\it via} direct 
integration with the Feynman parameters. 
The Master Integrals are Laurent-expanded around $D=4$ and 
the coefficients of the Laurent-expansion are then expressed in terms
of 1-dimensional harmonic polylogarithms (HPLs) 
\cite{Polylog,Polylog3}. As an ``empirical'' rule, we expanded all the 
MIs up to the order in $(D-4)$ which contains HPLs with maximum weight 
$w=4$; that is expected to be sufficient in order to express the 2-loop 
form factors up to their finite part in $(D-4)$ (but higher order terms, 
when needed, could be immediately provided by our method). 

The paper is structured as follows.

In Section \ref{reduction} we recall the main steps of the reduction to 
the MIs and their calculation with the differential equations method. 
In Section \ref{results} we list the results for the MIs. In Section 
\ref{6denom} we give, for completeness, the results of the 6-denominator
vertices which are not MIs, and in Section \ref{Asympt} the large 
momentum expansion of all the 6-denominator diagrams. In Appendix 
\ref{app1} we give the routing used for the explicit calculations and 
finally, in Appendix \ref{app2}, the results for the 1-loop subdiagrams, 
which enter in the calculation.

\section{Topologies and Master Integrals \label{reduction}}

The Feynman diagrams contributing to the order 
${\mathcal O}(\alpha_S^2)$ virtual corrections to $A_{FB}^{q \bar{q}}$,
are the 2-loop vertices of Fig.~\ref{fig1}, describing the 
annihilation of a quark and an antiquark, of incoming momenta $p_1$ and 
$p_2$, into a virtual boson of outgoing momentum $Q$.

The diagrams of Fig.~\ref{fig1} may involve: {\it 1)} a single massive 
flavour, as in the case of figures (a), (b), (c), (d), (e), (g), (h), 
and (i); {\it 2)} two flavours, one massive and the other taken as 
massless, as in figures (f), (j), (k), and (l). The 
double straight lines stand for the quark (antiquark) of mass $m$, while
the single straight lines for a lighter quark (antiquark), which in our 
approximation is treated as massless. All the external fermion lines 
are on the mass-shell: $p_1^2 = p_2^2 = -m^{2}$ in 
diagrams (a)--(j) and $p_1^2 = p_2^2 = 0$ in diagrams (k) and (l). 

In some details, the forward-backward asymmetry of the $t$-quark 
gets contributions only from the diagrams 
(a)--(j), where all the other lighter flavours running in internal loops 
are considered as massless, diagrams (f) and (j). 
In the case of the $b$-quark, 
the diagrams (a)--(j) account for the cases in which the $t$-quark is 
absent from the internal fermion loops, and $b$ is the massive flavour; 
in the diagrams (k) and (l), the fermion of the internal loop is a 
$t$-quark, and the external $b$-quark is considered as 
massless. In the latter approximation, contributions 
proportional to $m_b^2/m_t^2$ are neglected. 

By using suitably projectors (see for instance Section 2 of
\cite{RoPieRem2}), the (Lorentz invariant) form factors of any 
of the vertex graphs of Fig.~\ref{fig1} can be expressed in terms 
of several (typically a few hundreds) scalar integrals, whose integrands 
are combinations of scalar products of the external and loop 
momenta divided by the denominators appearing in the propagators of 
the internal lines. 
Following \cite{RoPieRem1}, from now on we will switch our attention 
from the Feynman diagrams to their topologies -- or combinations of 
different denominators.

\subsection{Reduction to MIs}

Skipping the diagrams (a)--(e) of Fig.~\ref{fig1}, which have already  
been considered in \cite{RoPieRem1}, the topologies corresponding to 
the diagrams (f)--(l) are the 6-denominator topologies shown in 
Fig.~\ref{fig1bis}, and the two 5-denominator topologies shown in
Fig.~\ref{fig2} (g) and (h).

The following graphical conventions 
apply: internal straight and wavy lines stand for propagators of mass 
$m$ and zero respectively; the mass-shell conditions are 
$p_1^2=p_2^2=-m^2$ for external straight lines and $p_1^2=p_2^2=0$ for 
external wavy lines; the dashed line on the right carries 
(space-like) momentum $Q=p_1+p_2$ with $Q^2>0$. 
According to these conventions, wavy lines correspond both to gluons 
(or photons) and light fermions, so that the topology (g) 
of Fig.~\ref{fig2}, for instance, 
refers to both Feynman diagrams (f) and (g) of Fig.~\ref{fig1}.

%%%%%%%%%%%%%%%%%%%% 6-den amplitudes %%%%%%%%%%%%%%%%%%%%%%%%%%%%%%%
\bfig
\bc
\subfigure[]{
\begin{fmfgraph*}(20,20)
\fmfleft{i1,i2}
\fmfright{o}
\fmf{plain}{i1,v1}
\fmf{plain}{i2,v2}
\fmf{dashes}{v5,o}
\fmf{photon,tension=.3}{v2,v3}
\fmf{plain,tension=.3}{v3,v5}
\fmf{photon,tension=.3}{v1,v4}
\fmf{plain,tension=.3}{v4,v5}
\fmf{plain,tension=0}{v2,v1}
\fmf{plain,tension=0}{v4,v3}
\end{fmfgraph*} }
\hspace{5mm}
\subfigure[]{
\begin{fmfgraph*}(20,20)
\fmfleft{i1,i2}
\fmfright{o}
\fmfforce{0.2w,0.93h}{v2}
\fmfforce{0.2w,0.07h}{v1}
\fmfforce{0.2w,0.5h}{v3}
\fmfforce{0.8w,0.5h}{v5}
\fmf{plain}{i1,v1}
\fmf{plain}{i2,v2}
\fmf{dashes}{v5,o}
\fmf{plain,tension=0}{v1,v5}
\fmf{photon,tension=0}{v4,v3}
\fmf{plain,tension=.4}{v2,v4}
\fmf{plain,tension=.4}{v4,v5}
\fmf{photon,tension=0}{v1,v3}
\fmf{photon,tension=0}{v2,v3}
\end{fmfgraph*}}
%
%%%%%%%%%%%%%%%%%%%%
%
\hspace{5mm}
\subfigure[]{
\begin{fmfgraph*}(20,20)
\fmfleft{i1,i2}
\fmfright{o}
\fmf{plain}{i1,v1}
\fmf{plain}{i2,v2}
\fmf{dashes}{v5,o}
\fmf{photon,tension=.3}{v2,v3}
\fmf{photon,tension=.3}{v3,v5}
\fmf{photon,tension=.3}{v1,v4}
\fmf{photon,tension=.3}{v4,v5}
\fmf{plain,tension=0}{v2,v1}
\fmf{photon,tension=0}{v4,v3}
\end{fmfgraph*} }
%
%%%%%%%%%%%%%%%%%%%%%
%
\hspace{5mm}
\subfigure[]{
\begin{fmfgraph*}(20,20)
\fmfleft{i1,i2}
\fmfright{o}
\fmf{photon}{i1,v1}
\fmf{photon}{i2,v2}
\fmf{dashes}{v5,o}
\fmf{photon,tension=.3}{v2,v3}
\fmf{plain,tension=.3}{v3,v5}
\fmf{photon,tension=.3}{v1,v4}
\fmf{plain,tension=.3}{v4,v5}
\fmf{photon,tension=0}{v2,v1}
\fmf{plain,tension=0}{v4,v3}
\end{fmfgraph*} }
%
%%%%%%%%%%%%%%%%%%%%
\vspace*{5mm}
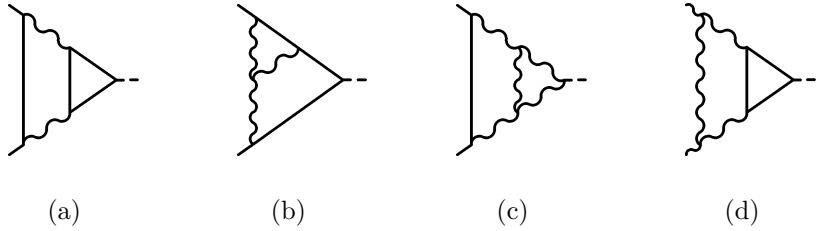
\caption{ \label{fig1bis} The four independent 6-denominator
topologies. Internal straight lines carry mass $m$, 
internal wavy lines are massless; the mass shell conditions are 
$p_1^2=p_2^2=-m^2$ for the external straight lines and $p_1^2=p_2^2=0$ 
for the external wavy lines; the dashed line on the right 
carries momentum $Q=p_1+p_2$, with $Q^2 > 0$ when $Q$ is space-like.} 
\ec
\efig
%%%%%%%%%%%%%%%%%%%%%%%%%%%%%%%%%%%%%%%%%%%%%%%%%%%%%%%%%%%%%%%%%%%%%%%%

%%%%%%%%%%%%%%%%%%%%%%%%%% 5 Denominatori
%
%%%%%%%%%%%%%%%%%%%%%%%%%% Independent diagrams %%%%%%%%%%%%%%%%%%%%%
\bfig
\bc
%%%%%%%%
\subfigure[]{
\begin{fmfgraph*}(20,20)
\fmfleft{i1,i2}
\fmfright{o}
\fmf{plain}{i1,v1}
\fmf{plain}{i2,v2}
\fmf{dashes}{v4,o}
\fmf{photon,tension=.4}{v2,v3}
\fmf{photon,tension=.2}{v3,v4}
\fmf{photon,tension=.15}{v1,v4}
\fmf{plain,tension=0}{v2,v1}
\fmf{photon,tension=0}{v1,v3}
\end{fmfgraph*} }
%
%%%%%%%%%%%%%%%%%%%%%%%
%
\subfigure[]{
\begin{fmfgraph*}(20,20)
\fmfleft{i1,i2}
\fmfright{o}
\fmf{photon}{i1,v1}
\fmf{photon}{i2,v2}
\fmf{dashes}{v4,o}
\fmf{photon,tension=.4}{v2,v3}
\fmf{plain,tension=.2}{v3,v4}
\fmf{plain,tension=.15}{v1,v4}
\fmf{photon,tension=0}{v2,v1}
\fmf{plain,tension=0}{v1,v3}
\end{fmfgraph*} }
%
%%%%%%%%%%%%%%%%%%%%%%%
%
%\hspace{3mm}
\subfigure[]{
\begin{fmfgraph*}(20,20)
\fmfleft{i1,i2}
\fmfright{o}
\fmfforce{0.2w,0.93h}{v2}
\fmfforce{0.2w,0.07h}{v1}
\fmfforce{0.2w,0.5h}{v3}
\fmfforce{0.8w,0.5h}{v4}
\fmf{plain}{i1,v1}
\fmf{plain}{i2,v2}
\fmf{dashes}{v4,o}
\fmf{photon,tension=0}{v1,v2}
\fmf{photon,tension=0}{v3,v4}
\fmf{plain,tension=0}{v2,v4}
\fmf{plain,tension=0}{v1,v4}
\end{fmfgraph*} }
%
%%%%%%%%%%%%%%%%%%%%%%%
%
%\hspace{3mm}
\subfigure[]{
\begin{fmfgraph*}(20,20)
\fmfleft{i1,i2}
\fmfright{o}
\fmf{plain}{i1,v1}
\fmf{plain}{i2,v2}
\fmf{dashes}{v4,o}
\fmf{photon,tension=.4}{v2,v3}
\fmf{photon,tension=.2}{v3,v4}
\fmf{photon,tension=.15}{v1,v4}
\fmf{plain,tension=0}{v2,v1}
\fmf{photon,tension=0,right=.6}{v4,v3}
\end{fmfgraph*} }
%
%%%%%%%%%%%%%%%%%%%%%%%
%
%\hspace{3mm}
\subfigure[]{
\begin{fmfgraph*}(20,20)
\fmfleft{i1,i2}
\fmfright{o}
\fmf{plain}{i1,v1}
\fmf{plain}{i2,v2}
\fmf{dashes}{v4,o}
\fmf{photon,tension=.4}{v2,v3}
\fmf{plain,tension=.2}{v3,v4}
\fmf{photon,tension=.15}{v1,v4}
\fmf{plain,tension=0}{v2,v1}
\fmf{plain,tension=0,right=.6}{v4,v3}
\end{fmfgraph*} }  \\
%
%%%%%%%%%%%%%%%%%%%%%%%
%
\subfigure[]{
\begin{fmfgraph*}(20,20)
\fmfleft{i1,i2}
\fmfright{o}
\fmf{photon}{i1,v1}
\fmf{photon}{i2,v2}
\fmf{dashes}{v4,o}
\fmf{photon,tension=.4}{v2,v3}
\fmf{plain,tension=.2}{v3,v4}
\fmf{photon,tension=.15}{v1,v4}
\fmf{photon,tension=0}{v2,v1}
\fmf{plain,tension=0,right=.6}{v4,v3}
\end{fmfgraph*} }
%
%%%%%%%%%%%%%%%%%%%%%%%
%
\subfigure[]{
\begin{fmfgraph*}(20,20)
\fmfleft{i1,i2}
\fmfright{o}
\fmfforce{0.2w,0.93h}{v2}
\fmfforce{0.2w,0.07h}{v1}
\fmfforce{0.2w,0.55h}{v3}
\fmfforce{0.2w,0.13h}{v5}
\fmfforce{0.8w,0.5h}{v4}
\fmf{plain}{i1,v1}
\fmf{plain}{i2,v2}
\fmf{dashes}{v4,o}
\fmf{photon}{v2,v3}
\fmf{photon,right=.5}{v3,v1}
\fmf{photon,right=.5}{v1,v3}
\fmf{plain}{v1,v4}
\fmf{plain}{v2,v4}
\end{fmfgraph*} }
% 
%%%%%%%%%%%%%%%%%%%%%%%
%
\subfigure[]{
\begin{fmfgraph*}(20,20)
\fmfleft{i1,i2}
\fmfright{o}
\fmfforce{0.2w,0.93h}{v2}
\fmfforce{0.2w,0.07h}{v1}
\fmfforce{0.2w,0.55h}{v3}
\fmfforce{0.2w,0.13h}{v5}
\fmfforce{0.8w,0.5h}{v4}
\fmf{photon}{i1,v1}
\fmf{photon}{i2,v2}
\fmf{dashes}{v4,o}
\fmf{photon}{v2,v3}
\fmf{plain,right=.5}{v3,v1}
\fmf{plain,right=.5}{v1,v3}
\fmf{photon}{v1,v4}
\fmf{photon}{v2,v4}
\end{fmfgraph*} }
% 
%%%%%%%%%%%%%%%%%%%%%%%
%
%\hspace{3mm}
\subfigure[]{
\begin{fmfgraph*}(20,20)
\fmfforce{0.5w,0.2h}{v3}
\fmfforce{0.5w,0.8h}{v2}
\fmfforce{0.2w,0.5h}{v1}
\fmfforce{0.8w,0.5h}{v4}
\fmfleft{i}
\fmfright{o}
\fmf{dashes}{i,v1}
\fmf{dashes}{v4,o}
\fmf{photon,left=.4}{v1,v2}
\fmf{photon,right=.4}{v1,v3}
\fmf{plain,left=.4}{v2,v4}
\fmf{plain,right=.4}{v3,v4}
\fmf{plain}{v2,v3}
\end{fmfgraph*} }
%
%%%%%%%%%%%%%%%%%%%%%%%
%
%\hspace{3mm}
\subfigure[]{
\begin{fmfgraph*}(20,20)
\fmfforce{0.5w,0.2h}{v3}
\fmfforce{0.5w,0.8h}{v2}
\fmfforce{0.2w,0.5h}{v1}
\fmfforce{0.8w,0.5h}{v4}
\fmfleft{i}
\fmfright{o}
\fmf{dashes}{i,v1}
\fmf{dashes}{v4,o}
\fmf{photon,left=.4}{v1,v2}
\fmf{photon,right=.4}{v1,v3}
\fmf{photon,left=.4}{v2,v4}
\fmf{photon,right=.4}{v3,v4}
\fmf{photon}{v2,v3}
\end{fmfgraph*} }  \\
%
%%%%%%%%%%%%%%%%%%%%%%%
%
%\hspace{3mm}
\subfigure[]{
\begin{fmfgraph*}(20,20)
\fmfforce{0.5w,0.2h}{v3}
\fmfforce{0.5w,0.8h}{v2}
\fmfforce{0.2w,0.5h}{v1}
\fmfforce{0.8w,0.5h}{v4}
\fmfleft{i}
\fmfright{o}
\fmf{plain}{i,v1}
\fmf{plain}{v4,o}
\fmf{photon,left=.4}{v1,v2}
\fmf{photon,left=.4}{v2,v4}
\fmf{plain,right=.4}{v1,v3}
\fmf{plain,right=.4}{v3,v4}
\fmf{photon}{v2,v3}
\end{fmfgraph*} }
%%%%%%%%%%%%%%%%%%%%%%%
%
%\hspace{3mm}
\subfigure[]{
\begin{fmfgraph*}(20,20)
\fmfleft{i1,i2}
\fmfright{o}
\fmf{plain}{i1,v1}
\fmf{plain}{i2,v2}
\fmf{dashes}{v4,o}
\fmf{photon,tension=.3}{v2,v3}
\fmf{photon,tension=.3}{v1,v3}
\fmf{plain,tension=0}{v2,v1}
\fmf{photon,tension=.2,left}{v3,v4}
\fmf{photon,tension=.2,right}{v3,v4}
\end{fmfgraph*} }
%
%%%%%%%%%%%%%%%%%%%%%%%%
%
\subfigure[]{
\begin{fmfgraph*}(20,20)
\fmfleft{i1,i2}
\fmfright{o}
\fmf{photon}{i1,v1}
\fmf{photon}{i2,v2}
\fmf{dashes}{v4,o}
\fmf{photon,tension=.3}{v2,v3}
\fmf{photon,tension=.3}{v1,v3}
\fmf{photon,tension=0}{v2,v1}
\fmf{plain,tension=.2,left}{v3,v4}
\fmf{plain,tension=.2,right}{v3,v4}
\end{fmfgraph*} }
%
%%%%%%%%%%%%%%%%%%%%%%%%
%
%\hspace{5mm}
\subfigure[]{
\begin{fmfgraph*}(20,20)
\fmfleft{i1,i2}
\fmfright{o}
\fmf{plain}{i1,v1}
\fmf{plain}{i2,v2}
\fmf{dashes}{v4,o}
\fmf{photon,tension=.3}{v2,v3}
\fmf{photon,tension=.3}{v1,v3}
\fmf{plain,tension=0}{v2,v1}
\fmf{plain,tension=.2,left}{v3,v4}
\fmf{plain,tension=.2,right}{v3,v4}
\end{fmfgraph*} }
%
%%%%%%%%%%%%%%%%%%%%%%%%
%\hspace{5mm}
\subfigure[]{
\begin{fmfgraph*}(30,20)
\fmfleft{i1,i2}
\fmfright{o}
\fmf{plain}{i1,v1}
\fmf{plain}{i2,v2}
\fmf{plain}{v7,o}
\fmf{plain,tension=.3}{v2,v3}
\fmf{plain,tension=.3}{v1,v3}
\fmf{photon,tension=0}{v2,v1}
\fmf{dashes}{v3,v4}
\fmf{phantom}{v4,v5}
\fmf{plain}{v5,v6}
\fmf{photon,tension=.2,left}{v6,v7}
\fmf{plain,tension=.2,right}{v6,v7}
\end{fmfgraph*} }
%%%%%%%%%%%%%%%%%%%%%%%%%%%%%

\caption{\label{fig2} 
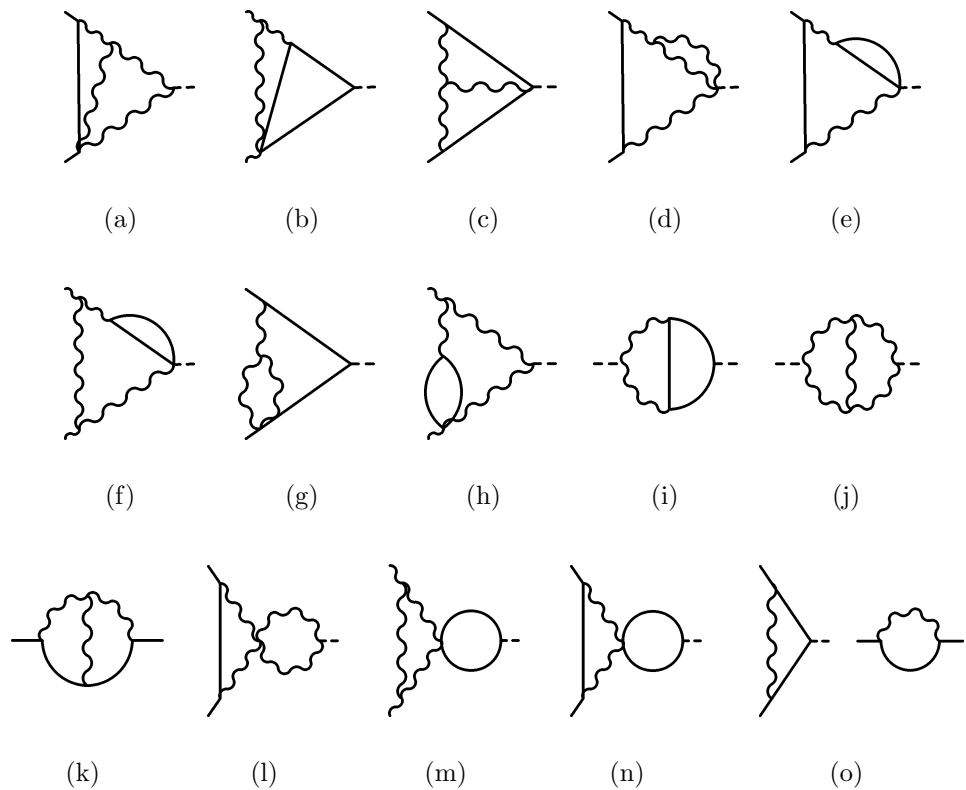
The set of the 15 independent 5-denominator topologies. 
The graphical conventions are the same as in Fig.~\ref{fig1bis}. } 
\ec
\efig
%%%%%%%%%%%%%%%%%%%%%%%%%%%%%%%%%%%%%%%%%%%%%%%%%%%%%%%%%%%%%%%%%%%%%

%%%%%%%%%%%%%%%%%%%%%%%%%% 4 Denominatori
%
%%%%%%%%%%%%%%%%%%%%%%%%%% Independent diagrams %%%%%%%%%%%%%%%%%%%%%
\bfig
\bc
%%%%%%%%%%%%%%%%%%%%%%%
\subfigure[]{
\begin{fmfgraph*}(20,20)
\fmfleft{i1,i2}
\fmfright{o}
\fmf{plain}{i1,v1}
\fmf{plain}{i2,v2}
\fmf{dashes}{v3,o}
\fmf{photon,tension=.3}{v2,v3}
\fmf{photon,tension=.3}{v1,v3}
\fmf{photon,tension=0,left=.5}{v2,v1}
\fmf{plain,tension=0,left=.5}{v1,v2}
\end{fmfgraph*} }
%
%%%%%%%%%%%%%%%%%%%%%%%
%
\subfigure[]{
\begin{fmfgraph*}(20,20)
\fmfleft{i1,i2}
\fmfright{o}
\fmf{photon}{i1,v1}
\fmf{photon}{i2,v2}
\fmf{dashes}{v3,o}
\fmf{plain,tension=.3}{v2,v3}
\fmf{plain,tension=.3}{v1,v3}
\fmf{photon,tension=0,left=.5}{v1,v2}
\fmf{plain,tension=0,left=.5}{v2,v1}
\end{fmfgraph*} }
%
%%%%%%%%%%%%%%%%%%%%%%%
%
\subfigure[]{
\begin{fmfgraph*}(20,20)
\fmfleft{i1,i2}
\fmfright{o}
\fmf{photon}{i1,v1}
\fmf{photon}{i2,v2}
\fmf{dashes}{v3,o}
\fmf{photon,tension=.3}{v2,v3}
\fmf{photon,tension=.3}{v1,v3}
\fmf{plain,tension=0,left=.5}{v2,v1}
\fmf{plain,tension=0,left=.5}{v1,v2}
\end{fmfgraph*} }
%
%%%%%%%%%%%%%%%%%%%%%%%
%
\subfigure[]{
\begin{fmfgraph*}(20,20)
\fmfleft{i1,i2}
\fmfright{o}
\fmf{plain}{i1,v1}
\fmf{plain}{i2,v2}
\fmf{dashes}{v3,o}
\fmf{photon,tension=.3}{v2,v3}
\fmf{photon,tension=.3}{v1,v3}
\fmf{plain,tension=0}{v2,v1}
\fmf{photon,tension=0,left=.5}{v2,v3}
\end{fmfgraph*} }
%
%%%%%%%%%%%%%%%%%%%%%%%
%
\subfigure[]{
\begin{fmfgraph*}(20,20)
\fmfleft{i1,i2}
\fmfright{o}
\fmf{photon}{i1,v1}
\fmf{photon}{i2,v2}
\fmf{dashes}{v3,o}
\fmf{plain,tension=.3}{v2,v3}
\fmf{photon,tension=.3}{v1,v3}
\fmf{photon,tension=0}{v2,v1}
\fmf{plain,tension=0,left=.5}{v2,v3}
\end{fmfgraph*} } \\
%
%%%%%%%%%%%%%%%%%%%%%%%
%
\subfigure[]{
\begin{fmfgraph*}(20,20)
\fmfforce{0.5w,0.2h}{v3}
\fmfforce{0.5w,0.8h}{v2}
\fmfforce{0.2w,0.5h}{v1}
\fmfforce{0.8w,0.5h}{v4}
\fmfleft{i}
\fmfright{o}
\fmf{dashes}{i,v1}
\fmf{dashes}{v4,o}
\fmf{photon,left=.4}{v1,v2}
\fmf{photon,right=.4}{v1,v3}
\fmf{photon,right=.4}{v3,v4}
\fmf{photon,left=.4}{v2,v4}
\fmf{photon,left=.6}{v3,v4}
\end{fmfgraph*} }
%
%%%%%%%%%%%%%%%%%%%%%%%
%
\subfigure[]{
\begin{fmfgraph*}(20,20)
\fmfforce{0.5w,0.2h}{v3}
\fmfforce{0.5w,0.8h}{v2}
\fmfforce{0.2w,0.5h}{v1}
\fmfforce{0.8w,0.5h}{v4}
\fmfleft{i}
\fmfright{o}
\fmf{dashes}{i,v1}
\fmf{dashes}{v4,o}
\fmf{photon,left=.4}{v1,v2}
\fmf{photon,right=.4}{v1,v3}
\fmf{photon,left=.4}{v2,v4}
\fmf{plain,left=.6}{v3,v4}
\fmf{plain,right=.4}{v3,v4}
\end{fmfgraph*} } 
%
%%%%%%%%%%%%%%%%%%%%%%%
%
\subfigure[]{
\begin{fmfgraph*}(20,20)
\fmfforce{0.5w,0.2h}{v3}
\fmfforce{0.5w,0.8h}{v2}
\fmfforce{0.2w,0.5h}{v1}
\fmfforce{0.8w,0.5h}{v4}
\fmfleft{i}
\fmfright{o}
\fmf{dashes}{i,v1}
\fmf{dashes}{v4,o}
\fmf{plain,left=.4}{v1,v2}
\fmf{plain,right=.4}{v1,v3}
\fmf{plain,left=.4}{v2,v4}
\fmf{photon,left=.6}{v3,v4}
\fmf{plain,right=.4}{v3,v4}
\end{fmfgraph*} } 
%
%%%%%%%%%%%%%%%%%%%%%%%
%
\subfigure[]{
\begin{fmfgraph*}(20,20)
\fmfforce{0.5w,0.2h}{v3}
\fmfforce{0.5w,0.8h}{v2}
\fmfforce{0.2w,0.5h}{v1}
\fmfforce{0.8w,0.5h}{v4}
\fmfleft{i}
\fmfright{o}
\fmf{photon}{i,v1}
\fmf{photon}{v4,o}
\fmf{photon,left=.4}{v1,v2}
\fmf{photon,right=.4}{v1,v3}
\fmf{photon,left=.4}{v2,v4}
\fmf{plain,left=.6}{v3,v4}
\fmf{plain,right=.4}{v3,v4}
\end{fmfgraph*} } 
%
%%%%%%%%%%%%%%%%%%%%%%%
%
\subfigure[]{
\begin{fmfgraph*}(20,20)
\fmfforce{0.5w,0.2h}{v3}
\fmfforce{0.5w,0.8h}{v2}
\fmfforce{0.2w,0.5h}{v1}
\fmfforce{0.8w,0.5h}{v4}
\fmfleft{i}
\fmfright{o}
\fmf{plain}{i,v1}
\fmf{plain}{v4,o}
\fmf{plain,left=.4}{v1,v2}
\fmf{photon,right=.4}{v1,v3}
\fmf{plain,left=.4}{v2,v4}
\fmf{photon,left=.6}{v3,v4}
\fmf{photon,right=.4}{v3,v4}
\end{fmfgraph*} }  \\
%
%%%%%%%%%%%%%%%%%%%%%%%
%
\subfigure[]{
\begin{fmfgraph*}(20,20)
\fmfleft{i1,i2}
\fmfright{o}
\fmf{plain}{i1,v1}
\fmf{plain}{i2,v2}
\fmf{dashes}{v3,o}
\fmf{photon,tension=.3}{v2,v3}
\fmf{photon,tension=.3}{v1,v3}
\fmf{plain,tension=0}{v2,v1}
\fmf{plain,right=45}{v3,v3}
\end{fmfgraph*} }
%
%%%%%%%%%%%%%%%%%%%%%%%
%
\subfigure[]{
\begin{fmfgraph*}(20,20)
\fmfleft{i1,i2}
\fmfright{o}
\fmf{photon}{i1,v1}
\fmf{photon}{i2,v2}
\fmf{dashes}{v3,o}
\fmf{photon,tension=.3}{v2,v3}
\fmf{photon,tension=.3}{v1,v3}
\fmf{photon,tension=0}{v2,v1}
\fmf{plain,right=45}{v3,v3}
\end{fmfgraph*} }
%
%%%%%%%%%%%%%%%%%%%%%%%
%
%\hspace{3mm}
\subfigure[]{
\begin{fmfgraph*}(20,20)
\fmfleft{i}
\fmfright{o}
\fmf{dashes}{i,v1}
\fmf{dashes}{v3,o}
\fmf{photon,tension=.15,left}{v1,v2}
\fmf{photon,tension=.15,right}{v1,v2}
\fmf{plain,tension=.15,left}{v2,v3}
\fmf{plain,tension=.15,right}{v2,v3}
\end{fmfgraph*} } 
%
%%%%%%%%%%%%%%%%%%%%%%%
%
%\hspace{3mm}
\subfigure[]{
\begin{fmfgraph*}(20,20)
\fmfleft{i}
\fmfright{o}
\fmf{dashes}{i,v1}
\fmf{dashes}{v3,o}
\fmf{photon,tension=.15,left}{v1,v2}
\fmf{photon,tension=.15,right}{v1,v2}
\fmf{photon,tension=.15,left}{v2,v3}
\fmf{photon,tension=.15,right}{v2,v3}
\end{fmfgraph*} }
%
%%%%%%%%%%%%%%%%%%%%%%%
%
%\hspace{3mm}
\subfigure[]{
\begin{fmfgraph*}(30,20)
\fmfleft{i}
\fmfright{o}
\fmf{plain}{i,v1}
\fmf{photon,tension=.15,left}{v1,v2}
\fmf{plain,tension=.15,right}{v1,v2}
\fmf{plain}{v2,v3}
\fmf{phantom}{v3,v4}
\fmf{dashes}{v4,v5}
\fmf{photon,tension=.15,left}{v5,v6}
\fmf{photon,tension=.15,right}{v5,v6}
\fmf{dashes}{v6,o}
\end{fmfgraph*} } 
%%%%%%%%%%%%%%%%%%%%%%%
\caption{\label{fig3} 
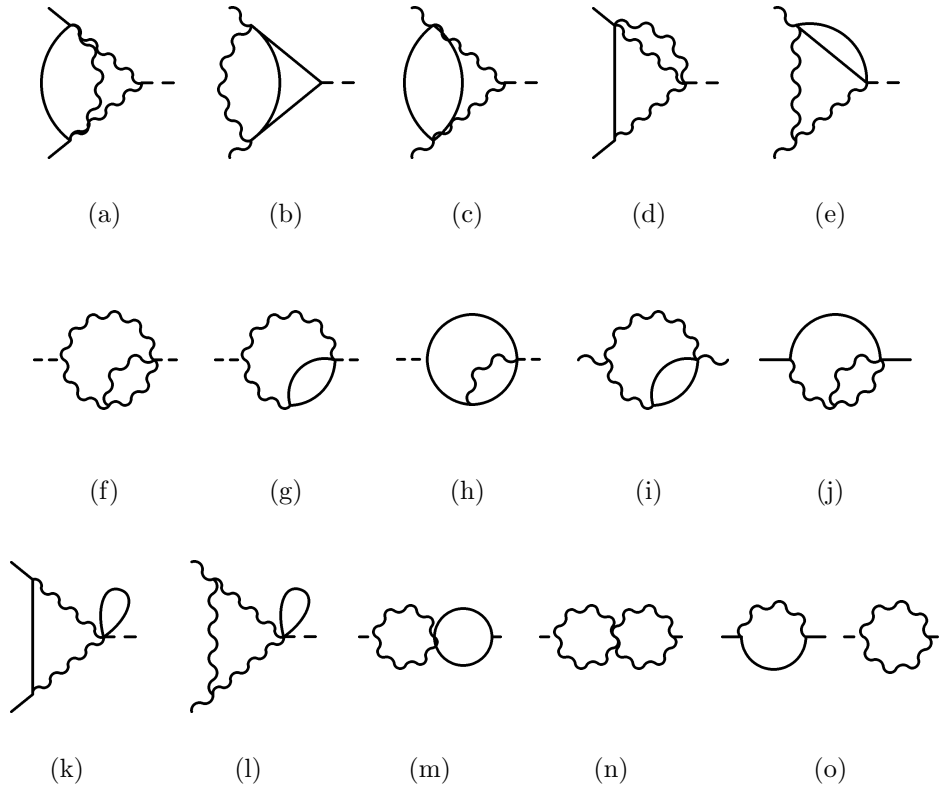
The set of the 15 independent 4-denominator topologies. 
The graphical conventions are the same as in Fig.~\ref{fig1bis}. } 
\ec
\efig
%%%%%%%%%%%%%%%%%%%%%%%%%%%%%%%%%%%%%%%%%%%%%%%%%%%%%%%%%%%%%%%%%%%%%

%%%%%%%%% 3 Denominatori
%
%%%%%%%%%%%%%%%%%%%%%%%%%% Independent diagrams %%%%%%%%%%%%%%%%%%%%%
\bfig
\bc
%%%%%%%%
\subfigure[]{
\begin{fmfgraph*}(20,20)
\fmfleft{i}
\fmfright{o}
\fmf{dashes}{i,v1}
\fmf{dashes}{v2,o}
\fmf{photon,tension=.15,left}{v1,v2}
\fmf{photon,tension=.15}{v1,v2}
\fmf{photon,tension=.15,right}{v1,v2}
\end{fmfgraph*} } 
%
%%%%%%%%%%%%%%%%%%%%%%%
%
\subfigure[]{
\begin{fmfgraph*}(20,20)
\fmfleft{i}
\fmfright{o}
\fmf{photon}{i,v1}
\fmf{photon}{v2,o}
\fmf{plain,tension=.15,left}{v1,v2}
\fmf{photon,tension=.15}{v1,v2}
\fmf{plain,tension=.15,right}{v1,v2}
\end{fmfgraph*} } 
%
%%%%%%%%%%%%%%%%%%%%%%%
%
\subfigure[]{
\begin{fmfgraph*}(20,20)
\fmfleft{i}
\fmfright{o}
\fmf{dashes}{i,v1}
\fmf{dashes}{v2,o}
\fmf{photon,tension=.22,left}{v1,v2}
\fmf{photon,tension=.22,right}{v1,v2}
\fmf{plain,right=45}{v2,v2}
\end{fmfgraph*} } 
%
%%%%%%%%%%%%%%%%%%%%%%%% 
%%%%%%%%%%%%%%%%%%%%%%%% 
\caption{\label{fig4} 
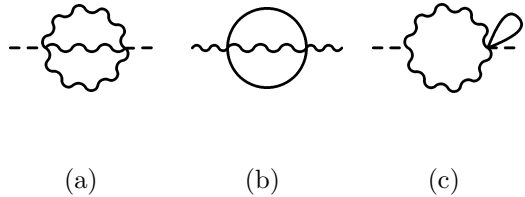
The set of the 2 independent 3-denominator topologies. 
The graphical conventions are the same as in Fig.~\ref{fig1bis}. } 
\ec
\efig
%%%%%%%%%%%%%%%%%%%%%%%%%%%%%%%%%%%%%%%%%%%%%%%%%%%%%%%%%%%%%%%%%%%%%

The tree of the subtopologies, generated top-down by removing the 
denominators one by one from the four topologies in Fig.~\ref{fig1bis} 
and the two topologies in Fig.~\ref{fig2} (g) and (h), overlaps partially
with the set of subtopologies already considered in \cite{RoPieRem1}. 
The new independent subtopologies are the 15 5-denominator ones
of Fig.~\ref{fig2}, the 15 4-denominator topologies 
shown in Fig.~\ref{fig3} and, finally, the 3 3-denominator 
topologies of Fig.~\ref{fig4}. The only non-trivial topology with 2 
denominators is the product of two tadpoles, already considered in 
\cite{RoPieRem1}.

As already said, the whole set of the scalar integrals, belonging to 
all the above topologies, is 
regularized within the $D$-continuous dimensional regularization
scheme, in which both IR and UV divergences show up as poles in $(D-4)$.

As already discussed at length in \cite{RoPieRem1}, thanks to the 
continuous dimensional regularization one can easily write several 
identities among the scalar integrals associated to a given topology 
(and its subtopologies), by using mainly the Integration by Parts 
method of \cite{Chet}, but also the Lorentz invariance identities 
\cite{Rem3} and the symmetry relations, that can occur in particular mass 
configurations. The identities can then be solved by standard techniques 
(Gauss substitution rule), whose implementation is, however, algebraically 
very demanding. Referring again for more details to \cite{RoPieRem1}, 
one can express in this way all the scalar integrals associated to the 
considered topologies as linear combinations of a small number of 
integrals, called the Master Integrals (MIs) for that topology. 

In our case, we find 18 new MIs, shown as ``decorated graphs'' in 
Fig.~\ref{fig6}. According to our graphical notations, a simple dot on a
propagator line, like in figures (e) and (k), means that, in the 
integrand of the concerned MI, the corresponding denominator is squared; 
a dot labeled by the number ``3'', like in figure (h), means that the 
corresponding denominator is raised to the third power; an explicitly 
written scalar product, like in (f) and (i), means that the 
corresponding integrand has that scalar product in the numerator.

\subsection{The differential equations} 

The calculation of the MIs is performed by means of the differential
equations method \cite{Kot,Rem1,Rem2}.

In our case, the MIs are functions of the squared momentum transfer $Q^2$, 
of the mass $m$ of the heavy quark, and of the dimension parameter $D$. 
According to~\cite{Rem1,Rem2}, the $h$ MIs of 
a given topology $M_i(D,m^2,Q^2),\ $ $i=1,..,h$, 
satisfy a system of $h$ coupled first-order linear 
differential equations in $Q^2$ of the form:
\bea
\frac{d}{d Q^2} M_{i}(D,m^2,Q^2) & = & 
\sum_{j=1}^{h} A_{ij}(D,m^2,Q^2) M_j(D,m^2,Q^2) \nn\\
& & \qquad \qquad + \sum_{ik} B_k(D,m^2,Q^2) N_k(D,m^2,Q^2) \ , 
\label{diffeq}
\eea
where the $N_k(D,m^2,Q^2)$ are MIs of the sub-topologies 
and the coefficients $A_{ij}(D,m^2,Q^2),\ $ 
$B_{ik}(D,m^2,Q^2)$ are ratios of polynomials in $D, m^2$ and $Q^2$. 

Obviously, the equations for the $M_i(D,m^2,Q^2)$ are not homogeneous, 
due to the presence of the $N_k(D,m^2,Q^2)$ corresponding to the 
sub-topologies. It is therefore natural to proceed bottom-up in the 
solution of the equations for the whole set of MIs, starting 
from the equations for the MIs of the simplest topologies and using 
their solutions within the equations for the MIs of the more complicated 
ones, whose non-homogeneous part can then be considered as known. 

%%% Solution in Laurent series of (D-4).

We search for a solution of the system (\ref{diffeq}) in Laurent
series of $(D-4)$:
\be
M_i(D,m^2,Q^2) = \sum_{j=-2}^{n} \, (D-4)^{j} M_i^{(j)}(m^2,Q^2) 
+ {\mathcal O} \bigl( (D-4)^{(n+1)} \bigr) \ , 
\ee
where $n$ is the required order in $(D-4)$. The solution of the system
proceeds order-by-order in $(D-4)$, as explained in \cite{RoPieRem1}; 
one obtains a set of chained systems, one for each power of $(D-4)$, 
all with the same homogeneous parts. 

In our case, 10 topologies out of 13 have a single MI, which means that 
they satisfy a single first-order linear differential equation;
the topology (c) of Fig.~\ref{fig3} has the two MIs of 
Fig.~\ref{fig6} (j)--(k), and finally 
the topologies (a) and (b) of Fig.~\ref{fig3} both have three MIs, shown 
in Fig.~\ref{fig6} (d)--(f) and (g)--(i). 

When the MIs for a given topology are two or more, the actual choice 
of the scalar integrals to be chosen as MIs, which is in principle 
arbitrary, can be of great help for simplifying the problem. It turns 
out, indeed, that with the choice of the three MIs shown in 
Fig.~\ref{fig6} (d)--(f), the system expanded in $(D-4)$ decouples into 
a two-by-two system for the two MIs (d), (f) and an equation involving 
the MI (e) only. The same happens for the three MIs corresponding to 
Fig.~\ref{fig6} (g)--(i). Likewise, the system for the two MIs (j) and 
(k) of Fig.~\ref{fig6} decouples, when expanded in $(D-4)$, in two 
first-order linear differential equations.

Following again \cite{Rem2,RoPieRem1}, the two-by-two first-order 
systems are transformed in the equivalent single equations of the second 
order, and all the resulting first and second-order single equations 
are solved by using Euler's method of the variation of the constants. 
The method requires the explicit knowledge of the solutions of the 
associated homogeneous equations; as in previous work, the solutions 
of all the homogeneous equations were simple algebraic functions, 
found immediately by inspecting the equations, so that we will not 
report them here too. 

As a last remark, the boundary conditions for the differential equations
were found by exploiting the known analytical properties of the MIs 
under consideration, imposing, typically, the regularity or the 
finiteness of the solution at the pseudo-thresholds of the MI. This 
qualitative information was completely sufficient for the quantitative 
determination of the otherwise arbitrary integration constants, which 
naturally arise when solving a system of differential equations. 

The equations for the MIs of the massless 2-loop sunrise, 
Fig.~\ref{fig6} (q), and for the MIs of the massless 1-loop bubbles of 
Fig.~\ref{fig6} (m), (o), (p) and (r) are entirely homogeneous, 
so that only the scaling dependence of the MIs on $Q^2$ could 
be derived by solving the differential equations. In those cases the 
complete result was easily obtained by direct integration with the 
Feynman parameters.

\section{Results for the MIs \label{results}}

%%%%%%%%% Master Integrals 2 %%%%%%%%%%%%%%%%%%%%%
\bfig
\bc
%%%%%%%%%%%%%%%%%%%%
%
\subfigure[]{
\begin{fmfgraph*}(20,20)
\fmfleft{i1,i2}
\fmfright{o}
\fmf{plain}{i1,v1}
\fmf{plain}{i2,v2}
\fmf{dashes}{v5,o}
\fmf{photon,tension=.3}{v2,v3}
\fmf{plain,tension=.3}{v3,v5}
\fmf{photon,tension=.3}{v1,v4}
\fmf{plain,tension=.3}{v4,v5}
\fmf{plain,tension=0}{v2,v1}
\fmf{plain,tension=0}{v4,v3}
\end{fmfgraph*} }
%
%%%%%%%%%%%%%%%%%%%%
%
%
%\hspace{10mm}
\subfigure[]{
\begin{fmfgraph*}(20,20)
\fmfforce{0.5w,0.2h}{v3}
\fmfforce{0.5w,0.8h}{v2}
\fmfforce{0.2w,0.5h}{v1}
\fmfforce{0.8w,0.5h}{v4}
\fmfleft{i}
\fmfright{o}
\fmf{dashes}{i,v1}
\fmf{dashes}{v4,o}
\fmf{photon,left=.4}{v1,v2}
\fmf{photon,right=.4}{v1,v3}
\fmf{plain,left=.4}{v2,v4}
\fmf{plain,right=.4}{v3,v4}
\fmf{plain}{v2,v3}
\end{fmfgraph*} } 
%
%%%%%%%%%%%%%%%%%%%%%%%
% 
%\hspace{10mm}
\subfigure[]{
\begin{fmfgraph}(20,20)
\fmfleft{i1,i2}
\fmfright{o}
\fmf{plain}{i1,v1}
\fmf{plain}{i2,v2}
\fmf{dashes}{v3,o}
\fmf{photon,tension=.3}{v2,v3}
\fmf{photon,tension=.3}{v1,v3}
\fmf{plain,tension=0}{v2,v1}
\fmf{photon,tension=0,left=.5}{v2,v3}
\end{fmfgraph} }
%
%%%%%%%%%%%%%%%%%%%%%%%
% 
%\hspace{10mm}
\subfigure[]{
\begin{fmfgraph*}(20,20)
\fmfleft{i1,i2}
\fmfright{o}
\fmf{plain}{i1,v1}
\fmf{plain}{i2,v2}
\fmf{dashes}{v3,o}
\fmf{photon,tension=.3}{v2,v3}
\fmf{photon,tension=.3}{v1,v3}
\fmf{photon,tension=0,left=.5}{v2,v1}
\fmf{plain,tension=0,left=.5}{v1,v2}
\end{fmfgraph*} }
%
%%%%%%%%%%%%%%%%%%%%%%%
%
%\hspace*{14mm}
\subfigure[]{
\begin{fmfgraph*}(20,20)
\fmfleft{i1,i2}
\fmfright{o}
\fmfforce{.4w,.5h}{d1}
%\fmfdot{d1}
\fmf{plain}{i1,v1}
\fmf{plain}{i2,v2}
\fmf{dashes}{v3,o}
\fmf{photon,tension=.3}{v2,v3}
\fmf{photon,tension=.3}{v1,v3}
\fmf{photon,tension=0,left=.5}{v2,v1}
\fmf{plain,tension=0,left=.5}{v1,v2}
\fmfv{decor.shape=circle,decor.filled=full,decor.size=.1w}{d1}
\end{fmfgraph*} }  \\
%
%%%%%%%%%%%%%%%%%%%%%%%
%
\hspace{-20mm}
\subfigure[]{
\begin{fmfgraph*}(20,20)
\fmfleft{i1,i2}
\fmfright{o}
\fmf{plain}{i1,v1}
\fmf{plain}{i2,v2}
\fmf{dashes}{v3,o}
\fmf{photon,tension=.3}{v2,v3}
\fmf{photon,tension=.3}{v1,v3}
\fmf{photon,tension=0,left=.5}{v2,v1}
\fmf{plain,tension=0,left=.5}{v1,v2}
\fmflabel{$(p_{1} \cdot k_{1})$ }{o}
\end{fmfgraph*} }
%
%%%%%%%%%%%%%%%%%%%%%%%
%
\hspace*{14mm}
\subfigure[]{
\begin{fmfgraph*}(20,20)
\fmfleft{i1,i2}
\fmfright{o}
\fmf{photon}{i1,v1}
\fmf{photon}{i2,v2}
\fmf{dashes}{v3,o}
\fmf{plain,tension=.3}{v2,v3}
\fmf{plain,tension=.3}{v1,v3}
\fmf{plain,tension=0,left=.5}{v2,v1}
\fmf{photon,tension=0,left=.5}{v1,v2}
\end{fmfgraph*} }
%
%%%%%%%%%%%%%%%%%%%%%%%
%
%\hspace*{14mm}
\subfigure[]{
\begin{fmfgraph*}(20,20)
\fmfleft{i1,i2}
\fmfright{o}
\fmfforce{.43w,.5h}{d1}
%\fmfdot{d1}
\fmf{photon}{i1,v1}
\fmf{photon}{i2,v2}
\fmf{dashes}{v3,o}
\fmf{plain,tension=.3}{v2,v3}
\fmf{plain,tension=.3}{v1,v3}
\fmf{plain,tension=0,left=.5}{v2,v1}
\fmf{photon,tension=0,left=.5}{v1,v2}
\fmfv{decor.shape=circle,decor.filled=full,decor.size=.1w}{d1}
\fmflabel{3}{d1}
\end{fmfgraph*} } 
%
%%%%%%%%%%%%%%%%%%%%%%%
%
\subfigure[]{
\begin{fmfgraph*}(20,20)
\fmfleft{i1,i2}
\fmfright{o}
\fmf{photon}{i1,v1}
\fmf{photon}{i2,v2}
\fmf{dashes}{v3,o}
\fmf{plain,tension=.3}{v2,v3}
\fmf{plain,tension=.3}{v1,v3}
\fmf{plain,tension=0,left=.5}{v2,v1}
\fmf{photon,tension=0,left=.5}{v1,v2}
\fmflabel{$(k_{1} \cdot k_{2})$ }{o}
\end{fmfgraph*} } \\
%
%%%%%%%%%%%%%%%%%%%%%%%
%
\subfigure[]{
\begin{fmfgraph*}(20,20)
\fmfleft{i1,i2}
\fmfright{o}
\fmf{photon}{i1,v1}
\fmf{photon}{i2,v2}
\fmf{dashes}{v3,o}
\fmf{photon,tension=.3}{v2,v3}
\fmf{photon,tension=.3}{v1,v3}
\fmf{plain,tension=0,left=.5}{v2,v1}
\fmf{plain,tension=0,left=.5}{v1,v2}
\end{fmfgraph*} }
%
%%%%%%%%%%%%%%%%%%%%%%%
%
\subfigure[]{
\begin{fmfgraph*}(20,20)
\fmfleft{i1,i2}
\fmfright{o}
\fmfforce{.5w,.65h}{d1}
\fmf{photon}{i1,v1}
\fmf{photon}{i2,v2}
\fmf{dashes}{v3,o}
\fmf{photon,tension=.3}{v2,v3}
\fmf{photon,tension=.3}{v1,v3}
\fmf{plain,tension=0,left=.5}{v2,v1}
\fmf{plain,tension=0,left=.5}{v1,v2}
%\fmflabel{$(p_{1} \cdot k_{1})$ }{o}
\fmfv{decor.shape=circle,decor.filled=full,decor.size=.1w}{d1}
\end{fmfgraph*} }
%
%%%%%%%%%%%%%%%%%%%%%%%
%
%\hspace{14mm}
\subfigure[]{
\begin{fmfgraph*}(20,20)
\fmfleft{i1,i2}
\fmfright{o}
\fmf{plain}{i1,v1}
\fmf{plain}{i2,v2}
\fmf{dashes}{v4,o}
\fmf{photon,tension=.3}{v2,v3}
\fmf{photon,tension=.3}{v1,v3}
\fmf{plain,tension=0}{v2,v1}
\fmf{plain,tension=.2,left}{v3,v4}
\fmf{plain,tension=.2,right}{v3,v4}
\end{fmfgraph*} }
%
%%%%%%%%%%%%%%%%%%%%%%%
%
%\hspace{5mm}
\subfigure[]{
\begin{fmfgraph*}(20,20)
\fmfleft{i1,i2}
\fmfright{o}
\fmf{plain}{i1,v1}
\fmf{plain}{i2,v2}
\fmf{dashes}{v4,o}
\fmf{photon,tension=.3}{v2,v3}
\fmf{photon,tension=.3}{v1,v3}
\fmf{plain,tension=0}{v2,v1}
\fmf{photon,tension=.2,left}{v3,v4}
\fmf{photon,tension=.2,right}{v3,v4}
\end{fmfgraph*} }
%
%%%%%%%%%%%%%%%%%%%%%%%
%
%\hspace{5mm}
\subfigure[]{
\begin{fmfgraph*}(20,20)
\fmfleft{i1,i2}
\fmfright{o}
\fmf{plain}{i1,v1}
\fmf{plain}{i2,v2}
\fmf{dashes}{v3,o}
\fmf{photon,tension=.3}{v2,v3}
\fmf{photon,tension=.3}{v1,v3}
\fmf{plain,tension=0}{v2,v1}
\fmf{plain,right=45}{v3,v3}
\end{fmfgraph*} } \\
%
%%%%%%%%%%%%%%%%%%%%%%%
%
\subfigure[]{
\begin{fmfgraph*}(20,20)
\fmfleft{i}
\fmfright{o}
\fmf{dashes}{i,v1}
\fmf{dashes}{v3,o}
\fmf{photon,tension=.15,left}{v1,v2}
\fmf{photon,tension=.15,right}{v1,v2}
\fmf{plain,tension=.15,left}{v2,v3}
\fmf{plain,tension=.15,right}{v2,v3}
\end{fmfgraph*} } 
%
%%%%%%%%%%%%%%%%%%%%%%%
%
%\hspace{10mm}
\subfigure[]{
\begin{fmfgraph*}(20,20)
\fmfleft{i}
\fmfright{o}
\fmf{dashes}{i,v1}
\fmf{dashes}{v3,o}
\fmf{photon,tension=.15,left}{v1,v2}
\fmf{photon,tension=.15,right}{v1,v2}
\fmf{photon,tension=.15,left}{v2,v3}
\fmf{photon,tension=.15,right}{v2,v3}
\end{fmfgraph*} } 
%
%%%%%%%%%%%%%%%%%%%%%%%
%\hspace{10mm}
\subfigure[]{
\begin{fmfgraph*}(20,20)
\fmfleft{i}
\fmfright{o}
\fmf{dashes}{i,v1}
\fmf{dashes}{v2,o}
\fmf{photon,tension=.15,left}{v1,v2}
\fmf{photon,tension=.15}{v1,v2}
\fmf{photon,tension=.15,right}{v1,v2}
\end{fmfgraph*} }
%
%%%%%%%%%%%%%%%%%%%%%%%
%
%\hspace{10mm}
\subfigure[]{
\begin{fmfgraph*}(20,20)
\fmfleft{i}
\fmfright{o}
\fmf{dashes}{i,v1}
\fmf{dashes}{v2,o}
\fmf{photon,tension=.22,left}{v1,v2}
\fmf{photon,tension=.22,right}{v1,v2}
\fmf{plain,right=45}{v2,v2}
\end{fmfgraph*} }
%
%%%%%%%%%%%%%%%%%%%%%%%
\caption{\label{fig6} 
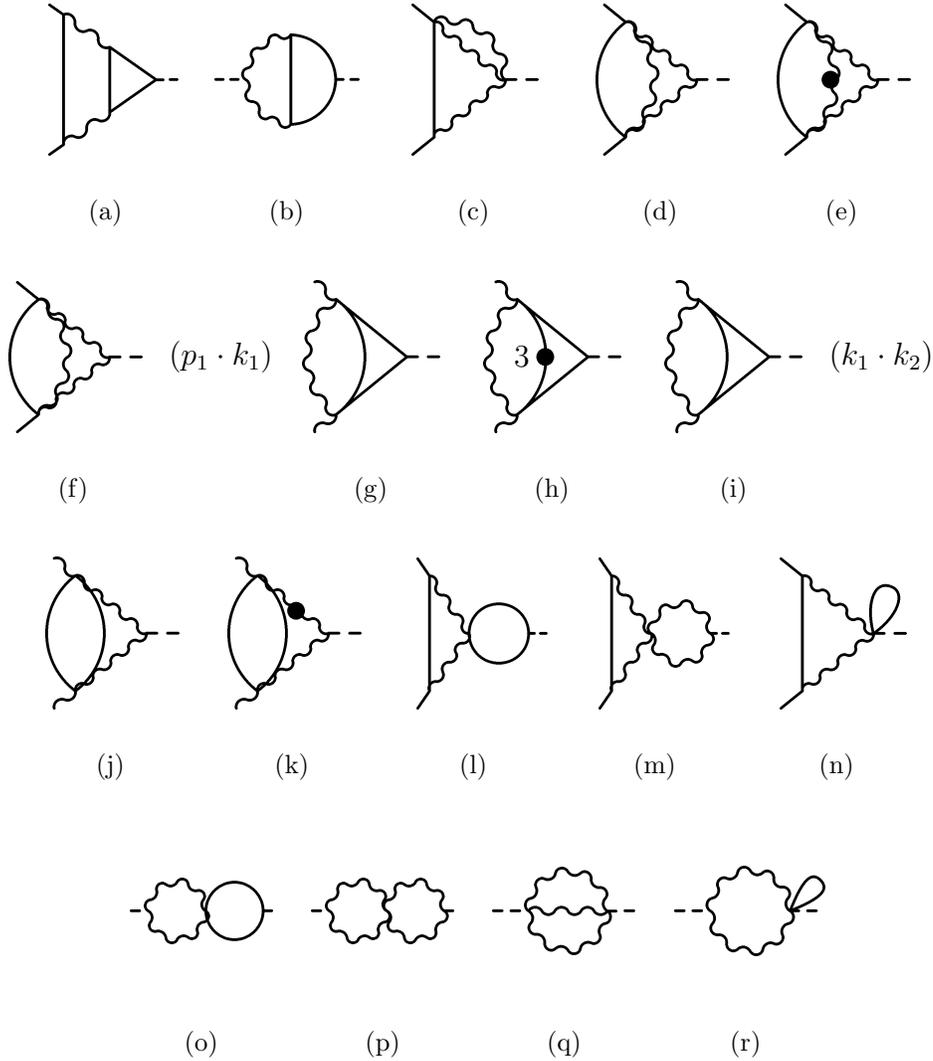
The 18 Master Integrals, calculated in the present work, 
represented as ``decorated graphs". A simple dot on a propagator 
line, like in figures (e) and (k), means that in the integrand 
the corresponding 
denominator is squared; a dot labeled by ``3'', figure (h), 
means that the corresponding denominator is raised to the third 
power; an explicitly written scalar product, as in graphs (f) and
(i), means that the corresponding integrand has that scalar product 
in the numerator.}
\ec
\efig
%%%%%%%%%%%%%%%%%%%%%%%%%%%%%%%%%%%%%%%%%%%%%%%%%%%%%%%%%%%%%%%%%%%%%

In this Section we give the results of the MIs of Fig.~\ref{fig6} as a 
Laurent series in $(D-4)$ and we express the coefficients of the series 
in terms of HPLs \cite{Polylog,Polylog3} of one variable. 

In all the cases, except the MIs (j) and (k) of Fig.~\ref{fig6}, 
coming from the reduction tree of the topology (h) of Fig.~\ref{fig2}, 
we express the result in terms of the variable $x$, defined as:
\be
x = \frac{\sqrt{Q^2+4m^2} - \sqrt{Q^2} }
           {\sqrt{Q^2+4m^2} + \sqrt{Q^2} } \ , \qquad
Q^2 = m^2 \frac{(1-x)^2}{x} \ .
\label{x}
\ee
When $Q$ is space-like and $Q^2$ varies from $0$ to $+\infty,\ $ $\ x$ 
varies from $1$ to $0$. When $Q$ is time-like, the analytic continuation 
is performed by putting $Q^2 = - (s+i\epsilon)$, with $s>0$. 
For $0 < s < 4m^2, \ $ $\ x $ varies in the upper unit circle; when 
$ 4m^2 < s < +\infty $, defining   
\be 
    y = \frac{ \sqrt{s}-\sqrt{s-4m^2} }{ \sqrt{s}+\sqrt{s-4m^2} } \ , 
    \qquad   s = m^2\frac{(1+y)^2}{y} \ , 
\label{y} 
\ee 
$y$ varies correspondingly from $1$ to $0$, and 
the analytic continuation is obtained by the replacement 
\be x = - y + i\epsilon \ . \label{xtoy} \ee 

For the MIs (j) and (k) of Fig.~\ref{fig6}, instead, we used 
the variable $\bar{x}$, defined as:
\be
\bar{x} = \frac{\sqrt{Q^2} - \sqrt{Q^2-4m^2}}
           {\sqrt{Q^2} + \sqrt{Q^2-4m^2}} \ , \qquad 
Q^2 = m^2 \frac{(1+ \bar{x})^2}{\bar{x}} \ .
\label{xbar}
\ee

When $Q$ is space-like and $Q^2$ varies from $+\infty$ and $4m^2$, 
$\bar{x}$ varies from $0$ to $1$; when $Q$ is still space-like, 
and $Q^2$ varies from $4m^2$ to $0$, we can give to $Q^2$ a negative 
imaginary part $-i\epsilon$ (anticipating the prescription for the 
continuation to time-like values) so that $\bar{x}$ varies in the upper 
unit circle; finally, when $Q$ is time-like and $Q^2=-(s+i\epsilon)$ 
with $s>0$, defining 
\be 
 \bar{y} = \frac{ \sqrt{s+4m^2}-\sqrt{s} }{ \sqrt{s+4m^2}+\sqrt{s} } \ , 
 \qquad  s = m^2\frac{(1-\bar{y})^2}{\bar{y}} \ , 
\label{ybar} 
\ee 
$\bar{y}$ varies from $1$, at $s=0$, to $0$, at $s=+\infty$, and the 
analytic continuation is given by the replacement 
\be 
\bar{x} = - \bar{y} + i\epsilon \ . \label{barxtobary} 
\ee

The denominators ${\mathcal D}'s$ appearing in the
formulas are given in Appendix \ref{app1}.
The loop integration measure in $D$ continuous dimensions is defined as 
\be 
\int{\mathfrak D}^Dk = \frac{m^{(4-D)}}{C(D)} 
\int \frac{d^D k}{(2\pi)^{(D-2)}} \, ,
\label{b00016a} 
\ee 
(corresponding to the energy scale $\mu_0=1$), where $C(D)$ is the 
following function of the continuous dimension $D$ 
\be
C(D) = (4 \pi)^{\frac{(4-D)}{2}} \Gamma \left( 3 - \frac{D}{2} \right) 
\label{b00018} \, ,
\ee
with the limiting value $C(4)=1$ for $D=4$.
With that choice, the 1-loop tadpole with mass $m$ reads 
\be 
\int{\mathfrak D}^Dk \ \frac{1}{k^2+m^2} = \frac {m^2}{(D-2)(D-4)} 
\ . 
\label{b00017a} 
\ee 

The explicit results follow in the next sections \cite{webpage}.

\subsection{2-loop Topologies with 3 denominators}
%%%%%%%%%%%%%%%%%%%%%%%%%%%%%%%%%%%%%%%%%%%%%%%%%%%%%%%%%%%%%%%%%%%%%
\bea
\parbox{25mm}{
\begin{fmfgraph*}(20,20)
\fmfleft{i}
\fmfright{o}
\fmf{dashes}{i,v1}
\fmf{dashes}{v2,o}
\fmf{photon,tension=.22,left}{v1,v2}
\fmf{photon,tension=.22,right}{v1,v2}
\fmf{plain,right=45}{v2,v2}
\end{fmfgraph*} }
& = & \, \int {\mathfrak{D}^D k_1} {\mathfrak{D}^D k_2}
\, \frac{1}{{\mathcal D}_{1}
            {\mathcal D}_{7}
            {\mathcal D}_{10}}  \nn\\ 
& = & {m^2} \sum_{i=-2}^{2} (D-4)^i \, F_{1}^{(i)}(x) 
  + {\mathcal O} \left( (D-4)^3 \right) \, , 
\eea 
where:
\bea
F^{(-2)}_{1}(x) & = & - \frac{1}{4} \, , \\
F^{(-1)}_{1}(x) & = & \frac{1}{8} \bigl[ 3 + H(0;x) + 2 H(1;x)  
\bigr] \, , \\
F^{(0)}_{1}(x)  & = & - \frac{1}{16} \bigl[ 7 
          - \zeta(2)
   + 3 H(0;x)
   + 6 H(1;x)
   + H(0,0;x)
   + 2 H(0,1;x) \nn\\
& &    + 2 H(1,0;x)
   + 4 H(1,1;x) 
\bigr] \, , \\
F^{(1)}_{1}(x)  & = & \frac{1}{32} \Bigl\{
            15 
          - 3  \zeta(2)
          - 2  \zeta(3)
   - (\zeta(2) -7 ) \bigl[ H(0;x)+ 2 H(1;x) \bigr] \nn\\
& & 
          + 3 H(0,0;x)  
          + 6 H(0,1;x) 
          + 6 H(1,0;x) 
          + 12 H(1,1;x)  \nn\\
& & 
          + H(0,0,0;x) 
          + 2 H(0,0,1;x)
          + 2 H(0,1,0;x)  \nn\\
& & 
          + 4 H(0,1,1;x) 
          + 2 H(1,0,0;x) 
          + 4 H(1,0,1;x)  \nn\\
& & 
          + 4 H(1,1,0;x) 
          + 16 H(1,1,1;x) \Bigr\} \, , \\
F^{(2)}_{1}(x)  & = & - \frac{1}{128} \biggl\{
            62 
          - 14  \zeta(2)
          - \frac{9}{5} \zeta(2)^2
          - 12  \zeta(3)
          + 2 ( 15 - 3 \zeta(2) \nn\\
& & - 2 \zeta(3) ) 
 \bigl[
     H(0;x) 
   + 2 H(1;x) \bigr]
        + 8 ( 7 - \zeta(2) ) 
 \bigl[
     4 H(0,0;x)  \nn\\
& & 
   + 2 H(0,1;x)
          + 2 H(1,0;x) 
   + H(1,1;x)  \bigr]
          + 6 H(0,0,0;x)  \nn\\
& & 
          + 12 H(0,0,1;x)    
          + 12 H(0,1,0;x) 
          + 24 H(0,1,1;x)  \nn\\
& & 
          + 12 H(1,0,0;x) 
          + 24 H(1,0,1;x) 
          + 24 H(1,1,0;x)  \nn\\
& & 
          + 48 H(1,1,1;x) 
          + 2 H(0,0,0,0;x) 
          + 4 H(0,0,0,1;x)  \nn\\
& & 
          + 4 H(0,0,1,0;x) 
          + 8 H(0,0,1,1;x) 
          + 4 H(0,1,0,0;x)  \nn\\
& & 
          + 8 H(0,1,0,1;x)
          + 8 H(0,1,1,0;x) 
          + 16 H(0,1,1,1;x)  \nn\\
& & 
          + 4 H(1,0,0,0;x) 
          + 8 H(1,0,0,1;x) 
          + 8 H(1,0,1,0;x)  \nn\\
& & 
          + 16 H(1,0,1,1;x)
          + 8 H(1,1,0,0;x) 
          + 16 H(1,1,0,1;x) \nn\\
& & 
          + 16 H(1,1,1,0;x) 
          + 32 H(1,1,1,1;x) \biggr\} \, .
\eea
%
%%%%%%%%%%%%%%%%%%%%%%%%%%%%%%%%%%%%%%%%%%%%%%%%%%%%%%%%%%%%%%%%%%%%%
%
\bea
\parbox{25mm}{
\begin{fmfgraph*}(20,20)
\fmfleft{i}
\fmfright{o}
\fmf{dashes}{i,v1}
\fmf{dashes}{v2,o}
\fmf{photon,tension=.15,left}{v1,v2}
\fmf{photon,tension=.15}{v1,v2}
\fmf{photon,tension=.15,right}{v1,v2}
\end{fmfgraph*} }
& = & \, \int {\mathfrak{D}^D k_1} {\mathfrak{D}^D k_2}
\, \frac{1}{{\mathcal D}_{1}
            {\mathcal D}_{9}
            {\mathcal D}_{23}}  \nn\\ 
& = & m^2 \sum_{i=-1}^{3} (D-4)^i \, F_{2}^{(i)}(x) + {\mathcal O} \left(
(D-4)^4 \right) \, ,
\eea
where:
\bea 
\hspace{-8mm}
F_2^{(-1)}(x) & = & \frac{1}{32} \bigg[
            \frac{1}{x}
          + x
          - 2
          \bigg] \, , \\
\hspace{-8mm}
F_2^{(0)}(x) & = & - \bigg[ \frac{1}{x} + x - 2 \bigg]
       \bigg[
            \frac{13}{128} 
   + \frac{1}{32} H(0;x) 
   + \frac{1}{16} H(1;x)
       \bigg] \, , \\
\hspace{-8mm}
F_2^{(1)}(x) & = & \frac{1}{128} 
\bigg[ \frac{1}{x} \! + \! x \! - \! 2 \bigg] \biggl\{ 
            \frac{115}{4} \! 
   - \! 2 \zeta(2)\! 
      + \! 13 \bigl[ H(0;x)\! 
          + \! 2 H(1;x) \bigr]\! 
          + \! 4 H(0,0;x) \nn\\
& &    + 8 H(0,1;x)
   + 8 H(1,0;x)
   + 16 H(1,1;x)
   \biggr\} \, , \\
\hspace{-8mm}
F_2^{(2)}(x) & = & - \frac{1}{128} 
\bigg[ \frac{1}{x} \! + \! x \! - \! 2 \bigg] \biggl\{ 
            \frac{865}{16}\! 
   - \frac{13}{2} \zeta(2) \! 
   - 5 \zeta(3)\! 
   + \biggl( \frac{115}{2}
   - \frac{\zeta(2)}{4} \biggr) 
 \bigl[ H(0;x) \nn\\
& & 
   + 2 H(1;x) \bigr]
   + 13 H(0,0;x)
   + 26 H(0,1;x)
   + 26 H(1,0;x) \nn\\
& & 
   + 52 H(1,1;x)
   + 8 H(0,0,0;x)
   + 16 H(0,0,1;x)
   + 16 H(0,1,0;x) \nn\\
& & 
   +  \! 32 H(0,1,1;x) \! 
   +  \! 16 H(1,0,0;x) \! 
   +  \! 32 H(1,0,1;x) \! 
   +  \! 32 H(1,1,0;x) \nn\\
& & 
   + 64 H(1,1,1;x)  \biggr\} \, , \\
\hspace{-8mm}
F_2^{(3)}(x) & = &  \frac{1}{512} 
\bigg[ \frac{1}{x} \! + \! x \! - \! 2 \bigg] \biggl\{
            \frac{5971}{8}
   - 115 \zeta(2) 
   - \frac{22}{5} \zeta^{2}(2) 
   - 130 \zeta(3) 
 + \biggl( 
     \frac{865}{4} \nn\\
& & 
   - 26 \zeta(2) 
   - 20 \zeta(3) \biggr) 
 \bigl[ H(0;x)
   + 2 H(1;x) \bigr]
   + ( 115 - 8 \zeta(2) ) 
 \bigl[ H(0,0;x) \nn\\
& & 
   + 2 H(0,1;x)
   + 2 H(1,0;x)
   + 4 H(1,1;x)
   \bigr]
 + 52 \bigl[ 
     H(0,0,0;x) \nn\\
& & 
          + 2 H(0,0,1;x)
          + 2 H(0,1,0;x)
          + 2 H(1,0,0;x)
          + 4 H(0,1,1;x) \nn\\
& & 
          + \! 4 H(1,0,1;x) \! 
          + \! 4 H(1,1,0;x)\! 
          + \! 8 H(1,1,1;x) \bigr]\! 
 + \! 16 \bigl[ 
     H(0,0,0,0;x) \nn\\
& & 
   + 2 H(0,0,0,1;x)
   + 2 H(0,0,1,0;x)
   + 4 H(0,0,1,1;x) \nn\\
& & 
   + 2 H(0,1,0,0;x)
   + 4 H(0,1,0,1;x)
   + 4 H(0,1,1,0;x) \nn\\
& & 
   + 8 H(0,1,1,1;x)
   + 2 H(1,0,0,0;x)
   + 4 H(1,0,0,1;x) \nn\\
& & 
   + 4 H(1,0,1,0;x)
   + 8 H(1,0,1,1;x)
   + 4 H(1,1,0,0;x) \nn\\
& & 
   + 8 H(1,1,0,1;x)
   + 8 H(1,1,1,0;x)
   + 16 H(1,1,1,1;x) \bigr] \biggr\} \, .
\eea
%
%%%%%%%%%%%%%%%%%%%%%%%%%%%%%%%%%%%%%%%%%%%%%%%%%%%%%%%%%%%%%%%%%%%%%

\subsection{2-loop Topologies with 4 denominators}

\bea
\parbox{25mm}{
\begin{fmfgraph*}(20,20)
\fmfleft{i}
\fmfright{o}
\fmf{dashes}{i,v1}
\fmf{dashes}{v3,o}
\fmf{photon,tension=.15,left}{v1,v2}
\fmf{photon,tension=.15,right}{v1,v2}
\fmf{photon,tension=.15,left}{v2,v3}
\fmf{photon,tension=.15,right}{v2,v3}
\end{fmfgraph*} }
 & = & \int {\mathfrak{D}^D k_1} {\mathfrak{D}^D k_2}
           \  \frac{1}{
                     {\mathcal D}_{1}
                     {\mathcal D}_{7}
                     {\mathcal D}_{9}
                     {\mathcal D}_{15}
                     }  \nn\\ 
& = & \sum_{i=-2}^{2} (D-4)^i \, F_{3}^{(i)}(x) + {\mathcal O} \left(
(D-4)^3 \right)  \, ,
\eea 
where: 
\bea 
\hspace{-6mm}
F_{3}^{(-2)}(x) & = & \frac{1}{4} \, , \\
\hspace{-6mm}
F_{3}^{(-1)}(x) & = & - \frac{1}{2} - \frac{1}{4} H(0;x) - 
\frac{1}{2} H(1;x) \, , \\
\hspace{-6mm}
F_{3}^{(0)}(x) & = & \frac{1}{8} \Bigl[ 
                 6
        - \zeta(2)
        + 4 H(0;x)
        + 8 H(1;x)
        + 2 H(0,0;x)
        + 4 H(0,1;x) \nn\\
& & 
        + 4 H(1,0;x)
        + 8 H(1,1;x) \Bigr] \, , \\
\hspace{-6mm}
F_{3}^{(1)}(x) & = &  - 1 + \frac{1}{8} \Bigl\{
                 2 \zeta(2)
        + \zeta(3)
        - ( 6 - \zeta(2) ) \bigl[
          H(0;x)
        + 2 H(1;x) \bigr] \nn\\
& & 
        - 4 H(0,0;x)
        - 8 H(0,1;x)
        - 8 H(1,0;x)
        - 16 H(1,1;x) \nn\\
& & 
        - 2 H(0,0,0;x)
        - 4 H(0,0,1;x)
        - 4 H(0,1,0;x)
        - 8 H(0,1,1;x) \nn\\
& & 
        - \!  4 H(1,0,0;x) \! 
        -  \! 8 H(1,0,1;x) \! 
        -  \! 8 H(1,1,0;x) \! 
        -  \! 16 H(1,1,1;x) \, , \\
\hspace{-6mm}
F_{3}^{(2)}(x) & = & \frac{5}{4} - \frac{3}{8} \zeta(2) - \frac{1}{80} 
                  \zeta^{2}(2) - \frac{1}{4} \zeta(3) 
           + \frac{1}{8} \Bigl[ 
          8
        - 2 \zeta(2)
        - \zeta(3) \Bigr] \bigl[ 
          H(0;x)  \nn\\
& & 
        + 2 H(1;x) \bigr]
           + \frac{1}{8} \Bigl[ 
          6
        - \zeta(2) \Bigr] \bigl[ 
          H(0,0;x)
        + 2 H(0,1;x)
        + 2 H(1,0;x) \nn\\
& & 
        + 4 H(1,1;x) \bigr]
    + \frac{1}{4} \Bigl[
                 2 H(0,0,0;x)
               + 4 H(0,0,1;x)
               + 8 H(0,1,1;x)\nn\\
& & 
               + 4 H(0,1,0;x)
               + 4 H(1,0,0;x)
               + 8 H(1,0,1;x)
               + 8 H(1,1,0;x)\nn\\
& & 
               + 16 H(1,1,1;x)
               + H(0,0,0,0;x)
               + 2 H(0,0,0,1;x)\nn\\
& & 
               + 2 H(0,0,1,0;x)
               + 4 H(0,0,1,1;x)
               + 2 H(0,1,0,0;x)\nn\\
& & 
               + 4 H(0,1,0,1;x)
               + 4 H(0,1,1,0;x)
               + 8 H(0,1,1,1;x)\nn\\
& & 
               + 2 H(1,0,0,0;x)
               + 4 H(1,0,0,1;x)
               + 4 H(1,0,1,0;x)\nn\\
& & 
               + 8 H(1,0,1,1;x)
               + 4 H(1,1,0,0;x)
               + 8 H(1,1,0,1;x)\nn\\
& & 
               + 8 H(1,1,1,0;x)
               + 16 H(1,1,1,1;x) \Bigr]  \, .
\eea

\bea
\parbox{25mm}{
\begin{fmfgraph*}(20,20)
\fmfleft{i}
\fmfright{o}
\fmf{dashes}{i,v1}
\fmf{dashes}{v3,o}
\fmf{photon,tension=.15,left}{v1,v2}
\fmf{photon,tension=.15,right}{v1,v2}
\fmf{plain,tension=.15,left}{v2,v3}
\fmf{plain,tension=.15,right}{v2,v3}
\end{fmfgraph*} }
 & = & \int {\mathfrak{D}^D k_1} {\mathfrak{D}^D k_2}
           \  \frac{1}{
                     {\mathcal D}_{1}
                     {\mathcal D}_{7}
                     {\mathcal D}_{10}
                     {\mathcal D}_{16}
                     }  \nn\\ 
& = & \sum_{i=-2}^{2} (D-4)^i \, F_{4}^{(i)}(x) + {\mathcal O} \left(
(D-4)^3 \right)  \, ,
\eea
where:
\bea
\hspace{-8mm}
F_{4}^{(-2)}(x) & = & \frac{1}{4} \, , \\
\hspace{-8mm}
F_{4}^{(-1)}(x) & = & - \frac{1}{2} - \frac{1}{4(1-x)} H(0;x) - 
\frac{1}{4} H(1;x) \, , \\
\hspace{-8mm}
F_{4}^{(0)}(x) & = & \frac{3}{4}
               - \frac{1}{8(1-x)} \Bigl[
          \zeta(2)
               - 4 H(0;x)
        - 3 H(0,0;x)
        + 2 H(-1,0;x) \nn\\
& & 
        - 2 H(0,1;x)
        - 2 H(1,0;x) \Bigr]
      + \frac{1}{8} \Bigl[ 
          4 H(1;x)
        - H(0,0;x) \nn\\
& & 
        + H(-1,0;x)
        + 2 H(1,1;x) \Bigr]
\, , \\
\hspace{-8mm}
F_{4}^{(1)}(x) & = & -1 
      + \frac{1}{16(1-x)} \Bigl\{
                 4 \zeta(2) \! 
        +  \! 8 \zeta(3) \! 
        - (12 \!  -  \! 3 \zeta(2)) H(0;x) \! 
        +  \! 2 \zeta(2) \bigl[ H(1;x) \nn\\
& & 
               -  \! H( \! -1;x) \bigr] \! 
        -  \! 12 H(0,0;x) \! 
        +  \! 8 H( \! -1,0;x) \! 
        -  \! 8 H(0,1;x) \! 
        -  \! 8 H(1,0;x) \nn\\
& & 
        -  \! 7 H(0,0,0;x) \! 
        -  \! 4 H( \! -1, \! -1,0;x) \! 
        +  \! 6 H( \! -1,0,0;x) \! 
        +  \! 4 H( \! -1,0,1;x) \nn\\
& & 
        + 4 H(-1,1,0;x)
        + 4 H(0,-1,0;x)
        - 6 H(0,0,1;x)
        - 6 H(0,1,0;x) \nn\\
& & 
        + 4 H(1,-1,0;x)
        - 6 H(1,0,0;x) 
        - 4 H(1,0,1;x)
        - 4 H(0,1,1;x) \nn\\
& & 
        - 4 H(1,1,0;x) \Bigr\}
      - \frac{1}{16} \Bigl\{
          \zeta(2) H(0;x)
        + 12 H(1;x)
        - \zeta(2) H(-1;x) \nn\\
& & 
               - 4 H(0,0;x)
               + 4 H(-1,0;x)
               + 8 H(1,1;x)
               - 3 \bigl[ H(0,0,0;x) \nn\\
& & 
               - H( \! -1,0,0;x) \bigr]
               +  \! 2 \bigl[ H( \! -1,0,1;x) \! 
               - H( \! -1, \! -1,0;x) \! 
               + \!  H( \! -1,1,0;x) \nn\\
& & 
               +  \! H(0, \! -1,0;x) \! 
               - H(0,0,1;x) \! 
               - H(0,1,0;x) \! 
               + \!  H(1, \! -1,0;x) \nn\\
& & 
               - H(1,0,0;x)
               + 2 H(1,1,1;x) \bigr] \Bigr\}
   \, , \\
\hspace{-8mm}
F_{4}^{(2)}(x) & = & \frac{5}{4}
               - \frac{1}{64} \zeta^{2}(2)
        - \frac{1}{8(1-x)} \biggl[ 
          3 \zeta(2)
   - \frac{1}{40} \zeta^{2}(2)
   + 2 \zeta(3) \biggr] 
 + \frac{1}{16} \biggl[ 
          2 \zeta(2) \nn\\
& & 
        + \zeta(3)
        + \frac{16-6 \zeta(2) -3 \zeta(3)}{(1-x)} 
      \biggr] H(0;x)
 - \frac{1}{16} \biggl[ 
          2 \zeta(2)
        + \zeta(3) \nn\\
& & 
        - \frac{2( 2 \zeta(2) + \zeta(3))}{(1-x)} 
      \biggr] H(-1;x)
 + \biggl[
          1
        - \frac{( 2 \zeta(2) + \zeta(3))}{8(1-x)} 
      \biggr] H(1;x) \nn\\
& & 
 - \frac{1}{32} \biggl[
          12 \! 
        - 3 \zeta(2) \! 
        - \frac{36  \! -  \! 7 \zeta(2)}{(1-x)} 
      \biggr] H(0,0;x) \! 
 + \!  \frac{1}{16} \biggl[
          \zeta(2)
        +  \! \frac{12  \! -  \! 3 \zeta(2)}{(1-x)} 
      \biggr] \times \nn\\
& &  
           \times \bigl[ 
        H(1,0;x) \! 
        + H(0,1;x) \bigr] \! 
        - \frac{3 \zeta(2)}{32} H(-1,0;x) \! 
 +  \! \frac{3}{16} \biggl[
           2 \! 
        -  \!  \frac{4  \! -  \! \zeta(2)}{(1-x)} 
      \biggr]  \times \nn\\
& &  
           \times H(-1,0;x)
 + \frac{\zeta(2)}{16} \biggl[
          1
        - \frac{2}{(1-x)} 
      \biggr]  \bigl[ H(-1,-1;x)
               - H(1,-1;x) \nn\\
& &  
               - H(0,-1;x)
        - H(-1,1;x) \bigr]
 + \frac{1}{4} \biggl[ 3 - \frac{1}{2(1-x)} \biggr] H(1,1;x) \nn\\
& &  
        - \frac{1}{8} \biggl[ 1  \! -  \! \frac{2}{(1-x)} \biggr]
   \Bigl\{ 3 \bigl[ H(0,0,0;x)  \! 
        -  \! H(-1,0,0;x) \bigr] \! 
        -  \! 2 H(-1,0,1;x) \nn\\
& &  
        -  \! 2 H( \! -1, \! 1, \! 0, \! x) \! 
        -  \! 2 H(0, \! -1, \! 0, \! x) \! 
        -  \! 2 H(1, \! -1, \! 0, \! x) \! 
        +  \! 2 H( \! -1, \! -1, \! 0, \! x) \Bigr\} \nn\\
& &  
        - \frac{1}{4} \biggl[ 1 - \frac{3}{(1-x)} \biggr]
   \Bigl\{ H(0,0,1;x)
        + H(0,1,0;x)
        + H(1,0,0;x) \Bigr\} \nn\\
& &  
        + \frac{1}{2} H(1,1,1;x)
        + \frac{1}{32(1-x)} \Bigl\{ 
                 4 \bigl[ H(0,0,0;x)
        + 4 H(1,0,1;x) \nn\\
& &  
        + 4 H(1,1,0;x) 
        + 4 H(0,1,1;x)\bigr]
        + H(0,0,0,0;x)
        + 2 H(0,0,0,1;x) \nn\\
& &  
        + 4 H(0,0,1,1;x)
        + 2 H(0,1,0,0;x)
        + 4 H(0,1,0,1;x) \nn\\
& &  
        + 4 H(0,1,1,0;x)
        + 16 H(0,1,1,1;x)
        -12 H(1,-1,0,0;x) \nn\\
& &  
        + 2 H(1,0,0,0;x)
        + 4 H(1,0,0,1;x)
        + 4 H(1,0,1,0;x) \nn\\
& &  
        + 8 H(1,0,1,1;x)
        + 4 H(1,1,0,0;x)
        + 8 H(1,1,0,1;x) \nn\\
& &  
        + 8 H(1,1,1,0;x) \Bigr\}
     + \frac{1}{16} \bigl[ 
          4 H(1,1,1,1;x)
        - 3 H(0,0,1,0;x)  \nn\\
& &  
        + 3 H(1,-1,0,0;x) \bigr]
   - \frac{1}{32} \biggl[ 1 - \frac{1}{(1-x)} \biggr]
       \Bigl\{
          7 \bigl[ H(0,0,0,0;x) \nn\\
& &  
        - H(-1,0,0,0;x) \bigr]
        + 4 \bigl[ H(-1,-1,0,1;x)
        + H(-1,-1,1,0;x) \nn\\
& &  
        + H(-1,0,-1,0;x)
        - H(-1,0,1,1;x)
        + H(-1,1,-1,0;x) \nn\\
& &  
        - H(-1,1,0,1;x)
        - H(-1,1,1,0;x)
        + H(0,-1,-1,0;x) \nn\\
& &  
        - H(0,-1,0,1;x)
        - H(0,-1,1,0;x)
        - H(0,0,-1,0;x) \nn\\
& &  
        + H(0,0,1,1;x)
        - H(0,1,-1,0;x)
        + H(0,1,0,1;x) \nn\\
& &  
        + H(0,1,1,0;x)
        + H(1,-1,-1,0;x)
        - H(-1,-1,-1,0;x) \nn\\
& &  
        - H(1,-1,0,1;x)
        - H(1,-1,1,0;x)
        - H(1,0,-1,0;x) \nn\\
& &  
        + H(1,0,0,1;x)
        + H(1,0,1,0;x)
        - H(1,1,-1,0;x) \nn\\
& &  
        + H(1,1,0,0;x) \bigr]
        + 6 \bigl[ H(-1,-1,0,0;x)
        - H(-1,0,0,1;x) \nn\\
& &  
        - H(-1,0,1,0;x)
        - H(-1,1,0,0;x)
        - H(0,-1,0,0;x) \nn\\
& &  
        + H(0,0,0,1;x)
        + H(0,1,0,0;x)
        + H(1,0,0,0;x) \bigr] \Bigr\} .
\eea

\bea
\parbox{25mm}{
\begin{fmfgraph*}(20,20)
\fmfleft{i1,i2}
\fmfright{o}
\fmf{plain}{i1,v1}
\fmf{plain}{i2,v2}
\fmf{dashes}{v3,o}
\fmf{photon,tension=.3}{v2,v3}
\fmf{photon,tension=.3}{v1,v3}
\fmf{plain,tension=0}{v2,v1}
\fmf{plain,right=45}{v3,v3}
\end{fmfgraph*} }
 & = & \int {\mathfrak{D}^D k_1} {\mathfrak{D}^D k_2}
           \  \frac{1}{
                     {\mathcal D}_{2}
                     {\mathcal D}_{3}
                     {\mathcal D}_{5}
                     {\mathcal D}_{10}
                     }  \nn\\ 
& = & \sum_{i=-1}^{1} (D-4)^i \, F_{5}^{(i)}(x) + {\mathcal O} \left(
(D-4)^2 \right)  \, ,
\eea
where:
\bea
\hspace{-8mm}
F_{5}^{(-1)}(x) & = & \frac{1}{16} \biggl[ \frac{1}{(1-x)}
- \frac{1}{(1+x)} \biggr] \bigl[ 4 \zeta(2) 
          + H(0,0;x)
   + 2 H(0,1;x) \bigr] \, , \\
\hspace{-8mm}
F_{5}^{(0)}(x) & = & - \frac{1}{32} \biggl[ \frac{1}{(1-x)}
- \frac{1}{(1+x)} \biggr] \Bigl\{ 4 \zeta(2)  \! 
          +  \! 5 \zeta(3) \! 
   - \zeta(2) \bigl[ H(0;x) \! 
   - 8 H(-1;x) \nn\\
& & 
   - 8 H(1;x) \bigr] \! 
   +  \! H(0,0;x) \! 
   +  \! 2 H(0,1;x) \! 
   +  \! H(0,0,0;x) \! 
   +  \! 2 H( \! -1,0,0;x) \nn\\
& & 
   + 4 H(-1,0,1;x)
   + 2 H(0,0,1;x)
   + 2 H(0,1,0;x)
   + 4 H(0,1,1;x) \nn\\
& & 
   + 2 H(1,0,0;x)
   + 4 H(1,0,1;x) \Bigr\}  \, , \\
\hspace{-8mm}
F_{5}^{(1)}(x) & = & \frac{1}{64} \biggl[ \frac{1}{(1-x)}
- \frac{1}{(1+x)} \biggr] \biggl\{ 
            4 \zeta(2)
     + \frac{59}{10} \zeta^{2}(2)
   + 5 \zeta(3)
   - \bigl[ \zeta(2) \nn\\
& & 
          + 2 \zeta(3) \bigr] H(0;x)
   + 2 \bigl[ 4 \zeta(2) + 5 \zeta(3) \bigr] 
     \bigl[ H(1;x) + H(-1;x) \bigr] \nn\\
& & 
   + \bigl[ 1 - \zeta(2) \bigr] 
     \bigl[ H(0,0;x) + 2 H(0,1;x) \bigr] 
   + 2 \zeta(2) \bigl[
     8 H(-1,1;x) \nn\\
& & 
   - H(-1,0;x)
   + 8 H(-1,-1;x)
   + 8 H(1,-1;x)
   - H(1,0;x) \nn\\
& & 
   + 8 H(1,1;x) \bigr]
   + H(0,0,0;x)
   + 2 H(0,0,1;x)
   + 2 H(-1,0,0;x) \nn\\
& & 
   + 4 H(-1,0,1;x)
   + 2 H(0,1,0;x)
   + 4 H(0,1,1;x)
   + 2 H(1,0,0;x) \nn\\
& & 
   + 4 H(1,0,1;x)
   + H(0,0,0,0;x)
   + 4 H(-1,-1,0,0;x) \nn\\
& & 
   + 8 H(-1,-1,0,1;x)
   + 2 H(-1,0,0,0;x)
   + 4 H(-1,0,0,1;x) \nn\\
& & 
   + 4 H(-1,0,1,0;x)
   + 8 H(-1,0,1,1;x)
   + 4 H(-1,1,0,0;x) \nn\\
& & 
   + 8 H(-1,1,0,1;x)
   + 2 H(0,0,0,1;x)
   + 2 H(0,0,1,0;x) \nn\\
& & 
   + 4 H(0,0,1,1;x)
   + 2 H(0,1,0,0;x)
   + 4 H(0,1,0,1;x) \nn\\
& & 
   + 4 H(0,1,1,0;x)
   + 8 H(0,1,1,1;x)
   + 4 H(1,-1,0,0;x) \nn\\
& & 
   + 8 H(1,-1,0,1;x)
   + 2 H(1,0,0,0;x)
   + 4 H(1,0,0,1;x) \nn\\
& & 
   + 4 H(1,0,1,0;x)
   + 8 H(1,0,1,1;x)
   + 4 H(1,1,0,0;x) \nn\\
& & 
   + 8 H(1,1,0,1;x)
     \biggr\}
    \, .
\eea

%%%%%%%%%%%%%%%%%%%%%%%%%%%%%%%%%%%%%%%%%%%%%%%%%%%%%%%%%%%%%%%%%%%%%
\bea
\parbox{25mm}{
\begin{fmfgraph*}(20,20)
\fmfleft{i1,i2}
\fmfright{o}
\fmf{plain}{i1,v1}
\fmf{plain}{i2,v2}
\fmf{dashes}{v3,o}
\fmf{photon,tension=.3}{v2,v3}
\fmf{photon,tension=.3}{v1,v3}
\fmf{plain,tension=0}{v2,v1}
\fmf{photon,tension=0,left=.5}{v2,v3}
\end{fmfgraph*} }
 & = & \int {\mathfrak{D}^D k_1} {\mathfrak{D}^D k_2}
           \  \frac{1}{
                     {\mathcal D}_{2}
                     {\mathcal D}_{5}
                     {\mathcal D}_{9}
                     {\mathcal D}_{19}
                     }  \nn\\ 
& = & \sum_{i=-2}^{2} (D-4)^i \, F_{6}^{(i)}(x) + {\mathcal O} \left(
(D-4)^3 \right)  \, ,
\eea
where:
\bea
\hspace{-8mm}
F_{6}^{(-2)}(x) & = & \frac{1}{8} \, , \\
\hspace{-8mm}
F_{6}^{(-1)}(x) & = & - \frac{5}{16} \, , \\
\hspace{-8mm}
F_{6}^{(0)}(x)  & = & \frac{1}{32} \Bigg\{
            19  
          +   4 \zeta(2)  
   +   2 \biggl[ 1   -   \frac{2}{(1+x)} \biggr]
        \Bigl[ 4 \zeta(2)  
   +   H(0,0;x)  \nn\\
& & 
   +   2 H(0,1;x) \Bigr]
       \Bigg\} , \\
\hspace{-8mm}
F_{6}^{(1)}(x) & = & - \frac{65}{64}
     - \frac{1}{16} \biggl[ 15 - \frac{20}{(1+x)} \biggr] \zeta(2) 
     - \frac{1}{32} \biggl[ 11 - \frac{14}{(1+x)} \biggr] \zeta(3) \nn\\
& & 
     - \frac{1}{32} \biggl[ 1 - \frac{2}{(1+x)} \biggr] 
     \bigl\{ \zeta(2) [ H(0;x)
            + 8 H(-1;x)
     + 16 H(1;x) ] \nn\\
& & 
     + 5 H(0,0;x)
     + 10 H(0,1;x)
     + 3 H(0,0,0;x)
     + 2 H(-1,0,0;x) \nn\\
& & 
     + 4 H(-1,0,1;x)
     + 6 H(0,0,1;x)
     + 4 H(0,1,0;x)
     + 8 H(0,1,1;x) \nn\\
& & 
            + 4 H(1,0,0;x)
     + 8 H(1,0,1;x) \bigr\}
   \, , \\
\hspace{-8mm}
F_{6}^{(2)}(x) & = & 
              \frac{211}{128}
       + \biggl[ 
              \frac{57}{32}
            - \frac{19}{8(1+x)} \biggr] \zeta(2)
       + \biggl[ 
              \frac{267}{640}
            - \frac{139}{320(1+x)} \biggr] \zeta^2(2)
       + \biggl[ 
              \frac{55}{64} \nn\\
& & 
            - \frac{35}{32(1+x)} \biggr] \zeta(3)
       + \frac{1}{64} \biggl[ 
              1
            - \frac{2}{(1+x)} \biggr] \Bigl\{
       (5 \zeta(2) - 2 \zeta(3) ) H(0;x) \nn\\
& & 
     + (40 \zeta(2) + 14 \zeta(3) ) H(-1;x)
     + (80 \zeta(2) + 28 \zeta(3) ) H(1;x)
     + (19  \nn\\
& & 
            - \zeta(2) ) H(0,0;x)
     + (38 + 12 \zeta(2) ) H(0,1;x)
     + \zeta(2) \bigl[ 
       16 H(-1,-1;x) \nn\\
& & 
     + 64 H(0,-1;x)
     + 2 H(-1,0;x)
     + 32 H(1,-1;x)
     + 4 H(1,0;x) \nn\\
& & 
     + 64 H(1,1;x)
     + 32 H(-1,1;x) \bigr] 
     + 15 H(0,0,0;x)
     + 10 H(-1,0,0;x) \nn\\
& & 
     + 20 H(-1,0,1;x) \! 
     +  \! 30 H(0,0,1;x)  \! 
     +  \! 20 H(0,1,0;x) \! 
     +  \! 40 H(0,1,1;x) \nn\\
& & 
     + 20 H(1,0,0;x)
     + 40 H(1,0,1;x)
     + 7 H(0,0,0,0;x) \nn\\
& & 
     + 4 H(-1,-1,0,0;x)
     + 8 H(-1,-1,0,1;x) 
     + 6 H(-1,0,0,0;x) \nn\\
& & 
     + 12 H(-1,0,0,1;x)
     + 8 H(-1,0,1,0;x)
     + 16 H(-1,0,1,1;x) \nn\\
& & 
     + 8 H(-1,1,0,0;x)
     + 16 H(-1,1,0,1;x)
     + 2 H(0,-1,0,0;x) \nn\\
& & 
     + 4 H(0,-1,0,1;x)
     + 14 H(0,0,0,1;x)
     + 12 H(0,0,1,0;x) \nn\\
& & 
     + 24 H(0,0,1,1;x)
     + 12 H(0,1,0,0;x)
     + 24 H(0,1,0,1;x) \nn\\
& & 
     + 16 H(0,1,1,0;x)
     + 32 H(0,1,1,1;x)
     + 8 H(1,-1,0,0;x) \nn\\
& & 
     + 16 H(1,-1,0,1;x)
     + 12 H(1,0,0,0;x)
     + 24 H(1,0,0,1;x) \nn\\
& & 
     + 16 H(1,0,1,0;x)
     + 32 H(1,0,1,1;x)
     + 16 H(1,1,0,0;x) \nn\\
& & 
     + 32 H(1,1,0,1;x)
   \Bigr\}  \, .
\eea
%%%%%%%%%%%%%%%%%%%%%%%%%%%%%%%%%%%%%%%%%%%%%%%%%%%%%%%%%%%%%%%%%%%%%

%%%%%%%%%%%%%%%%%%%%%%%%%%%%%%%%%%%%%%%%%%%%%%%%%%%%%%%%%%%%%%%%%%%%%
\bea
\parbox{35mm}{
\begin{fmfgraph*}(20,20)
\fmfleft{i1,i2}
\fmfright{o}
\fmf{plain}{i1,v1}
\fmf{plain}{i2,v2}
\fmf{dashes}{v3,o}
\fmf{photon,tension=.3}{v2,v3}
\fmf{photon,tension=.3}{v1,v3}
\fmf{photon,tension=0,left=.5}{v2,v1}
\fmf{plain,tension=0,left=.5}{v1,v2}
\end{fmfgraph*} }
& = & \int {\mathfrak{D}^D k_1} {\mathfrak{D}^D k_2}
           \frac{1}{
                     {\mathcal D}_{1}
                     {\mathcal D}_{10}
                     {\mathcal D}_{19}
                     {\mathcal D}_{21}
                     } \nn\\ 
& = & \sum_{i=-2}^{1} (D-4)^i \, F_{7}^{(i)}(x) + {\mathcal O} \left(
(D-4)^2 \right) , \\
\parbox{20mm}{
\begin{fmfgraph*}(20,20)
\fmfleft{i1,i2}
\fmfright{o}
\fmf{plain}{i1,v1}
\fmf{plain}{i2,v2}
\fmf{dashes}{v3,o}
\fmf{photon,tension=.3}{v2,v3}
\fmf{photon,tension=.3}{v1,v3}
\fmf{photon,tension=0,left=.5}{v2,v1}
\fmf{plain,tension=0,left=.5}{v1,v2}
%\fmflabel{$(p_{1} \cdot k_{1})$ }{o}
\end{fmfgraph*} } \ (p_{1} \cdot k_{1})
& = & \int {\mathfrak{D}^D k_1} {\mathfrak{D}^D k_2}
           \frac{(p_1 \cdot k_1)}{
                     {\mathcal D}_{1}
                     {\mathcal D}_{10}
                     {\mathcal D}_{19}
                     {\mathcal D}_{21}
                     } \nn\\ 
& = & m^2 \sum_{i=-2}^{1} (D-4)^i \, F_{8}^{(i)}(x) + {\mathcal O} \left(
(D-4)^2 \right) , \\
\parbox{35mm}{
\begin{fmfgraph*}(20,20)
\fmfleft{i1,i2}
\fmfright{o}
\fmfforce{.4w,.5h}{d1}
\fmf{plain}{i1,v1}
\fmf{plain}{i2,v2}
\fmf{dashes}{v3,o}
\fmf{photon,tension=.3}{v2,v3}
\fmf{photon,tension=.3}{v1,v3}
\fmf{photon,tension=0,left=.5}{v2,v1}
\fmf{plain,tension=0,left=.5}{v1,v2}
\fmfv{decor.shape=circle,decor.filled=full,decor.size=.1w}{d1}
\end{fmfgraph*} }
& = & \int {\mathfrak{D}^D k_1} {\mathfrak{D}^D k_2}
           \frac{1}{
                     {\mathcal D}^2_{1}
                     {\mathcal D}_{10}
                     {\mathcal D}_{19}
                     {\mathcal D}_{21}
                     } \nn\\ 
& = & \frac{1}{m^2} \sum_{i=-1}^{1} (D-4)^i \, F_{9}^{(i)}(x) 
  + {\mathcal O} \left( (D-4)^2 \right) ,
\eea
where:
\bea
\hspace*{-6mm} 
F_{7}^{(-2)}(x) & = & \frac{1}{8} \, , \\
\hspace*{-6mm} 
F_{7}^{(-1)}(x) & = &    - \frac{5}{16}
          - \frac{1}{8} H(0;x)
          - \frac{1}{4} H(1;x) \, , \\
\hspace*{-6mm} 
F_{7}^{(0)}(x) & = & \frac{1}{32} \Bigg\{
            19 \! 
          - 8 \zeta(2) \! 
          +  \! 10 H(0;x) \! 
          +  \! 20 H(1;x) \! 
          +  \! 2 H(0,0;x) \! 
          +  \! 4 H(0,1;x) \! 
          \nn \\
\hspace*{-6mm}           & &
          + 6 H(1,0;x)
          + 12 H(1,1;x)
          + \frac{1}{(1+x)} \Big[
               8 \zeta(2)
             + 4 H(0,1;x) 
          \nn \\
\hspace*{-6mm}           & &
             + 2 H(0,0;x) 
          \Big]
          + \Bigg[ \frac{1}{(1-x)} - \frac{1}{(1+x)}
          \Bigg] \Big[
          - 2 \zeta(3) 
          + 2 \zeta(2) H(0;x) 
          \nn \\
\hspace*{-6mm}           & &
          +  \! H(0,0,0;x)  \! 
          +  \! 2 H(0,0,1;x)  \! 
          +  \! 2 H(0,1,0;x)  \! 
          +  \! 4 H(0,1,1;x) 
          \Big]
     \Bigg\} , \\
\hspace*{-6mm} F_{7}^{(1)}(x) & = &  
          - \frac{65}{64}
          + \frac{1}{8} \zeta(3)
     - \frac{1}{32} \Bigl[ 
             19 H(0;x)
    + (38 - 9 \zeta(2)) H(1;x) 
    + 10 H(0,0;x) \nn\\
\hspace*{-6mm} & & 
    + 20 H(0,1;x)
    + 15 H(1,0;x)
    + 30 H(1,1;x)
    + 4 H(0,0,0;x) \nn\\
\hspace*{-6mm} & & 
    + 8 H(0,0,1;x)
    + 10 H(1,1,0;x)
    + 6 H(0,1,0;x)
    + 12 H(0,1,1;x) \nn\\
\hspace*{-6mm} & & 
    + 20 H(1,1,1;x)
    + 2 H(-1,0,0;x)
    + 7 H(1,0,0;x)
    + 2 H(-1,0,1;x) \nn\\
\hspace*{-6mm} & & 
    + 14 H(1,0,1;x) \Bigr]
     + \frac{1}{32} \biggl[ 1 - \frac{1}{(1+x)} \biggr] 
        \Bigl\{ 
      20 \zeta(2)
    + 9 \zeta(3)
    - \zeta(2) \bigl[ 
       H(0;x) \nn\\
\hspace*{-6mm} & & 
    +  \! 4 H(-1;x)  \! 
    +  \! 10 H(1;x) \bigr] \! 
    +  \! 5 H(0,0;x) \! 
    +  \! 10 H(0,1;x) \! 
    +  \! 2 H(0,0,0;x) \nn\\
\hspace*{-6mm} & & 
    + 4 H(0,0,1;x)
    + 2 H(0,1,0;x)
    + 4 H(0,1,1;x)
    + 2 H(-1,0,0;x) \nn\\
\hspace*{-6mm} & & 
    + 4 H(1,0,0;x)
    + 4 H(-1,0,1;x)
    + 8 H(1,0,1;x) \Bigr\} \nn\\
\hspace*{-6mm} & & 
     + \frac{1}{64} \biggl[ \frac{1}{(1-x)}  - \frac{1}{(1+x)} \biggr] 
        \biggl\{ \frac{32}{5} \zeta^{2}(2)
    + 2 \zeta(3)
    - ( 2 \zeta(2) - 3 \zeta(3) ) H(0;x) \nn\\
\hspace*{-6mm} & & 
    + 4 \zeta(3) H(-1;x)
    - \zeta(2) \bigl[ 
       H(0,0;x)
    + 4 H(-1,0;x)
    + 2 H(0,1;x) \bigr] \nn\\
\hspace*{-6mm} & & 
    - H(0,0,0;x)
    + 2 H(0,0,1;x)
    + 2 H(0,1,0;x)
    + 4 H(0,1,1;x) \nn\\
\hspace*{-6mm} & & 
    + 3 H(0,0,0,0;x)
    + 2 H(-1,0,0,0;x)
    + 4 H(-1,0,0,1;x) \nn\\
\hspace*{-6mm} & & 
    + 4 H(-1,0,1,0;x)
    + 8 H(-1,0,1,1;x)
    + 6 H(0,0,0,1;x) \nn\\
\hspace*{-6mm} & & 
    + 4 H(0,0,1,0;x)
    + 8 H(0,0,1,1;x)
    + 6 H(0,1,0,0;x) \nn\\
\hspace*{-6mm} & & 
    + 12 H(0,1,0,1;x)
    + 12 H(0,1,1,0;x)
    + 24 H(0,1,1,1;x) \biggr\}
,  \\
\hspace*{-6mm} F_{8}^{(-2)}(x) & = & - \frac{1}{64} \Bigg[
            \frac{1}{x}
          + x
          + 2
          \Bigg] \, , \\
\hspace*{-6mm} F_{8}^{(-1)}(x) & = & \frac{1}{256} \bigg[
            \frac{1}{x}
          + x
          + 2
          \bigg] \Big[
            9 
          + 4 H(0;x)
          + 8 H(1;x)
          \Big] \, , \\
\hspace*{-6mm} F_{8}^{(0)}(x) & = & \frac{1}{1024} \Bigg\{
          -  \! 126 \! 
          -  \! 63 \frac{1}{x} \! 
          -  \! 63 x \! 
          -  \! 16 \zeta(2) \! 
          +  \! 32 \zeta(2) x  \! 
          +  \! \frac{128 \zeta(2)}{(1+x)}  \! 
          - \Bigg[
               32  \! 
             +  \! 16 \frac{1}{x}
          \nn \\
          & &
             +  \! 8  x \! 
             -  \! 32 \frac{1}{(1+x)}
            \Bigg] 
            \big[ H(0,0;x)  \! +  \! 2 H(0,1;x) \big]
          - \bigg[
            40  \! +  \! \frac{36}{x}  \! +  \! 36 x  
            \bigg] 
            \big[
                  H(0;x) 
          \nn \\
          & &
              + 2 H(1;x)
            \big]
          - \bigg[
            32 + \frac{24}{x} + 24 x  
            \bigg] \big[
                  H(1,0;x) 
              + 2 H(1,1;x)
            \big]
            \nn \\
            & &
          + \Bigg[   \frac{1}{(1-x)} 
                    - \frac{1}{(1+x)} 
            \Bigg]  
            \big[
               16 \zeta(3) 
             - 16 \zeta(2) H(0;x) 
             - 8 H(0,0,0;x) 
            \nn \\
             & &
             - 16 H(0,0,1;x)
             - 16 H(0,1,0;x)
             - 32 H(0,1,1;x)
            \big]      
        \Bigg\} \, , \\
\hspace*{-6mm} F_{8}^{(1)}(x) & = &   
         \frac{405}{4096} \biggl[ 2 +  \frac{1}{x} + x \biggr] 
       + \biggl[ \frac{23}{256} - \frac{9}{128}x  \biggr] \zeta(2)
       + \frac{1}{64} \biggl[ \frac{1}{2} -  \frac{1}{x} - \frac{13}{4}
         x \biggr]  \zeta(3) \nn\\
& & 
             - \frac{1}{4(1+x)} \zeta(2)
       + \frac{1}{256} \biggl[ \frac{1}{(1-x)} - \frac{29}{(1+x)}  
         \biggr] \zeta(3)
       + \frac{1}{1024}  \biggl[ 
            30
          + \frac{63}{x}  \nn\\
& & 
          + ( 63 + 4 \zeta(2) ) x \biggr] H(0;x)
       + \frac{1}{64}  \biggl[ 
            4
          + \frac{1}{x} 
          - x \biggr] \zeta(2) H(-1;x)
       + \frac{1}{512}  \biggl[ 
            30  \nn\\
& & 
          + 56 \zeta(2)
          + \frac{( 63 + 10 \zeta(2) )}{x}  
          + ( 63 - 22 \zeta(2) ) x \biggr] H(1;x)  \nn\\
& & 
       - \frac{1}{64(1+x)} \Bigl\{ 
         \zeta(2) \bigl[ H(0;x)
   + 8 H(-1;x)
   + 16 H(1;x) \bigr]
   + 4 H(0,0;x)  \nn\\
& & 
   + 8 H(0,1;x) \Bigr\}
       + \frac{3}{512} \biggl[ 8 + \frac{6}{x} + 3 x \biggr]
          \bigl[ H(0,0;x) 
   + 2 H(0,1;x) \bigr]  \nn\\
& & 
       + \frac{1}{512} \biggl[ 16 + \frac{27}{x} + 27 x \biggr]
          \bigl[ H(1,0;x) 
   + 2 H(1,1;x) \bigr]
     + \frac{1}{512} \biggl[ 16 + \frac{8}{x} + 4x   \nn\\
& & 
                            - \frac{1}{(1-x)}
                            - \frac{23}{(1+x)} \biggr] 
   \bigl[ H(0,0,0;x)
         + 2 H(0,0,1;x) \bigr]
     + \frac{1}{256} \biggl[ 12 + \frac{12}{x}   \nn\\
& & 
                            + 4x - \frac{2}{(1-x)}
                            - \frac{30}{(1+x)} \biggr] 
   \bigl[ H(0,1,0;x)
         + 2 H(0,1,1;x) \bigr]
     + \frac{1}{256} \biggl[ 4   \nn\\
& & 
                    + \frac{1}{x} - x - \frac{8}{(1+x)} 
         \biggr] 
   \bigl[ H(-1,0,0;x)
         + 2 H(-1,0,1;x) \bigr]
     + \frac{1}{256} \biggl[ 12 + \frac{7}{x}   \nn\\
& & 
                    + 3x - \frac{16}{(1+x)} 
         \biggr] 
   \bigl[ H(1,0,0;x)
         + 2 H(1,0,1;x) \bigr] 
     + \frac{1}{128} \biggl[ 4 + \frac{5}{x} + 5x 
         \biggr] \times \nn\\
& & 
   \times \bigl[ H(1,1,0;x)
         + 2 H(1,1,1;x) \bigr] 
    - \frac{1}{256} \biggl[ \frac{1}{(1-x)} - \frac{1}{(1+x)} \biggr]
 \biggl\{  
     \frac{5}{32} \zeta^{2}(2) \nn\\
& & 
    + ( \zeta(2) + 3 \zeta(3) )H(0;x)
    + 4 \zeta(3) H(-1;x)
    - \zeta(2) \bigl[ H(0,0;x) \nn\\
& & 
    + 4 H(-1,0;x)
    - H(0,1;x) \bigr]
    - 3 H(0,0,0,0;x)
    - 2 H(-1,0,0,0;x) \nn\\
& & 
    - 4 H(-1,0,0,1;x)
    - 4 H(-1,0,1,0;x)
    - 8 H(-1,0,1,1;x) \nn\\
& & 
    - 6 H(0,0,0,1;x)
    - 4 H(0,0,1,0;x)
    - 8 H(0,0,1,1;x) \nn\\
& & 
    - 6 H(0,1,0,0;x)
    - 12 H(0,1,0,1;x)
    - 12 H(0,1,1,0;x) \nn\\
& & 
    - 24 H(0,1,1,1;x)
    \biggr\}   \, , \\
\hspace*{-6mm} F_{9}^{(-1)}(x) & = & \frac{1}{16} 
         \Bigg[ \frac{1}{(1-x)} - \frac{1}{(1+x)}  
         \Bigg] \big[ 
            4 \zeta(2)
          + H(0,0;x)
          + 2 H(0,1;x)
         \big]  \, , \\
\hspace*{-6mm} F_{9}^{(0)}(x) & = & 
     - \frac{1}{32} 
         \Bigg[ \frac{1}{(1-x)} - \frac{1}{(1+x)}  
         \Bigg] \big[ 
            9 \zeta(3) 
          + 8 \zeta(2) H(-1;x)  
          - \zeta(2)  H(0;x) 
          \nn \\
          & &
          +  \! 16 \zeta(2) H(1;x) \! 
          +  \! 2 H(-1,0,0;x)  \! 
          + \!  4 H(-1,0,1;x)  \! 
          +  \! 2 H(0,0,0;x)  \! 
          \nn \\
& &
          + 4 H(0,0,1;x) 
          + 2 H(0,1,0;x) 
          + 4 H(0,1,1;x) 
          + 4 H(1,0,0;x) 
          \nn \\
& &
          + 8 H(1,0,1;x) 
         \big] \, , \\
\hspace*{-6mm} F_{9}^{(1)}(x) & = &  \frac{1}{64} \biggl[ \frac{1}{(1-x)}
- \frac{1}{(1+x)} \biggr] \biggl\{ 
            \frac{287}{10} \zeta^{2}(2)
    - \zeta(3) \Bigl[ 3 H(0;x)
   - 32 H(1;x) \nn \\
& &
   - 18 H(-1;x) \Bigr]
    + \zeta(2) \Bigl[ 14 H(0,1;x)
          + 64 H(1,1;x)
   + 8 H(0,-1;x) \nn \\
& &
   - 2 H(-1,0;x)
   + 32 H(-1,1;x)
   + 32 H(1,-1;x)
   + 16 H(-1,-1;x) \Bigr] \nn \\
& &
   + 4 H(0,0,0,0;x)
   + 4 H(-1,-1,0,0;x)
   + 8 H(-1,-1,0,1;x) \nn \\
& &
   + 4 H(-1,0,0,0;x)
   + 8 H(-1,0,0,1;x)
   + 4 H(-1,0,1,0;x) \nn \\
& &
   + 8 H(-1,0,1,1;x)
   + 8 H(-1,1,0,0;x)
   + 16 H(-1,1,0,1;x) \nn \\
& &
   + 2 H(0,-1,0,0;x)
   + 4 H(0,-1,0,1;x)
   + 8 H(0,0,0,1;x) \nn \\
& &
   + 6 H(0,0,1,0;x)
   + 12 H(0,0,1,1;x)
   + 6 H(0,1,0,0;x) \nn \\
& &
   + 12 H(0,1,0,1;x)
   + 4 H(0,1,1,0;x)
   + 8 H(0,1,1,1;x) \nn \\
& &
   + 8 H(1,-1,0,0;x)
   + 16 H(1,-1,0,1;x)
   + 10 H(1,0,0,0;x) \nn \\
& &
   + 20 H(1,0,0,1;x)
   + 12 H(1,0,1,0;x)
   + 24 H(1,0,1,1;x) \nn \\
& &
   + 16 H(1,1,0,0;x)
   + 32 H(1,1,0,1;x) \biggr\}  \, .
\eea
%%%%%%%%%%%%%%%%%%%%%%%%%%%%%%%%%%%%%%%%%%%%%%%%%%%%%%%%%%%%%%%%%%%%%

%%%%%%%%%%%%%%%%%%%%%%%%%%%%%%%%%%%%%%%%%%%%%%%%%%%%%%%%%%%%%%%%%%%%%
\bea
\parbox{35mm}{
\begin{fmfgraph*}(20,20)
\fmfleft{i1,i2}
\fmfright{o}
\fmf{photon}{i1,v1}
\fmf{photon}{i2,v2}
\fmf{dashes}{v3,o}
\fmf{plain,tension=.3}{v2,v3}
\fmf{plain,tension=.3}{v1,v3}
\fmf{plain,tension=0,left=.5}{v2,v1}
\fmf{photon,tension=0,left=.5}{v1,v2}
\end{fmfgraph*} }
& = & \int {\mathfrak{D}^D k_1} {\mathfrak{D}^D k_2}
           \frac{1}{
                     {\mathcal D}_{1}
                     {\mathcal D}_{10}
                     {\mathcal D}_{20}
                     {\mathcal D}_{21}
                     } \nn\\ 
& = & \sum_{i=-2}^{1} (D-4)^i \, F_{10}^{(i)}(x) + {\mathcal O} \left(
(D-4)^2 \right) , \\
\parbox{20mm}{
\begin{fmfgraph*}(20,20)
\fmfleft{i1,i2}
\fmfright{o}
\fmf{photon}{i1,v1}
\fmf{photon}{i2,v2}
\fmf{dashes}{v3,o}
\fmf{plain,tension=.3}{v2,v3}
\fmf{plain,tension=.3}{v1,v3}
\fmf{plain,tension=0,left=.5}{v2,v1}
\fmf{photon,tension=0,left=.5}{v1,v2}
%\fmflabel{$(k_{1} \cdot k_{2})$ }{o}
\end{fmfgraph*} } \ (k_{1} \cdot k_{2})
& = & \int {\mathfrak{D}^D k_1} {\mathfrak{D}^D k_2}
           \frac{(k_1 \cdot k_2)}{
                     {\mathcal D}_{1}
                     {\mathcal D}_{10}
                     {\mathcal D}_{20}
                     {\mathcal D}_{21}
                     } \nn\\ 
& = & m^2 \sum_{i=-2}^{1} (D-4)^i \, F_{11}^{(i)}(x) + {\mathcal O} \left(
(D-4)^2 \right) , \\
\parbox{35mm}{
\begin{fmfgraph*}(20,20)
\fmfleft{i1,i2}
\fmfright{o}
\fmfforce{.43w,.5h}{d1}
\fmf{photon}{i1,v1}
\fmf{photon}{i2,v2}
\fmf{dashes}{v3,o}
\fmf{plain,tension=.3}{v2,v3}
\fmf{plain,tension=.3}{v1,v3}
\fmf{plain,tension=0,left=.5}{v2,v1}
\fmf{photon,tension=0,left=.5}{v1,v2}
\fmfv{decor.shape=circle,decor.filled=full,decor.size=.1w}{d1}
\fmflabel{3}{d1}
\end{fmfgraph*} }
& = & \int {\mathfrak{D}^D k_1} {\mathfrak{D}^D k_2}
           \frac{1}{
                     {\mathcal D}^3_{1}
                     {\mathcal D}_{10}
                     {\mathcal D}_{20}
                     {\mathcal D}_{21}
                     } \nn\\ 
& = & \frac{1}{m^4} \sum_{i=0}^{2} (D-4)^i \, F_{12}^{(i)}(x) 
  + {\mathcal O} \left( (D-4)^3 \right) ,
\eea
where:
\bea
\hspace*{-7mm} F_{10}^{(-2)}(x) & = & \frac{1}{8} \, , \\
\hspace*{-7mm} F_{10}^{(-1)}(x) & = &    - \frac{5}{16}
          + \frac{1}{8} \Biggl[ 1 - \frac{2}{(1-x)} \Biggr] H(0;x) \, , 
   \\
\hspace*{-7mm} F_{10}^{(0)}(x) & = & \frac{19}{32} 
           - \frac{1}{16(1-x)} \Biggl[ 1 - \frac{1}{(1-x)} \Biggr] 
      \Bigl\{ 4 \zeta(3) \! 
     + \! 2 \zeta(2) H(0;x) \! 
     + \! H(0,0,0;x) \nn\\
& & 
     + 2 H(0,1,0;x) \Bigr\} 
    - \frac{1}{16} \Biggl[ 1 - \frac{2}{(1-x)} \Biggr] 
      \Bigl\{ 5 H(0;x)
      + H(0,0;x) \nn\\
& & 
      - 2 H(-1,0;x)
      + H(1,0;x) \Bigr\}
    + \frac{1}{16(1-x)} H(0,0;x) \, , \\
\hspace*{-7mm} F_{10}^{(1)}(x) & = & - \frac{65}{64}  \! 
          +  \! \frac{1}{32(1-x)} \Bigl\{ 
     5 \zeta(3) \! 
   +  \! \zeta(2) H(0;x) \! 
   - 5 H(0,0;x)
   - 2 H(0,0,0;x) \nn\\
& & 
   + 2 H(0,-1,0;x) \Bigr\} 
   + \frac{1}{16} H(0,1,0;x)
  + \frac{1}{32} \Biggl[ 1 - \frac{2}{(1-x)} \Biggr]
    \Bigl\{ 
    \zeta(3) \nn\\
& & 
  + ( 19 + \zeta(2) ) H(0;x)
  + \zeta(2) H(1;x)
  + 5 H(0,0;x)
  + 5 H(1,0;x) \nn\\
& & 
  - 10 H( \! -1,0;x) \! 
  + \! 2 H(0,0,0;x) \! 
  + \!  4 H( \! -1, \! -1,0;x) \! 
  -  \! 3 H( \! -1,0,0;x) \nn\\
& & 
  - 2 H(-1,1,0;x) \! 
  -  \! 2 H(0, \! -1,0;x) \! 
  -  \! 2 H(1, \! -1,0;x) \! 
  +  \! H(1,0,0;x) \nn\\
& & 
  + 2 H(1,1,0;x) \Bigr\} \! 
   +  \! \frac{1}{32(1-x)} \Biggl[ 1 \! - \! \frac{1}{(1-x)} \Biggr]
    \Biggl\{ 
    4 \zeta(3) \! 
  - \! \frac{22}{5} \zeta^2(2) \! 
  + \! ( 2\zeta(2) \nn\\
& & 
         - 3 \zeta(3) ) H(0;x)
  - 8 \zeta(3) H(1;x)
  + \! \zeta(2) \Bigl[ 
    H(0,0;x)\! 
         + \! 2 H(0,1;x) \nn\\
& & 
  - 4 H(0,-1;x)
  - 4 H(1,0;x) \Bigr]
  + H(0,0,0;x)
  - 2 H(0,1,0;x) \nn\\
& & 
  + 3 H(0,0,0,0;x)
  - 2 H(0,-1,0,0;x)
  - 4 H(0,-1,1,0;x) \nn\\
& & 
  - 2 H(0,0,-1,0;x)
  + 2 H(0,0,1,0;x)
  - 4 H(0,1,-1,0;x) \nn\\
& & 
  + 4 H(0,1,0,0;x)
  + 4 H(0,1,1,0;x)
  - 2 H(1,0,0,0;x) \nn\\
& & 
  - 4 H(1,0,1,0;x) \Biggr\} \, , \\
\hspace*{-7mm} F_{11}^{(-2)}(x) & = & - \frac{1}{16} 
          + \frac{1}{32} 
         \Biggl[ x 
  + \frac{1}{x} \Biggr] \, , \\
\hspace*{-7mm} F_{11}^{(-1)}(x) & = &   \frac{9}{64}
          - \frac{1}{128} 
         \Biggl[ x 
  + \frac{1}{x} \Biggr] \Bigl\{ 
    9
  + 4 H(0;x) \Bigr\}
  - \frac{1}{16} \Biggl[
    1
  - \frac{2}{(1-x)} \Biggr] H(0;x) \, , \\
\hspace*{-7mm} F_{11}^{(0)}(x) & = & 
          - \frac{71}{256}
          + \frac{63}{256} x
   + \frac{1}{64} \Biggl[ \frac{1}{x} - \frac{2}{(1-x)} \Biggr] 
     H(0,0;x)
          + \frac{1}{512} \Biggl[ x - \frac{1}{x} \Biggr]
     \Bigl\{
     63 \nn\\
& & 
   + 18 H(0;x)
   - 8 H(0,0;x)
   + 16 H(-1,0;x)
   - 8 H(1,0;x) \Bigr\} \nn\\
& & 
   + \frac{1}{32} \Biggl[ 1  -  \frac{2}{(1-x)} \Biggr]
     \Bigl\{ 
     4 H(0;x)  
   +  H(0,0;x) 
   -  2 H(-1,0;x)  \nn\\
& & 
   +   H(1,0;x) \Bigr\}
   + \frac{1}{32(1-x)} \Biggl[ 1 \! - \! \frac{1}{(1-x)} \Biggr]
    \Bigl\{ 
     4 \zeta(3)
   + 2 \zeta(2) H(0;x) \nn\\
& & 
   + H(0,0,0;x)
   + 2 H(0,1,0;x) \Bigr\} \, , \\
\hspace*{-7mm} F_{11}^{(1)}(x) & = & 
          + \frac{525}{1024}
          - \frac{1}{64} \zeta(3) 
   - \frac{1}{64} \Biggl[ 1 - \frac{1}{(1-x)^2} \Biggr] \zeta(2)
      H(0;x)
   - \frac{x}{1024} \Bigl[ 405  \nn\\
& & 
   - 40 \zeta(3) 
   - 8 \zeta(2) H(0;x)
   + 36 H(0,0;x)
   + 16 H(0,0,0;x) \nn\\
& & 
   + 16 H(1,0,0;x)
   - 16 H(0,-1,0;x) \Bigr]
   - \frac{1}{128(1-x)} \Bigl[
     6 \zeta(3) \nn\\
& & 
   - 11 H(0,0;x)
   - 4 H(0,0,0;x)
          - 4 H(1,0,0;x)
   + 4 H(0,-1,0;x) \Bigr] \nn\\
& & 
   + \frac{1}{2048} \Biggl[ x - \frac{1}{x} \Biggr] \Bigl\{
     405
   - 4 \Bigl[ 
     16 \zeta(3)
   - 63 H(0;x)
   - 4 \zeta(2) H(1;x) \nn\\
& & 
   - 36 H(0,0;x)
   - 18 H(1,0;x)
   + 36 H(-1,0;x) \Bigr]
   - 16 \Bigl[
     4 H(0,0,0;x) \nn\\
& & 
   + 2 H(0,1,0;x)
   - 3 H(-1,0,0;x)
   - 2 H(-1,1,0;x)
   + 3 H(1,0,0;x) \nn\\
& & 
   + \! 2 H(1, \! 1, \! 0;x) \! 
   + \! 4 H( \! -1,\! -1, \! 0;x) \! 
   - \! 2 H(1,\! -1, \! 0;x) \! 
   - \! 4 H(0,\! -1, \! 0;x) \Bigr] \Bigr\} \nn\\
& & 
          - \frac{1}{128} \Biggl[ 1 - \frac{2}{(1-x)} \Biggr] 
     \Bigl\{
     \zeta(2) \! 
   + 27 H(0;x) \! 
   + 2 \zeta(2) H(1;x) \! 
   + 7 H(0,0;x) \nn\\
& & 
   + 9 H(1,0;x)
   - 16 H(-1,0;x)
   + 4 H(0,0,0;x)
   + 4 H(0,1,0;x) \nn\\
& & 
   - 6 H(-1,0,0;x) \! 
   - 4 H(-1,1,0;x) \! 
   +  \! 2 H(1,0,0;x) \! 
   +  \! 4 H(1,1,0;x) \nn\\
& & 
   + 8 H(-1,-1,0;x)
   - 4 H(1,-1,0;x)
   - 4 H(0,-1,0;x) \Bigr\} \nn\\
& & 
   - \frac{1}{64(1 \! - \! x)} \Biggl[ 1 \! - \! \frac{1}{(1 \! - \! x)} \Biggr]
     \Biggl\{ \! 
     2 \zeta(3) \! 
   -  \! \frac{22}{5} \zeta^2(2) \! 
   -  \! \zeta(3) \bigl[ 3 H(0;x) \! 
   +  \! 8 H(1;x) \bigr] \nn\\
& & 
   + \zeta(2) \bigl[
     H(0,0;x)
   - 4 H(1,0;x)
   + 2 H(0,1;x)
   - 4 H(0,-1;x) \bigr] \nn\\
& & 
   + \frac{1}{2} H(0,0,0;x)
   + H(0,1,0;x)
   + 3 H(0,0,0,0;x) \nn\\
& & 
   - 2 H(0,-1,0,0;x)
   - 4 H(0,-1,1,0;x)
   - 2 H(0,0,-1,0;x) \nn\\
& & 
   + 2 H(0,0,1,0;x)
   - 4 H(0,1,-1,0;x)
   + 4 H(0,1,0,0;x) \nn\\
& & 
   + 4 H(0,1,1,0;x)
   - 2 H(1,0,0,0;x)
   - 4 H(1,0,1,0;x) \Biggr\}  , \\
\hspace*{-7mm} F_{12}^{(0)}(x) & = & 
        - \frac{1}{32(1-x)} \Biggl[ 1 \! - \! \frac{1}{(1-x)} \Biggr]
   H(0,0;x) \, , \\
\hspace*{-7mm} F_{12}^{(1)}(x) & = & 
        - \frac{1}{64(1-x)} \Biggl[ 1  -  \frac{1}{(1-x)} \Biggr]
   \Bigl\{
   9 \zeta(3)
 + 3 \zeta(2) H(0;x)
 - H(0,0,0;x) \nn\\
& & 
 + 2 H(0,-1,0;x)
 + 2 H(0,1,0;x)
 + 2 H(1,0,0;x) \Bigr\} \, , \\
\hspace*{-7mm} F_{12}^{(2)}(x) & = & 
        - \frac{1}{256(1-x)} 
 \Biggl[ 1-\frac{1}{(1-x)} \Biggr]
   \Bigl\{
           11 \zeta^2(2)  
 +  12 \zeta(3) H(1;x)  \nn\\
& & 
 -  2 \zeta(2) \Bigl[
   H(0,0;x)
 - 6 \bigl( H(0,-1;x) + H(1,0;x) \bigr)
 + 2 H(0,1;x) \Bigr] \nn\\
& & 
 + 2 H(0,0,0,0;x)
 - 4 \Bigl[ 
   H(0,0,-1,0;x)
 + H(0,1,0,0;x) \Bigr] \nn\\
& & 
 + 8 \Bigl[
   H(0,-1,-1,0;x)
 + H(0,-1,1,0;x)
 + H(0,1,-1,0;x) \nn\\
& & 
 - H(0,1,1,0;x)
 - H(1,0,-1,0;x)
 + 2 H(1,0,0,0;x) \nn\\
& & 
 + 2 H(1,0,1,0;x)
 + H(1,1,0,0;x) \Bigr] \Bigr\} \, ,
\eea

%%%%%%%%%%%%%%%%%%%%%%%%%%%%%%%%%%%%%%%%%%%%%%%%%%%%%%%%%%%%%%%%%%%%%

As we already pointed out, the following MIs, involved 
only in the calculation of the topology (h) in Fig. \ref{fig2}, are 
given in terms of HPLs of the variable $\bar{x}$ defined in 
Eq.~(\ref{xbar}). In agreement with \cite{UgoRo}, we find:

\bea
\parbox{25mm}{
\begin{fmfgraph*}(20,20)
\fmfleft{i1,i2}
\fmfright{o}
\fmf{photon}{i1,v1}
\fmf{photon}{i2,v2}
\fmf{dashes}{v3,o}
\fmf{photon,tension=.3}{v2,v3}
\fmf{photon,tension=.3}{v1,v3}
\fmf{plain,tension=0,left=.5}{v2,v1}
\fmf{plain,tension=0,left=.5}{v1,v2}
\end{fmfgraph*} }
& = & \int {\mathfrak{D}^D k_1} {\mathfrak{D}^D k_2}
           \frac{1}{
                     {\mathcal D}_{2}
                     {\mathcal D}_{10}
                     {\mathcal D}_{19}
                     {\mathcal D}_{21}
                     } \nn\\ 
& = & \sum_{i=-2}^{1} (D-4)^i \, F_{13}^{(i)}(\bar{x}) + {\mathcal O} \left(
(D-4)^2 \right) , 
\label{UGO1} \\
\parbox{25mm}
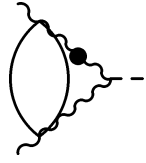
{
\begin{fmfgraph*}(20,20)
\fmfleft{i1,i2}
\fmfright{o}
\fmfforce{.5w,.65h}{d1}
\fmf{photon}{i1,v1}
\fmf{photon}{i2,v2}
\fmf{dashes}{v3,o}
\fmf{photon,tension=.3}{v2,v3}
\fmf{photon,tension=.3}{v1,v3}
\fmf{plain,tension=0,left=.5}{v2,v1}
\fmf{plain,tension=0,left=.5}{v1,v2}
%\fmflabel{$(p_{1} \cdot k_{1})$ }{o}
\fmfv{decor.shape=circle,decor.filled=full,decor.size=.1w}{d1}
\end{fmfgraph*} } 
& = & \int {\mathfrak{D}^D k_1} {\mathfrak{D}^D k_2}
           \frac{1}{
                     {\mathcal D}_{2}
                     {\mathcal D}_{10}
                     {\mathcal D}_{19}^{2}
                     {\mathcal D}_{21}
                     } \nn\\ 
& = & \frac{1}{m^2} \sum_{i=-2}^{2} (D-4)^i \, F_{14}^{(i)}(\bar{x}) 
  + {\mathcal O} \left( (D-4)^3 \right) , 
\label{UGO2} 
\eea
where:
\bea
\hspace*{-5mm} F_{13}^{(-2)}(\bar{x}) & = & \frac{1}{8}  \, , \\
\hspace*{-5mm} F_{13}^{(-1)}(\bar{x}) & = & - \frac{5}{16} 
               - \frac{1}{8} \Bigl[
           H(0;\bar{x})
  - 2 H(-1;\bar{x}) \Bigr] \, , \\
\hspace*{-5mm} F_{13}^{(0)}(\bar{x}) & = &  \frac{19}{32} 
          - \frac{1}{8} \Biggl[ 
     1 
   - \frac{1}{(1+\bar{x})} \Biggr] \zeta(2)
   + \frac{5}{16} \Bigl[
           H(0;\bar{x})
  - 2 H(-1;\bar{x}) \Bigr] \nn\\
& & 
   - \frac{1}{8} \Bigl[
           H(-1,0;\bar{x})
  - 2 H(-1,-1;\bar{x}) \Bigr]
   + \frac{1}{8(1+\bar{x})} \Bigl[
           H(0,0;\bar{x}) \nn\\
& & 
  - 2 H(0,-1;\bar{x}) \Bigr]
          - \frac{1}{8(1+\bar{x})} \Biggl[  
     1 
   - \frac{1}{(1+\bar{x})} \Biggr] \Bigl[
                  2 \zeta(3)
         - \zeta(2) H(0;\bar{x}) \nn\\
& & 
  - H(0,0,0;\bar{x})
  + 2 H(0,0,-1;\bar{x}) \Bigr]  \, , \\
\hspace*{-5mm} F_{13}^{(1)}(\bar{x}) & = &  - \frac{1}{64} \Bigl\{
            65 \! 
   + \! 38 \bigl[ 
     H(0;\bar{x}) \! 
   - \! 2 H(-1;\bar{x}) \bigr] \! 
   + \! 8 \zeta(2)  H(1;\bar{x}) \! 
   - \! 20 H(\! -1,0;\bar{x}) \nn\\
& & 
   + 40 H(-1,-1;\bar{x})
   + 8 H(-1,-1,0;\bar{x})
   - 16 H(-1,-1,-1;\bar{x}) \Bigr\} \nn\\
& & 
        + \frac{1}{16(1+\bar{x})} \Bigl[
     2 \zeta(3)
   - 5 H(0,0;\bar{x})
   + 10 H(0,-1;\bar{x})
   - 2 H(0,0,0;\bar{x}) \nn\\
& & 
   - 4 H(-1,0,  -1;\bar{x})  
   + 2 H(-1,0,0;\bar{x})  
   + 4 H(0,0,  -1;\bar{x})   \nn\\
& & 
   + 4 H(1,0,  -1;\bar{x})
   - 2 H(1,0,0;\bar{x})
   - 4 H(0,-1,-1;\bar{x}) \nn\\
& & 
   + \! 2 H(0,\! -1,0;\bar{x}) \Bigr]
        + \! \frac{1}{16} \Biggl[ 1 - \frac{1}{(1\! +\! \bar{x})} \Biggr] 
   \Bigl\{
     5 \zeta(2)\! 
   - \! 2 \zeta(3) \! 
   + \! 2 \zeta(2) \Bigl[ 
       H(0;\bar{x}) \nn\\
& & 
     - H(-1;\bar{x})
     + 2 H(1;\bar{x}) \Bigr]
   + 2 H(0,0,0;\bar{x})
   - 4 H(0,0,-1;\bar{x}) \nn\\
& & 
   - 4 H(1,0, \! -1;\bar{x}) \! 
   +  \! 2 H(1,0,0;\bar{x}) \Bigr\}
        - \frac{1}{32(1 \! + \! \bar{x})} \Biggl[ 1 - \frac{1}{(1 \! + \! \bar{x})} 
   \Biggr] \Bigl\{
     \zeta^2(2) \nn\\
& & 
   - 12 \zeta(3)
   + 6 \zeta(2) H(0;\bar{x})
   - 4 \zeta(3) \Bigl[
       H(0;\bar{x}) 
     + 2 H(-1;\bar{x}) \Bigr] \nn\\
& & 
   + 2 \zeta(2) \Bigl[ 
       H(0,0;\bar{x})
     - 2 H(0,-1;\bar{x})
     + 2 H(-1,0;\bar{x})
     + 4 H(0,1;\bar{x}) \Bigr] \nn\\
& & 
   + 6 H(0,0,0;\bar{x})
   - 12 H(0,0,-1;\bar{x})
   + 6 H(0,0,0,0;\bar{x}) \nn\\
& & 
   - 8 H(-1,0,0,-1;\bar{x})
   + 4 H(-1,0,0,0;\bar{x})
   + 8 H(0,-1,0,-1;\bar{x}) \nn\\
& & 
   - 4 H(0,-1,0,0;\bar{x})
   + 8 H(0,0,-1,-1;\bar{x})
   - 4 H(0,0,-1,0;\bar{x}) \nn\\
& & 
   - 12 H(0,0,0,-1;\bar{x})
   - 16 H(0,1,0,-1;\bar{x})
   + 8 H(0,1,0,0;\bar{x}) \Bigr\}  \, , \\
\hspace*{-5mm} F_{14}^{(-2)}(\bar{x}) & = & 
   - \frac{1}{4(1+\bar{x})} 
                 \Biggl[ 
   1 
   - \frac{1}{(1+\bar{x})} \Biggr] \, , \\
\hspace*{-5mm} F_{14}^{(-1)}(\bar{x}) & = & 
     \frac{1}{8(1+\bar{x})} 
                 \Biggl[ 
   1 
        - \frac{1}{(1+\bar{x})} \Biggr] \Bigl[
           H(0;\bar{x})
  - 2 H(-1;\bar{x}) \Bigr] \, , \\
\hspace*{-5mm} F_{14}^{(0)}(\bar{x}) & = & 
     \frac{1}{8(1+\bar{x})} 
                 \Biggl[ 
   1 \! 
        - \! \frac{2}{(1+\bar{x})} 
        + \! \frac{1}{(1+\bar{x})^2} \Biggr] \zeta(2)
     - \frac{1}{8(1+\bar{x})} 
                 \Biggl[ 
   1 \! 
        - \! \frac{1}{(1+\bar{x})} \Biggr] \Bigl\{
          2 \nn\\
& & 
        + H(0;\bar{x})
        - 2 H(-1;\bar{x})
        + 2 H(-1,-1;\bar{x})
        - H(-1,0;\bar{x}) \Bigl\} \nn\\
& & 
     - \frac{1}{8(1+\bar{x})^2} 
                 \Biggl[ 
   1 
        - \frac{1}{(1+\bar{x})} \Biggr] \Bigl\{
          H(0,0;\bar{x})
        - 2 H(0,-1;\bar{x}) \Bigr\}  \, , \\
\hspace*{-5mm} F_{14}^{(1)}(\bar{x}) & = & 
       \frac{1}{8(1+\bar{x})} 
                 \Biggl[ 
   1 
        - \frac{1}{(1+\bar{x})} \Biggr] \Bigl\{
          H(0;\bar{x})
        - 2 H(-1;\bar{x})
        + 2 H(-1,-1;\bar{x}) \nn\\
& & 
        - \! H(-1,0;\bar{x}) 
        -  \! 2 H(-1, \! -1, \! -1;\bar{x}) \Bigr\}
     -  \! \frac{1}{8(1+\bar{x})^2} 
                 \Biggl[ 
   1  \! 
        -  \! \frac{1}{(1+\bar{x})} \Biggr] \Bigl\{
          \zeta(3) \nn\\
& & 
        - \zeta(2) H(1;\bar{x})
        - H(0,0;\bar{x})
        + 2 H(0,-1;\bar{x})
        - H(0,0,0;\bar{x}) \nn\\
& & 
        - H(1,0,0;\bar{x})
        - H(\! -1, \! -1,0;\bar{x})\! 
        + \! 2 H(0,0,\! -1;\bar{x})\! 
        + \! 2 H(1,0,\! -1;\bar{x}) \nn\\
& & 
        - 2 H(-1,0,-1;\bar{x})
        + H(-1,0,0;\bar{x})
        - 2 H(0,-1,-1;\bar{x}) \nn\\
& & 
        + H(0,-1,0;\bar{x})
    \Bigr\}
     - \frac{1}{8(1+\bar{x})} 
                 \Biggl[ 
   1 
        - \frac{2}{(1+\bar{x})} 
        + \frac{1}{(1+\bar{x})^2} \Biggr] \Bigl\{
          \zeta(3) \nn\\
& & 
        + \zeta(2) \Bigl[
          H(0;\bar{x})
        - H(-1;\bar{x})
        + H(1;\bar{x}) \Bigr]
        + H(0,0,0;\bar{x}) \nn\\
& & 
        - 2 H(0,0,-1;\bar{x})
        - 2 H(1,0,-1;\bar{x})
        + H(1,0,0;\bar{x})
    \Bigr\}   \, , \\
\hspace*{-5mm} F_{14}^{(2)}(\bar{x}) & = &  
                                      - \frac{1}{20(1+\bar{x})} \zeta^2(2)
       - \frac{1}{16(1+\bar{x})} \Biggl[ 1 - \frac{1}{(1+\bar{x})} \Biggr]
     \Biggl\{
     4
   - \frac{3}{5} \zeta^2(2)
   - 2 \zeta(3) \nn\\
& & 
   + 2 \bigl( 1 - \zeta(3) \bigr) H(0;\bar{x})
   - 4  H(-1;\bar{x})
   + 2 \bigl( \zeta(2) + 2 \zeta(3) \bigr) H(1;\bar{x}) \nn\\
& & 
   - \zeta(2) H(-1,0;\bar{x})
   - 2 H(-1,0;\bar{x})
   + 2 H(1,0;\bar{x})
   - 2 H(1,-1;\bar{x}) \nn\\
& & 
   + 4 H(1,1;\bar{x})
   - 2 H(-1,1;\bar{x})
   + 4 H(-1,-1;\bar{x})
   + 2 H(0,0,0;\bar{x}) \nn\\
& & 
   -  \! 4 H(0,0, \! -1;\bar{x}) \! 
   -  \! 4 H(1,0, \! -1;\bar{x}) \! 
   +  \! 2 H(1,0,0;\bar{x}) \! 
   +  \! 2 H( \! -1, \! -1,0;\bar{x}) \nn\\
& & 
   - 4 H(-1,-1,-1;\bar{x})
   - 4 H(0,0,0,0;\bar{x})
   - 2 H(-1,-1,-1,0;\bar{x}) \nn\\
& & 
   +  \! 4 H(-1, \! -1, \! -1, \! -1;\bar{x}) \Biggr\} \! 
      -  \! \frac{1}{16(1 \! + \! \bar{x})^2} \Biggl[ 1  \! 
              -  \! \frac{1}{(1 \! + \! \bar{x})} \Biggr]
         \Bigl\{
     4 \zeta(3) H( \! -1;\bar{x}) \nn\\
& & 
   + 2 H(0,0;\bar{x})
   - 4 H(0,-1;\bar{x})
   + 2 H(0,1;\bar{x})
   - 2 H(0,-1,0;\bar{x}) \nn\\
& & 
   + 4 H(0,-1,-1;\bar{x})
   - 2 H(-1,0,0;\bar{x})
   + 4 H(-1,0,-1;\bar{x}) \nn\\
& & 
   + 6 H(0,0,0,0;\bar{x})
   - 4 H(-1,-1,0,-1;\bar{x})
   + 2 H(-1,-1,0,0;\bar{x}) \nn\\
& & 
   -  \! 4 H(-1,0, \! -1, \! -1;\bar{x}) \! 
   +  \! 2 H(-1,0, \! -1,0;\bar{x}) \! 
   -  \! 4 H(0, \! -1, \! -1, \! -1;\bar{x}) \nn\\
& & 
   + 2 H(0,-1,-1,0;\bar{x})
   + 6 H(-1,0,0,-1;\bar{x})
   - 3 H(-1,0,0,0;\bar{x}) \nn\\
& & 
   + 4 H(-1,1,0,-1;\bar{x})
   + 4 H(0,-1,0,-1;\bar{x})
   - 2 H(-1,1,0,0;\bar{x}) \nn\\
& & 
   - 2 H(0,-1,0,0;\bar{x})
   + 4 H(0,0,-1,-1;\bar{x})
   - 2 H(0,0,-1,0;\bar{x}) \nn\\
& & 
   - 4 H(0,1,0,-1;\bar{x})
   - 4 H(0,0,0,-1;\bar{x})
   + 2 H(0,1,0,0;\bar{x}) \nn\\
& & 
   + 4 H(1,-1,0,-1;\bar{x})
   - 2 H(1,-1,0,0;\bar{x})
   + 4 H(1,0,-1,-1;\bar{x}) \nn\\
& & 
   - 2 H(1,0,-1,0;\bar{x})
   - 8 H(1,0,0,-1;\bar{x})
   + 4 H(1,0,0,0;\bar{x}) \nn\\
& & 
   - 8 H(1,1,0,-1;\bar{x})
   + 4 H(1,1,0,0;\bar{x}) \Bigr\}
      + \frac{1}{32(1+\bar{x})} \Biggl[ 1 - \frac{2}{(1+\bar{x})}  \nn\\
& & + \frac{1}{(1+\bar{x})^2}
        \Biggr] \Bigl\{
     4 \zeta(2)
   + \zeta^2(2)
   - 8 \zeta(3)
   + 4 \bigl[ \zeta(2) 
     - 2 \zeta(3) \bigr] \bigl[ 
       H(0;\bar{x}) \nn\\
& & 
     + 2  H(1;\bar{x}) \bigr]
   - 4 \zeta(2) H(-1;\bar{x})
   + 2 \zeta(2) \bigl[ 
     H(0,0;\bar{x})
   - 2 H(-1,0;\bar{x}) \nn\\
& & 
   - 2 H(0,-1;\bar{x})
   + 2 H(-1,-1;\bar{x})
   + 4 H(1,0;\bar{x})
   + 4 H(0,1;\bar{x}) \nn\\
& & 
   - 4 H(1,-1;\bar{x})
   + 8 H(1,1;\bar{x})
   - 4 H(-1,1;\bar{x}) \bigr]
   + 8 H(0,0,0;\bar{x}) \nn\\
& & 
   - 16 H(0,0,-1;\bar{x})
   - 16 H(1,0,-1;\bar{x})
   + 8 H(1,0,0;\bar{x}) \nn\\
& & 
   - 2 H(0,0,0,0;\bar{x})
   + 4 H(-1,0,0,-1;\bar{x})
   - 2 H(-1,0,0,0;\bar{x}) \nn\\
& & 
   + 8 H(-1,1,0,-1;\bar{x})
   - 4 H(-1,1,0,0;\bar{x})
   + 8 H(0,-1,0,-1;\bar{x}) \nn\\
& & 
   - 4 H(0,-1,0,0;\bar{x})
   + 8 H(0,0,-1,-1;\bar{x})
   - 4 H(0,0,-1,0;\bar{x}) \nn\\
& & 
   - 12 H(0,0,0,-1;\bar{x})
   - 16 H(0,1,0,-1;\bar{x})
   + 8 H(0,1,0,0;\bar{x}) \nn\\
& & 
   + 8 H(1,-1,0,-1;\bar{x})
   - 4 H(1,-1,0,0;\bar{x})
   + 8 H(1,0,-1,-1;\bar{x}) \nn\\
& & 
   - 4 H(1,0,-1,0;\bar{x})
   - 16 H(1,0,0,-1;\bar{x})
   + 4 H(1,0,0,0;\bar{x}) \nn\\
& & 
   - 16 H(1,1,0,-1;\bar{x})
   + 8 H(1,1,0,0;\bar{x}) \Bigr\}  \, .
\eea

\subsection{2-loop Topologies with 5 denominators}

\bea
\parbox{25mm}{
\begin{fmfgraph*}(20,20)
\fmfleft{i1,i2}
\fmfright{o}
\fmf{plain}{i1,v1}
\fmf{plain}{i2,v2}
\fmf{dashes}{v4,o}
\fmf{photon,tension=.3}{v2,v3}
\fmf{photon,tension=.3}{v1,v3}
\fmf{plain,tension=0}{v2,v1}
\fmf{photon,tension=.2,left}{v3,v4}
\fmf{photon,tension=.2,right}{v3,v4}
\end{fmfgraph*} }
 & = & \int {\mathfrak{D}^D k_1} {\mathfrak{D}^D k_2}
           \  \frac{1}{
                     {\mathcal D}_{2}
                     {\mathcal D}_{3}
                     {\mathcal D}_{5}
                     {\mathcal D}_{9}
                     {\mathcal D}_{15}
                     } \nn\\  
& = & \frac{1}{m^2} \sum_{i=-1}^{1} (D-4)^i \, F_{15}^{(i)}(x) 
  + {\mathcal O} \left( (D-4)^2 \right)  \, , 
\eea
where:
\bea
\hspace{-5mm}
F_{15}^{(-1)}(x) & = &  - \frac{1}{16} \biggl[ \frac{1}{(1-x)}
- \frac{1}{(1+x)} \biggr] \bigl[ 4 \zeta(2) 
          + H(0,0;x)
   + 2 H(0,1;x) \bigr] \, , \\
\hspace{-5mm}
F_{15}^{(0)}(x) & = & \frac{1}{32} \biggl[ \frac{1}{(1 \! - \! x)} \! 
- \!  \frac{1}{(1 \! + \! x)} \biggr] \Bigl\{ 8 \zeta(2)  \! 
          + \! 5 \zeta(3) \! 
   +  \! \zeta(2) \bigl[ 3 H(0;x) \! 
   + \!  8 H(-1;x) \nn\\
& & 
   + 16 H(1;x) \bigr]
   + 2 H(0,0;x)
   + 4 H(0,1;x)
   + 4 H(0,0,0;x) \nn\\
& & 
   + 2 H(-1,0,0;x) \! 
   +  \! 4 H(-1,0,1;x) \! 
   +  \! 8 H(0,0,1;x) \! 
   +  \! 6 H(0,1,0;x) \nn\\
& & 
   + 12 H(0,1,1;x)
   + 4 H(1,0,0;x)
   + 8 H(1,0,1;x) \Bigr\}   \, , \\
\hspace{-5mm}
F_{15}^{(1)}(x) & = &  - \frac{1}{64} \biggl[ \frac{1}{(1-x)}
- \frac{1}{(1+x)} \biggr] \biggl\{ 
            16 \zeta(2)
     + \frac{19}{10} \zeta^{2}(2)
   + 10 \zeta(3)
   + 3 \bigl[ 2 \zeta(2) \nn\\
& & 
          + \zeta(3) \bigr] H(0;x)
   + \bigl[ 16 \zeta(2) + 10 \zeta(3) \bigr] 
     \bigl[ 2 H(1;x) + H(-1;x) \bigr]  \nn\\
& & 
   + 4 H(0,0;x)
   + 2 \bigl[ 4 + 5 \zeta(2) \bigr] H(0,1;x)
   + 2 \zeta(2) \bigl[
     6 H(1,0;x) \nn\\
& & 
   + 3 H(-1,0;x)
   + 4 H(0,-1;x)
   + 32 H(1,1;x)
   + 16 H(-1,1;x) \nn\\
& & 
   +  \! 16 H(1,-1;x) \! 
   +  \! 8 H(-1,-1;x) \bigr] \! 
   +  \! 4 \bigl[
     2 H(0,0,0;x) \! 
   +  \! 4 H(0,0,1;x) \nn\\
& & 
   + 2 H(1,0,0;x)
   + 3 H(0,1,0;x)
   + H(-1,0,0;x)
   + 4 H(1,0,1;x) \nn\\
& & 
   + 6 H(0,1,1;x)
   + 2 H(-1,0,1;x) \bigr]
   + 11 H(0,0,0,0;x) \nn\\
& & 
   + 4 H(-1,-1,0,0;x)
   + 8 H(-1,-1,0,1;x)
   + 8 H(-1,0,0,0;x) \nn\\
& & 
   + 16 H(-1,0,0,1;x)
   + 12 H(-1,0,1,0;x)
   + 24 H(-1,0,1,1;x) \nn\\
& & 
   + 8 H(-1,1,0,0;x)
   + 16 H(-1,1,0,1;x)
   + 2 H(0,-1,0,0;x) \nn\\
& & 
   + 4 H(0,-1,0,1;x)
   + 22 H(0,0,0,1;x)
   + 20 H(0,0,1,0;x) \nn\\
& & 
   + 40 H(0,0,1,1;x)
   + 18 H(0,1,0,0;x)
   + 36 H(0,1,0,1;x) \nn\\
& & 
   + 28 H(0,1,1,0;x)
   + 56 H(0,1,1,1;x)
   + 8 H(1,-1,0,0;x) \nn\\
& & 
   + 16 H(1,-1,0,1;x)
   + 16 H(1,0,0,0;x)
   + 32 H(1,0,0,1;x) \nn\\
& & 
   + 24 H(1,0,1,0;x)
   + 48 H(1,0,1,1;x)
   + 16 H(1,1,0,0;x) \nn\\
& & 
   + 32 H(1,1,0,1;x)
   \biggl\}
         \, .
\eea

\bea
\parbox{25mm}{
\begin{fmfgraph*}(20,20)
\fmfleft{i1,i2}
\fmfright{o}
\fmf{plain}{i1,v1}
\fmf{plain}{i2,v2}
\fmf{dashes}{v4,o}
\fmf{photon,tension=.3}{v2,v3}
\fmf{photon,tension=.3}{v1,v3}
\fmf{plain,tension=0}{v2,v1}
\fmf{plain,tension=.2,left}{v3,v4}
\fmf{plain,tension=.2,right}{v3,v4}
\end{fmfgraph*} }
 & = & \int {\mathfrak{D}^D k_1} {\mathfrak{D}^D k_2}
           \  \frac{1}{
                     {\mathcal D}_{2}
                     {\mathcal D}_{3}
                     {\mathcal D}_{5}
                     {\mathcal D}_{10}
                     {\mathcal D}_{16}
                     }  \nn\\ 
& = & \frac{1}{m^2} \sum_{i=-1}^{1} (D-4)^i \, F_{16}^{(i)}(x) 
  + {\mathcal O} \left( (D-4)^2 \right)  \, ,
\eea
where:
\bea
\hspace{-5mm}
F_{16}^{(-1)}(x) & = &   - \frac{1}{16} \biggl[ \frac{1}{(1-x)}
- \frac{1}{(1+x)} \biggr] \bigl[ 4 \zeta(2) 
          + H(0,0;x)
   + 2 H(0,1;x) \bigr] \, , \\
\hspace{-5mm}
F_{16}^{(0)}(x) & = & - \frac{1}{16(1-x)} \biggl[ 1 - \frac{1}{(1-x)}
                  \biggr]
     \Bigr[ 4 \zeta(2) H(0;x)
   + 3 H(0,0,0;x) \nn\\
& & 
   + 4 H(0,0,1;x)
   + 2 H(0,1,0;x) \Bigl]
   + \frac{1}{32} \biggl[ \frac{1}{(1-x)}
- \frac{1}{(1+x)} \biggr] \Bigl\{ 
            8 \zeta(2) \nn\\
& & 
   + 5 \zeta(3)
   + \zeta(2) \bigl[ H(0;x)
   + 8 H(-1;x)
   + 8 H(1;x) \bigr]
   + 2 H(0,0;x) \nn\\
& & 
   + 4 H(0,1;x)
   + H(0,0,0;x)
   + 2 H(0,0,1;x)
   + 2 H(0,1,0;x) \nn\\
& & 
   +  \! 2 H(-1,0,0;x) \! 
   +  \! 4 H(-1,0,1;x) \! 
   +  \! 4 H(0,1,1;x) \! 
   +  \! 2 H(1,0,0;x) \nn\\
& & 
   + 4 H(1,0,1;x)
           \Bigl\}  \, , \\
\hspace{-5mm}
F_{16}^{(1)}(x) & = & - \frac{1}{32(1-x)} \biggl[ 1 - \frac{1}{(1-x)}
         \biggr] \Bigl\{
     4 \zeta^{2}(2)
   - \bigl[ 8 \zeta(2) + 5 \zeta(3) \bigr] H(0;x) \nn\\
& & 
   - \zeta(2) \bigl[
     H(0,0;x)
   + 6 H(0,1;x)
   + 8 H(1,0;x)
   + 8 H(0,-1;x) \bigr] \nn\\
& & 
   - 6 H(0,0,0;x) \! 
   - 8 H(0,0,1;x) \! 
   - 4 H(0,1,0;x) \! 
   +  \! 2 H(0, \! -1,0,0;x) \nn\\
& & 
   + 4 H(0,-1,1,0;x)
   + 2 H(0,0,-1,0;x)
   + 4 H(0,1,-1,0;x) \nn\\
& & 
   - 10 H(0,0,0,0;x)
   - 12 H(0,0,0,1;x)
   - 10 H(0,0,1,0;x) \nn\\
& & 
   - 8 H(0,0,1,1;x)
   - 8 H(0,1,0,0;x)
   - 8 H(0,1,0,1;x) \nn\\
& & 
   - 4 H(0,1,1,0;x)
   - 6 H(1,0,0,0;x)
   - 8 H(1,0,0,1;x) \nn\\
& & 
   - 4 H(1,0,1,0;x) \Bigr\} 
 - \frac{1}{64} \biggl[ \frac{1}{(1-x)}
             - \frac{1}{(1+x)} \biggr] 
    \biggl\{ 
     16 \zeta(2)
   + \frac{59}{10} \zeta^{2}(2) \nn\\
& & 
   + 10 \zeta(3)
   - 2 \bigl[ \zeta(2) + \zeta(3) \bigr] H(0;x)
   + 2 \bigl[ 8 \zeta(2) + 5 \zeta(3) \bigr] \bigl[
     H(1;x) \nn\\
& & 
   + H(-1;x) \bigr]
   + \bigl[ 4 - \zeta(2) \bigr] \bigl[ 
     H(0,0;x)
          + 2 H(0,1;x) \bigr] \nn\\
& & 
   + 2 \zeta(2) \bigl[ 
     8 H(1,-1;x)
   - H(1,0;x)
   + 8 H(1,1;x)
   - H(-1,0;x) \nn\\
& & 
   + 8 H(-1,1;x)
   + 8 H(-1,-1;x) \bigr]
   + 2 H(0,0,0;x)
   + 4 H(0,0,1;x) \nn\\
& & 
   + 4 H(0,1,0;x)
   + 8 H(0,1,1;x)
   + 4 H(1,0,0;x)
   + 8 H(1,0,1;x) \nn\\
& & 
   + 4 H(-1,0,0;x)
   + 8 H(-1,0,1;x)
   + H(0,0,0,0;x) \nn\\
& & 
   + 2 H(0,0,0,1;x)
   + 2 H(0,0,1,0;x)
   + 4 H(0,0,1,1;x) \nn\\
& & 
   + 2 H(0,1,0,0;x)
   + 4 H(0,1,0,1;x)
   + 4 H(0,1,1,0;x) \nn\\
& & 
   + 2 H(1,0,0,0;x)
   + 4 H(1,0,0,1;x)
   + 4 H(1,0,1,0;x) \nn\\
& & 
   + 4 H(-1,-1,0,0;x)
   + 8 H(-1,-1,0,1;x)
   + 2 H(-1,0,0,0;x) \nn\\
& & 
   + 4 H(-1,0,0,1;x)
   + 4 H(-1,0,1,0;x)
   + 8 H(-1,0,1,1;x) \nn\\
& & 
   + 4 H(-1,1,0,0;x)
   + 8 H(-1,1,0,1;x)
   + 8 H(0,1,1,1;x) \nn\\
& & 
   + 4 H(1,-1,0,0;x)
   + 8 H(1,-1,0,1;x)
   + 8 H(1,0,1,1;x) \nn\\
& & 
   + 4 H(1,1,0,0;x)
   + 8 H(1,1,0,1;x) \biggr\}
        \, .
\eea

%%%%%%%%%%%%%%%%%%%%%%%%%%%%%%%%%%%%%%%%%%%%%%%%%%%%%%%%%%%%%%%%%%%%%
\bea
\parbox{25mm}{
\begin{fmfgraph*}(20,20)
\fmfforce{0.5w,0.2h}{v3}
\fmfforce{0.5w,0.8h}{v2}
\fmfforce{0.2w,0.5h}{v1}
\fmfforce{0.8w,0.5h}{v4}
\fmfleft{i}
\fmfright{o}
\fmf{dashes}{i,v1}
\fmf{dashes}{v4,o}
\fmf{photon,left=.4}{v1,v2}
\fmf{photon,right=.4}{v1,v3}
\fmf{plain,left=.4}{v2,v4}
\fmf{plain,right=.4}{v3,v4}
\fmf{plain}{v2,v3}
\end{fmfgraph*} }
 & = & \int {\mathfrak{D}^D k_1} {\mathfrak{D}^D k_2}
           \frac{1}{
                     {\mathcal D}_{1}
                     {\mathcal D}_{7}
                     {\mathcal D}_{10}
                     {\mathcal D}_{18}
                     {\mathcal D}_{24}
                     } \nn\\ 
& = & \frac{1}{m^2} \sum_{i=0}^{1} (D-4)^i \, F_{17}^{(i)}(x) 
  + {\mathcal O} \left( (D-4)^2 \right) , 
\label{BOLLA}
\eea
where:
\bea
F_{17}^{(0)}(x) & = &  - \frac{1}{8} 
          \Bigg[ \frac{1}{(1-x)} - \frac{1}{(1-x)^2}
          \Bigg] \Bigg[
            3 \zeta(3)
          + H(0,0,1;x)
          + H(0,1,0;x)
          \nn \\
          & &
          - 2 H(1,0,0;x)
          \Bigg] \, , \\
F_{17}^{(1)}(x) & = &  \frac{1}{80}
          \Bigg[ \frac{1}{(1-x)} - \frac{1}{(1-x)^2}
          \Bigg] \Big[
            30 \zeta(3) 
          + 9  \zeta^2(2)
          + 30 \zeta(3) H(0;x)
          \nn \\
          & & 
          + 60 \zeta(3) H(1;x)  
          - 5 \zeta(2) H(0,1;x)  
          + 10 \zeta(2) H(1,0;x)  
          \nn \\
          & & 
          - 20 H(1,0,0;x) 
          + 10 H(0,1,0;x) 
          + 10 H(0,0,1;x) 
          \nn \\
          & & 
          - 10 H(0,-1,0,0;x) 
          - 10 H(0,-1,0,1;x) 
          - 10 H(0,-1,1,0;x) 
          \nn \\
          & & 
          + 10 H(0,0,1,0;x) 
          + 10 H(0,0,1,1;x) 
          - 10 H(0,1,-1,0;x) 
          \nn \\
          & & 
          + 20 H(0,0,0,1;x) 
          + 15 H(0,1,0,0;x) 
          + 10 H(0,1,0,1;x) 
          \nn \\
          & & 
          + 10 H(0,1,1,0;x) 
          + 60 H(1,0,-1,0;x) 
          + 25 H(0,0,-1,0;x) 
          \nn \\
          & & 
          - 30 H(1,0,0,0;x) 
          + 10 H(1,0,0,1;x) 
          - 10 H(1,0,1,0;x) 
          \Big] \, .
\eea
%%%%%%%%%%%%%%%%%%%%%%%%%%%%%%%%%%%%%%%%%%%%%%%%%%%%%%%%%%%%%%%%%%%%%

Here, the zeroth order term is in agreement with $I_{12}$ of 
\cite{Fleischer} and the first order in $(D-4)$ is in agreement with
$F_{10101}$ of \cite{Davydychev}. 
For carrying out the comparison, it is necessary to account for the 
difference in normalization and notation; the results are found to agree 
by multiplying the formulas of~\cite{Fleischer,Davydychev} by the overall 
factor 
\be 
\left( \frac{i}{4} \right)^2 \Bigl( 1- \zeta(2) \epsilon^2 
+ {\mathcal O}( \epsilon^3 ) \Bigr) \, ,
\ee 
by replacing their variable $y$ by our $x$, and finally by recalling 
$(D-4) = -2 \epsilon$.

\subsection{2-loop Topologies with 6 denominators}

We find only one MI with six propagators, associated to Fig.~\ref{fig1bis} 
(a); all the integrals associated to the other six propagator 
topologies can be expressed in terms of MIs of their subtopologies. 
The six propagator MI is 

%%%%%%%%%%%%%%%%%%%%%%%%%%%%%%%%%%%%%%%%%%%%%%%%%%%%%%%%%%%%%%%%%%%%%
\bea
\parbox{20mm}{
\begin{fmfgraph*}(20,20)\fmfkeep{Primo}
\fmfleft{i1,i2}
\fmfright{o}
\fmf{plain}{i1,v1}
\fmf{plain}{i2,v2}
\fmf{dashes}{v5,o}
\fmf{photon,tension=.3}{v2,v3}
\fmf{plain,tension=.3}{v3,v5}
\fmf{photon,tension=.3}{v1,v4}
\fmf{plain,tension=.3}{v4,v5}
\fmf{plain,tension=0}{v2,v1}
\fmf{plain,tension=0}{v4,v3}
\end{fmfgraph*} }
 & = & \int {\mathfrak{D}^D k_1} {\mathfrak{D}^D k_2}
           \frac{1}{
                     {\mathcal D}_{2}
                     {\mathcal D}_{3}
                     {\mathcal D}_{5}
                     {\mathcal D}_{10}
                     {\mathcal D}_{20}
                     {\mathcal D}_{22}
                     } \nn\\ 
& = & \frac{1}{m^4} F_{18}^{(0)}(x) + {\mathcal O}(D-4),
\eea
where:
\bea
F_{18}^{(0)}(x) & = & \frac{1}{128} \bigg[
            \frac{4}{(1-x)^3}
          - \frac{6}{(1-x)^2}
          + \frac{1}{(1-x)}
          + \frac{1}{(1+x)}
          \bigg] \Bigg\{
            \frac{6}{5}  \zeta^2(2)
          \nn \\
          & &
          + 2  \zeta(3) H(0;x) 
          + 4  \zeta(2) H(0,0;x)
          + 8  H(0,0,-1,0;x)
          \nn \\
          & &
          + H(0,0,0,0;x)
          + 6  H(0,0,0,1;x) 
          + 4  H(0,1,0,0;x)
          \Bigg\} \, .
\eea
%%%%%%%%%%%%%%%%%%%%%%%%%%%%%%%%%%%%%%%%%%%%%%%%%%%%%%%%%%%%%%%%%%%%%

\section{6-denominator reducible integrals \label{6denom}}

Besides the 6-denominator MI of the previous Section, there are no other
MIs corresponding to the remaining 6-denominator topologies 
of Fig.~\ref{fig1bis}, {\it i.e.} it turns out that all the related 
scalar integrals can be entirely expressed in terms of the MIs of their 
subtopologies, given in the previous sections. 

For completeness and for easiness of comparison with the results of 
other authors, we list now the explicit values of a few reducible 
6-denominator integrals ({\it i.e.} integrals which can be expressed in
terms of MIs of their subtopologies).

%%%%%%%%%%%%%%%%%%%%%%%%%%%%%%%%%%%%%%%%%%%%%%%%%%%%%%%%%%%%%%%%%%%%%
\bea
\parbox{20mm}{
\begin{fmfgraph*}(20,20)\fmfkeep{Secondo}
\fmfleft{i1,i2}
\fmfright{o}
\fmfforce{0.2w,0.93h}{v2}
\fmfforce{0.2w,0.07h}{v1}
\fmfforce{0.2w,0.5h}{v3}
\fmfforce{0.8w,0.5h}{v5}
\fmf{plain}{i1,v1}
\fmf{plain}{i2,v2}
\fmf{dashes}{v5,o}
\fmf{plain,tension=0}{v1,v5}
\fmf{photon,tension=0}{v4,v3}
\fmf{plain,tension=.4}{v2,v4}
\fmf{plain,tension=.4}{v4,v5}
\fmf{photon,tension=0}{v1,v3}
\fmf{photon,tension=0}{v2,v3}
\end{fmfgraph*}}
 & = & \int {\mathfrak{D}^D k_1} {\mathfrak{D}^D k_2}
           \frac{1}{
                     {\mathcal D}_{1}
                     {\mathcal D}_{2}
                     {\mathcal D}_{9}
                     {\mathcal D}_{17}
                     {\mathcal D}_{20}
                     {\mathcal D}_{22}
                     } \nn\\ 
& = & \frac{1}{m^4} \sum_{i=-2}^{0} (D-4)^i \, F_{19}^{(i)}(x) 
  + {\mathcal O}(D-4),
\eea
where:
\bea
\hspace*{-8mm} F_{19}^{(-2)}(x) & = &  \frac{1}{32} \bigg[ \frac{1}{(1-x)} 
           - \frac{1}{(1+x)} \bigg] H(0;x) \, , \\
\hspace*{-8mm} F_{19}^{(-1)}(x) & = &  \frac{1}{16} \Bigg\{
             \Bigg[ \frac{1}{(1\! -\! x)} \! 
      - \! \frac{1}{(1 \! + \! x)} \Bigg] H(0;x) \! 
             + \! \Bigg[ \frac{1}{(1 \! + \! x)} \! 
                  - \! \frac{1}{(1 \! + \! x)^2} \Bigg] H(0,0;x)
     \Bigg\} , \\
\hspace*{-8mm} F_{19}^{(0)}(x) & = &  \frac{1}{64} \Bigg\{
         2 \zeta(3) \Bigg[    \frac{1}{(1\! -\! x)} \! 
                            + \! \frac{1}{(1\! +\! x)^2} \! 
                            - \! \frac{2}{(1\! +\! x)} 
                    \Bigg]
       + \Bigg[    \frac{1}{(1-x)} \! 
                 - \! \frac{1}{(1+x)} 
         \Bigg] \times
                 \nn \\
\hspace*{-8mm} & & \times \big[
            8 H(0;x)
          - \! 2 \zeta(2) H(-1;x) \! 
          + \! 7 \zeta(2) H(0;x)\! 
          - \! 4 H(\! -1,\! -1,0;x)
         \big] 
          \nn \\
\hspace*{-8mm} & &
       + \bigg[   \frac{1}{(1+x)} 
                 - \frac{1}{(1+x)^2} 
         \bigg] \big[
            8 H(0,0;x) 
          - 4 H(0,1,0;x) 
         \bigg]
       + \bigg[   \frac{2}{(1-x)} 
          \nn \\
\hspace*{-8mm} & &
                - \frac{8}{(1+x)^2} 
                + \frac{6}{(1+x)} 
         \bigg] \big[   H(-1,0,0;x) 
                       + H(0,-1,0;x) \big]
                         \nn \\
\hspace*{-8mm} & &
       + \Bigg[   \frac{7}{(1-x)} 
                + \frac{10}{(1+x)^2} 
                - \frac{17}{(1+x)} 
         \Bigg] H(0,0,0;x)    
      \Bigg\} \, .
\eea
%%%%%%%%%%%%%%%%%%%%%%%%%%%%%%%%%%%%%%%%%%%%%%%%%%%%%%%%%%%%%%%%%%%%%

%%%%%%%%%%%%%%%%%%%%%%%%%%%%%%%%%%%%%%%%%%%%%%%%%%%%%%%%%%%%%%%%%%%%%
\bea
\parbox{20mm}{
\begin{fmfgraph*}(20,20)\fmfkeep{Terzo}
\fmfleft{i1,i2}
\fmfright{o}
\fmfforce{0.2w,0.93h}{v2}
\fmfforce{0.2w,0.07h}{v1}
\fmfforce{0.2w,0.3h}{v3}
\fmfforce{0.2w,0.7h}{v4}
\fmfforce{0.8w,0.5h}{v5}
\fmf{plain}{i1,v1}
\fmf{plain}{i2,v2}
\fmf{dashes}{v5,o}
\fmf{plain}{v2,v5}
\fmf{photon}{v1,v3}
\fmf{photon}{v2,v4}
\fmf{plain}{v1,v5}
\fmf{photon,left}{v3,v4}
\fmf{photon,left}{v4,v3}
\end{fmfgraph*} } 
 & = & \int {\mathfrak{D}^D k_1} {\mathfrak{D}^D k_2}
           \frac{1}{
                     {\mathcal D}_{1}^2
                     {\mathcal D}_{4}
                     {\mathcal D}_{6}
                     {\mathcal D}_{9}
                     {\mathcal D}_{17}
                     } \nn\\ 
& = & \frac{1}{m^4} \sum_{i=-1}^{0} (D-4)^i \, F_{20}^{(i)}(x) 
  + {\mathcal O}(D-4),
\eea
where:
\bea 
F_{20}^{(-1)}(x) & = &  
      \frac{3}{128} 
       \Bigg\{
           4 \Bigg[ \frac{1}{(1+x)^2} - \frac{1}{(1+x)} \Bigg]
         + \Bigg[ \frac{1}{(1-x)} 
                + \frac{4}{(1+x)^3} 
                \nn \\
                & &
                - \frac{6}{(1+x)^2} 
                + \frac{1}{(1+x)} 
         \Bigg] H(0;x)
       \Bigg\} \, , \\
F_{20}^{(0)}(x) & = &  
       \frac{3}{128} 
       \Bigg\{
           8 \Bigg[ \frac{1}{(1+x)^2} \! - \! \frac{1}{(1+x)} \Bigg]\! 
         + \! \Bigg[ \frac{1}{(1-x)} \! 
                + \! \frac{4}{(1+x)^3} \! 
                - \! \frac{6}{(1+x)^2} 
                \nn \\
                & &
                + \frac{1}{(1+x)} 
         \Bigg] \Bigg[
              \zeta(2) 
             + 2 H(-1,0;x)
             - H(0,0;x)
         \Bigg]
       \Bigg\} \, .
\eea
%%%%%%%%%%%%%%%%%%%%%%%%%%%%%%%%%%%%%%%%%%%%%%%%%%%%%%%%%%%%%%%%%%%%%

%%%%%%%%%%%%%%%%%%%%%%%%%%%%%%%%%%%%%%%%%%%%%%%%%%%%%%%%%%%%%%%%%%%%%
\bea
\parbox{20mm}{
\begin{fmfgraph*}(20,20)\fmfkeep{Quarto}
\fmfleft{i1,i2}
\fmfright{o}
\fmf{plain}{i1,v1}
\fmf{plain}{i2,v2}
\fmf{dashes}{v5,o}
\fmf{photon,tension=.3}{v2,v3}
\fmf{photon,tension=.3}{v3,v5}
\fmf{photon,tension=.3}{v1,v4}
\fmf{photon,tension=.3}{v4,v5}
\fmf{plain,tension=0}{v2,v1}
\fmf{photon,tension=0}{v4,v3}
\end{fmfgraph*} }
 & = & \int {\mathfrak{D}^D k_1} {\mathfrak{D}^D k_2}
           \frac{1}{
                     {\mathcal D}_{2}
                     {\mathcal D}_{3}
                     {\mathcal D}_{5}
                     {\mathcal D}_{9}
                     {\mathcal D}_{19}
                     {\mathcal D}_{21}
                     } \nn\\ 
& = & \frac{1}{m^4} F_{21}^{(0)}(x) + {\mathcal O}(D-4),
\eea
where:
\bea
F_{21}^{(0)}(x) & = &  
    \frac{1}{128} 
         \Bigg[
                  \frac{4}{(1-x)^3} 
                - \frac{6}{(1-x)^2}
                + \frac{1}{(1-x)} 
                + \frac{1}{(1+x)}
         \Bigg] \Bigg[
             \frac{56}{5} \zeta^2(2)
           \nn \\
           & &
           + 4 \zeta(2) H(0,0;x)
           + H(0,0,0,0;x) 
           + 2 H(0,0,0,1;x)
         \Bigg] \, .
\eea
%%%%%%%%%%%%%%%%%%%%%%%%%%%%%%%%%%%%%%%%%%%%%%%%%%%%%%%%%%%%%%%%%%%%%

%%%%%%%%%%%%%%%%%%%%%%%%%%%%%%%%%%%%%%%%%%%%%%%%%%%%%%%%%%%%%%%%%%%%%

\bea
\parbox{20mm}{
\begin{fmfgraph*}(20,20)\fmfkeep{Quinto}
\fmfleft{i1,i2}
\fmfright{o}
\fmf{photon}{i1,v1}
\fmf{photon}{i2,v2}
\fmf{dashes}{v5,o}
\fmf{photon,tension=.3}{v2,v3}
\fmf{plain,tension=.3}{v3,v5}
\fmf{photon,tension=.3}{v1,v4}
\fmf{plain,tension=.3}{v4,v5}
\fmf{photon,tension=0}{v2,v1}
\fmf{plain,tension=0}{v4,v3}
\end{fmfgraph*} }
 & = & \int {\mathfrak{D}^D k_1} {\mathfrak{D}^D k_2}
           \frac{1}{
                     {\mathcal D}_{1}
                     {\mathcal D}_{3}
                     {\mathcal D}_{5}
                     {\mathcal D}_{10}
                     {\mathcal D}_{20}
                     {\mathcal D}_{22}
                     } \nn\\ 
& = & \frac{1}{m^4} \sum_{i=-2}^{0} (D-4)^i \, F_{22}^{(i)}(x) 
  + {\mathcal O}(D-4),
\label{23591921} 
\eea
where:
\bea
\hspace{-5mm}
F_{22}^{(-2)}(x) & = & \frac{1}{4(1-x)^2} \Biggl[ 
              1 
     - \frac{2}{(1-x)}
     + \frac{1}{(1-x)^2}
            \Biggr]  H(0,0;x)  \, , 
\label{Dav1} \\
\hspace{-5mm}
F_{22}^{(-1)}(x) & = &  - \frac{1}{8(1-x)^2} \Biggl[ 
              1 
     - \frac{2}{(1-x)}
     + \frac{1}{(1-x)^2}
            \Biggr] \Bigl\{
       3 \zeta(3)
     + \zeta(2) H(0;x) \nn\\
& & 
     + 2 \Bigl[ H(0,0,0;x)
     +  H(0,-1,0;x)
     +  H(0,0,1;x)
     +  H(0,1,0;x) \nn\\
& & 
     +  H(1,0,0;x) \Bigr] \Bigr\}  \, , 
\label{Dav2} \\
\hspace{-5mm}
F_{22}^{(0)}(x) & = &  \frac{1}{16(1-x)^2} \Biggl[ 
              1 
     - \frac{2}{(1-x)}
     + \frac{1}{(1-x)^2}
            \Biggr] \Biggl\{
              \frac{27}{10} \zeta^2(2)
     + \zeta(3) \Bigl[ 
       11 H(0;x) \nn\\
& & 
     + 6 H(1;x) \Bigr]
     + 2 \zeta(2) \Bigl[
       2 H(0,0;x)
     + H(0,1;x)
     + H(1,0;x) \nn\\
& & 
     - 3 H(0,-1;x) \Bigr]
     + 2 \Bigl[
       2 H(0,0,0,0;x)
     - 14 H(0,-1,-1,0;x) \nn\\
& & 
     + 7 H(0,-1,0,0;x)
     - 2 H(0,-1,0,1;x)
     + 2 H(0,-1,1,0;x) \nn\\
& & 
     + 9 H(0,0,-1,0;x)
     + 4 H(0,0,0,1;x)
     + 2 H(0,0,1,0;x) \nn\\
& & 
     + 2 H(0,0,1,1;x)
     + 2 H(0,1,-1,0;x)
     + 2 H(0,1,0,0;x) \nn\\
& & 
     + 2 H(0,1,0,1;x)
     + 2 H(0,1,1,0;x)
     + 2 H(1,0,-1,0;x) \nn\\
& & 
     + 2 H(1,0,0,0;x)
     + 2 H(1,0,0,1;x)
     + 2 H(1,0,1,0;x) \nn\\
& & 
     + 2 H(1,1,0,0;x) \Bigr] \Biggr\} \, .
\label{Dav3} 
\eea

The imaginary part of the Feynman diagram in Fig.~\ref{fig1} (k), 
relative to the topology in Fig.~\ref{fig1bis} (e), was firstly 
evaluated in \cite{Kuehn}.

The corresponding scalar diagram Eq.~(\ref{23591921}) was firstly 
calculated in \cite{Fleischer1} using the small momentum expansion, then
in \cite{Fleischer} in terms of binomial sums. The same diagram was 
also considered in \cite{Davydychev}, and expressed in terms of 
generalized Nielsen's polylogarithms and harmonic polylogarithms. 

Our expression, given in Eqs.~(\ref{Dav1}--\ref{Dav3}), agrees with
the expression of the amplitude $P_{126}$ given in Eq. (4.11) of 
\cite{Davydychev} (for the comparison see the discussion following the
diagram in Eq.~(\ref{BOLLA})) and with the numerical checks provided 
by the TOPSIDE collaboration \cite{Passa}. \\

%%%%%%%%%%%%%%%%%%%%%%%%%%%%%%%%%%%%%%%%%%%%%%%%%%%%%%%%%%%%%%%%%%%%%

The next scalar diagram is expressed in terms of the
variable $\bar{x}$, as in the case of the 4-denominator MIs of 
Fig.~\ref{fig6} (j) and (k). In agreement with \cite{UgoRo}, we find:

\bea
\parbox{20mm}{
\begin{fmfgraph*}(20,20)\fmfkeep{Sesto}
\fmfleft{i1,i2}
\fmfright{o}
\fmfforce{0.2w,0.93h}{v2}
\fmfforce{0.2w,0.07h}{v1}
\fmfforce{0.2w,0.3h}{v3}
\fmfforce{0.2w,0.7h}{v4}
\fmfforce{0.8w,0.5h}{v5}
\fmf{photon}{i1,v1}
\fmf{photon}{i2,v2}
\fmf{dashes}{v5,o}
\fmf{photon}{v2,v5}
\fmf{photon}{v1,v3}
\fmf{photon}{v2,v4}
\fmf{photon}{v1,v5}
\fmf{plain,left}{v3,v4}
\fmf{plain,left}{v4,v3}
\end{fmfgraph*} }
 & = & \int {\mathfrak{D}^D k_1} {\mathfrak{D}^D k_2}
           \frac{1}{
                     {\mathcal D}_{1}^2
                     {\mathcal D}_{3}
                     {\mathcal D}_{5}
                     {\mathcal D}_{10}
                     {\mathcal D}_{18}
                     } \nn\\ 
& = & \frac{1}{m^4} \sum_{i=-2}^{0} (D-4)^i \, F_{23}^{(i)}(\bar{x}) 
  + {\mathcal O}(D-4),
\label{UGO3}
\eea
where:
\bea
\hspace{-5mm}
F_{23}^{(-2)}(\bar{x}) & = &  - \frac{1}{24(1+ \bar{x} )} \Biggl[ 
             1
    + \frac{11}{(1+ \bar{x} )}
    - \frac{24}{(1+ \bar{x} )^2}
    + \frac{12}{(1+ \bar{x} )^3} \Biggr] \, , \\
\hspace{-5mm}
F_{23}^{(-1)}(\bar{x}) & = &   - \frac{1}{4(1+ \bar{x} )^2} \Biggl[ 
             1
    - \frac{2}{(1+ \bar{x} )} 
    + \frac{1}{(1+ \bar{x} )^2} \Biggr]
    - \frac{1}{24(1+ \bar{x} )} \Biggl[ 
             1
    + \frac{11}{(1+ \bar{x} )} \nn\\
& & 
    - \frac{24}{(1+ \bar{x} )^2}
    + \frac{12}{(1+ \bar{x} )^3} \Biggr] \Bigl\{
      H(0; \bar{x} ) 
    - 2 H(-1; \bar{x} ) \Bigr\} \, , \\
\hspace{-5mm}
F_{23}^{(0)}(\bar{x}) & = &   \frac{1}{(1+ \bar{x} )^2} \Biggl[ 
             \frac{3}{16}
    - \frac{7}{16(1+ \bar{x} )} 
    + \frac{1}{3(1+ \bar{x} )^2}  
    - \frac{1}{12(1+ \bar{x} )^3} \Biggr] \zeta(2) \nn\\
& & 
    + \frac{1}{6(1+ \bar{x} )} \Biggl[ 
             \frac{13}{36}
    - \frac{19}{9(1+ \bar{x} )} 
    + \frac{7}{2(1+ \bar{x} )^2}  
    - \frac{7}{4(1+ \bar{x} )^3} \Biggr]  \nn\\
& & 
    + \! \frac{1}{72(1 \! +  \! \bar{x} )} \Biggl[ 
             2 \! 
    +  \! \frac{1}{(1 \! +  \! \bar{x} )}  \! 
    -  \! \frac{6}{(1 \! +  \! \bar{x} )^2}   \! 
    +  \! \frac{3}{(1 \! +  \! \bar{x} )^3} \Biggr] \Bigl\{
      H(0; \bar{x} ) \! 
    -  \! 2 H( \! -1; \bar{x} ) \Bigr\} \nn\\
& & 
    + \frac{1}{48(1+ \bar{x} )} \Biggl[ 
             1
    + \frac{11}{(1+ \bar{x} )} 
    - \frac{24}{(1+ \bar{x} )^2}  
    + \frac{12}{(1+ \bar{x} )^3} \Biggr] \Bigl\{
      H(-1,0; \bar{x} ) \nn\\
& & 
    - 2 H(-1, \! -1; \bar{x} ) \Bigr\}
    - \frac{1}{48(1+ \bar{x} )} \Biggl[ 
             1 \! 
    +  \! \frac{2}{(1+ \bar{x} )}  \! 
    - \frac{3}{(1+ \bar{x} )^2}   \! 
    - \frac{4}{(1+ \bar{x} )^3}  \nn\\
& &  
    + \frac{4}{(1+ \bar{x} )^4} \Biggr] \Bigl\{
      H(0,0; \bar{x} )
    - 2 H(0,-1; \bar{x} ) \Bigr\} \, .
\eea
   
\end{fmffile}  
\begin{fmffile}{2NONab2}

\section{Expansion for $Q^2 \gg m^2$ \label{Asympt}}

We list, in this Section, the asymptotic expansion of the 6-denominator 
vertex diagrams given in the previous sections, in order to show their 
behaviour for momentum transfer larger than the mass.

Putting $L = \ln{(Q^2/m^2)} $ and keeping only the leading term in 
$m^2/Q^2$, we find: 
\bea
\hspace{-5mm}
m^4 \parbox{20mm}{\fmfreuse{Primo}}& \simeq & 
      \left( \frac{m^2}{Q^2} \right)^2 \Biggl\{ 
         \frac{3}{40} \zeta^2(2)
       - \frac{1}{8} \zeta(3) \, L
       + \frac{1}{8} \zeta(2) \, L^2
       + \frac{1}{384} \, L^4 \Biggr\} \, . \\
& & \nn\\
& & \nn\\
\hspace{-5mm}
m^4 \parbox{20mm}{\fmfreuse{Secondo}} & \simeq &
- \frac{1}{(D-4)^2}  \left( \frac{m^2}{Q^2} \right) \, \frac{1}{16}  L 
       - \frac{1}{(D-4)} \left( \frac{m^2}{Q^2} \right) \Biggl\{ 
         \frac{1}{8} \, L
       - \frac{1}{32} \, L^2  \Biggr\}  \nn\\
& & 
       +  \left( \frac{m^2}{Q^2} \right) \Biggl\{ 
         \frac{1}{32} \zeta(3)
       - \frac{1}{4} \Biggl[ 1 + \frac{7}{8} \zeta(2) \Biggr]  \, L
       + \frac{1}{16} \zeta(2) \, L^2 \nn\\
& & 
       - \frac{1}{96} \, L^3 \Biggr\} \, . \\
& & \nn\\
& & \nn\\
\hspace{-5mm}
m^4 \ \parbox{20mm}{\fmfreuse{Terzo}} & \simeq &
 - \frac{1}{(D-4)} \left( \frac{m^2}{Q^2} \right)   \frac{3}{32} 
 - \left( \frac{m^2}{Q^2} \right)  \frac{3}{16}  \, . \\
& & \nn\\
& & \nn\\
\hspace{-5mm}
m^4 \parbox{20mm}{\fmfreuse{Quarto}} & \simeq &
 \left( \frac{m^2}{Q^2} \right)^2 \Biggl\{ 
         \frac{7}{10} \zeta^2(2)
       + \frac{1}{8} \zeta(2) \, L^2
       + \frac{1}{384} \, L^4 \Biggr\} \, . \\
& & \nn\\
& & \nn\\
\hspace{-5mm}
m^4 \parbox{20mm}{\fmfreuse{Quinto}} & \simeq &
      \frac{1}{(D-4)^2} \left( \frac{m^2}{Q^2} \right)^2 
             \frac{1}{8}  \, L^2  \nn\\
& &  - \frac{1}{(D-4)} \left( \frac{m^2}{Q^2} \right)^2 \Biggl\{ 
         \frac{3}{8} \zeta(3)
       - \frac{1}{8} \zeta(2) \, L
       - \frac{1}{24} \, L^3 \Biggr\} \nn\\
& &  + \left( \frac{m^2}{Q^2} \right)^2 \Biggl\{ 
         \frac{27}{160} \zeta^2(2)
       - \frac{11}{16} \zeta(3) \, L
       + \frac{1}{8} \zeta(2) \, L^2 \nn\\
& & 
       + \frac{1}{96} \, L^4 \Biggr\}  \, . \\
& & \nn\\
& & \nn\\
\hspace{-5mm}
m^4 \ \parbox{20mm}{\fmfreuse{Sesto}} & \simeq &
- \frac{1}{(D-4)^2} \left( \frac{m^2}{Q^2} \right) \frac{1}{24} 
- \frac{1}{(D-4)} \left( \frac{m^2}{Q^2}\right) \frac{1}{48}  L \nn\\
& & + \left( \frac{m^2}{Q^2} \right) \Biggl\{ 
         \frac{13}{216} 
       + \frac{1}{48} \zeta(2)
       - \frac{1}{36} \, L  \Biggr\}  \, .
\eea

\section{Summary}

We presented in this paper the explicit analytic values of all the 
MIs occurring in the 2-loop QCD corrections to the forward-backward 
asymmetry of the production of a quark-antiquark pair in $e^+ e^-$ 
annihilation processes. We keep the full dependence on the squared 
momentum transfer $Q^2$ and on the mass $m$ of the heavier quark of
each diagram. The results are given for space-like $Q$; the time-like 
region can be recovered by standard analytic continuation. 

Out of the 35 MIs of the problem, 17 were already evaluated
in a previous paper. Therefore, in this work we 
gave only the results concerning the 18 new MIs, which come from the 
the topologies related to the non-abelian Feynman 
diagrams and from the diagrams contributing to the axial form factors.

All the integrals are regularized within the continuous $D$-dimensional
regularization scheme, where both UV and IR divergences appear as poles 
in $(D-4)$.

The method used for the reduction to the MIs is based on the IBPs, LI
and general symmetry relations, while the calculation of the MIs was
performed by means of the differential equations method. The results 
were given as a Laurent series expansion around $D=4$ up to the term in 
$(D-4)$ containing 1-dimensional harmonic polylogarithms with maximum 
weight $w=4$.

For completeness and easiness of comparison with other results in the 
literature, we presented also the result for a few 6-denominator 
scalar integrals, although they are not genuine MIs, but can be expressed 
in terms of MIs of their subtopologies, giving as well their 
expansions for large momentum transfer.

\section{Acknowledgments} 
We kindly acknowledge discussions with W. Bernreuter and T. Gehrmann, 
R. Heinesch and T. Leineweber. 
We wish to thank the TOPSIDE collaboration, and in particular S. 
Uccirati, for useful discussions and for the numerical checks of some 
diagrams.
We are grateful to J. Vermaseren for his kind assistance in the use
of the algebra manipulating program {\tt FORM}~\cite{FORM}, by which
all our calculations were carried out. 
P.M. wishes to thank B. Feucht and A. V. Kotikov for useful 
discussions. We acknowledge a complete agreement with U. Aglietti for
the results of the MIs in Eqs.~(\ref{UGO1},\ref{UGO2}) and of the scalar
6-denominator diagram in Eq.~(\ref{UGO3}).

\appendix

\section{Propagators \label{app1}}

We list here the denominators of the integral expressions appeared in
the paper.
\bea
{\mathcal D}_{1} & = &  k_{1}^{2} \, , \\
{\mathcal D}_{2} & = & [k_{1}^{2}+m^2] \, , \\
{\mathcal D}_{3} & = &  (p_{1}-k_{1})^{2} \, , \\
{\mathcal D}_{4} & = & [(p_{1}-k_{1})^{2}+m^2] \, , \\
{\mathcal D}_{5} & = &  (p_{2}+k_{1})^{2} \, , \\
{\mathcal D}_{6} & = & [(p_{2}+k_{1})^{2}+m^2] \, , \\
{\mathcal D}_{7} & = &  (p_{1}+p_{2}-k_{1})^{2} \, , \\
{\mathcal D}_{8} & = & [(p_{1}+p_{2}-k_{1})^{2}+m^2] \, , \\
{\mathcal D}_{9} & = &  k_{2}^{2} \, , \\
{\mathcal D}_{10} & = & [k_{2}^{2}+m^2] \, , \\
{\mathcal D}_{11} & = &  (p_{1}-k_{2})^{2} \, , \\
{\mathcal D}_{12} & = & [(p_{1}-k_{2})^{2}+m^2] \, , \\
{\mathcal D}_{13} & = &  (p_{2}+k_{2})^{2} \, , \\
{\mathcal D}_{14} & = & [(p_{2}+k_{2})^{2}+m^2] \, , \\
{\mathcal D}_{15} & = &  (p_{1}+p_{2}-k_{2})^{2} \, , \\
{\mathcal D}_{16} & = & [(p_{1}+p_{2}-k_{2})^{2}+m^2] \, , \\
{\mathcal D}_{17} & = &  (k_{1}+k_{2})^{2} \, , \\
{\mathcal D}_{18} & = & [(k_{1}+k_{2})^{2}+m^2] \, , \\
{\mathcal D}_{19} & = &  (p_{1}-k_{1}-k_{2})^{2} \, , \\
{\mathcal D}_{20} & = & [(p_{1}-k_{1}-k_{2})^{2}+m^2] \, , \\
{\mathcal D}_{21} & = &  (p_{2}+k_{1}+k_{2})^{2} \, , \\
{\mathcal D}_{22} & = & [(p_{2}+k_{1}+k_{2})^{2}+m^2] \, , \\
{\mathcal D}_{23} & = &  (p_{1}+p_{2}-k_{1}-k_{2})^{2} \, , \\
{\mathcal D}_{24} & = & [(p_{1}+p_{2}-k_{1}-k_{2})^{2}+m^2] \, .
\eea

\section{One-loop topologies \label{app2}}

In this Appendix we give the results of the 1-loop scalar integrals
entering in the computation of the MIs of Fig. \ref{fig6} (l), (m),
(n), (o) and (p). The tadpole and the 1-loop bubble with two equal
masses, which enter as well in the MIs (l) and (n), are given in 
\cite{RoPieRem1}.

\be
\hspace*{-10mm} \parbox{22mm}{
\begin{fmfgraph*}(20,20)
\fmfleft{i}
\fmfright{o}
\fmf{dashes}{i,v1}
\fmf{dashes}{v2,o}
\fmf{photon,tension=.22,left}{v1,v2}
\fmf{photon,tension=.22,right}{v1,v2}
%\fmf{plain,right=45}{v2,v2}
\end{fmfgraph*} } 
= \hspace*{2mm}  \int {\mathfrak{D}^D k_1}
          \ \frac{1}{
                     {\mathcal D}_{1}
                     {\mathcal D}_{7}
                     } = 
\sum_{i=-1}^{3} (D-4)^i \, G_{1}^{(i)}(x) + {\mathcal O} 
\left( (D-4)^4 \right) \, , 
\ee
where:
\bea
G_{1}^{(-1)}(x) & = & 
       - \frac{1}{2} \, , \\
G_{1}^{(0)}(x) & = & 
         \frac{1}{2}
       + \frac{1}{4} H(0;x)
       + \frac{1}{2} H(1;x) \, , \\
G_{1}^{(1)}(x) & = & 
          - \frac{1}{2}
          + \frac{1}{8} \zeta(2)
          - \frac{1}{4} H(0;x)
          - \frac{1}{2} H(1;x)
          - \frac{1}{8} H(0,0;x)
          - \frac{1}{4} H(0,1;x)
          \nn \\
          & & 
          - \frac{1}{4} H(1,0;x)
          - \frac{1}{2} H(1,1;x) \, , \\
G_{1}^{(2)}(x) & = &  \frac{1}{16} \Bigl\{ 
            8
   - 2 ( \zeta(2)
   + \zeta(3) )
   + ( 4 - \zeta(2) ) \bigl[
     H(0;x) 
   + 2 H(1;x) \bigr] \nn\\
& & 
   +  \! 2 H(0,0;x) \! 
          +  \! 4 H(1,0;x) \! 
          +  \! 4 H(0,1;x) \! 
          +  \! 8 H(1,1;x) \! 
          +  \! H(0,0,0;x) \nn\\
& & 
          + 2 H(0,0,1;x)
          + 2 H(0,1,0;x)
          + 4 H(0,1,1;x)
          + 2 H(1,0,0;x) \nn\\
& & 
          + 4 H(1,0,1;x)
          + 4 H(1,1,0;x)
          + 8 H(1,1,1;x)  \Bigr\} \, , \\
G_{1}^{(3)}(x) & = & \frac{1}{32} \biggl\{ 
            \frac{9}{10} \zeta^{2}(2)
          - \bigr[ 4
          - \zeta(2)
          - \zeta(3) \bigr] \bigl[ 4
   + 2 H(0;x)
   + 4 H(1;x) \bigr] \nn\\
& & 
   - ( 4 + \zeta(2) ) \bigl[
     H(0,0;x)
          + 2 H(0,1;x)
          + 2 H(1,0;x)
          + 4 H(1,1;x) \bigr] \nn\\
& & 
          - 2 H(0,0,0;x)
          - 4 H(0,0,1;x)
          - 4 H(0,1,0;x)
          - 8 H(0,1,1;x) \nn\\
& & 
          - 4 H(1,0,0;x)
          - 8 H(1,0,1;x)
          - 8 H(1,1,0;x)
          - 16 H(1,1,1;x) \nn\\
& & 
          - H(0,0,0,0;x)
          - 2 H(0,0,0,1;x)
          - 2 H(0,0,1,0;x) \nn\\
& & 
          - 4 H(0,0,1,1;x)
          - 2 H(0,1,0,0;x)
          - 4 H(0,1,0,1;x) \nn\\
& & 
          - 4 H(0,1,1,0;x)
          - 8 H(0,1,1,1;x)
          - 2 H(1,0,0,0;x) \nn\\
& & 
          - 4 H(1,0,0,1;x)
          - 4 H(1,0,1,0;x)
          - 8 H(1,0,1,1;x) \nn\\
& & 
          - 4 H(1,1,0,0;x)
          - 8 H(1,1,0,1;x)
          - 8 H(1,1,1,0;x) \nn\\
& & 
          - 16 H(1,1,1,1;x)
    \biggr\} \, .
\eea
%%%%%%%%%%%%%%%%%%%%%%%%%%%%%%%%%%%%%%%%%%%%%%%%%%%%%%%%%%%%%%%%%%%%%

%%%%%%%%%%%%%%%%%%%%%%%%%%%%%%%%%%%%%%%%%%%%%%%%%%%%%%%%%%%%%%%%%%%%%
\be
\hspace*{-10mm} \parbox{22mm}{
\begin{fmfgraph*}(20,20)
\fmfleft{i1,i2}
\fmfright{o}
\fmf{plain}{i1,v1}
\fmf{plain}{i2,v2}
\fmf{dashes}{v3,o}
\fmf{photon,tension=.3}{v2,v3}
\fmf{photon,tension=.3}{v1,v3}
\fmf{plain,tension=0}{v2,v1}
%\fmf{plain,right=45}{v3,v3}
\end{fmfgraph*} }
= \hspace*{2mm}  \int {\mathfrak{D}^D k_1}
          \ \frac{1}{
                     {\mathcal D}_{2}
                     {\mathcal D}_{3}
                     {\mathcal D}_{5}
                     } = 
\frac{1}{m^2} \sum_{i=0}^{2} (D-4)^i \, G_{2}^{(i)}(x) + {\mathcal O} 
\left( (D-4)^3 \right) \, , 
\ee
where:
\bea
G_{2}^{(0)}(x) & = & 
      \frac{1}{8} \bigg[ \frac{1}{(1-x)} - \frac{1}{(1+x)} \bigg]
        \Big[
            4 \zeta(2)
          + H(0,0;x)
          + 2 H(0,1;x)
        \Big] \, , \\
G_{2}^{(1)}(x) & = & 
      - \frac{1}{16} \Bigg[ \frac{1}{(1-x)} - \frac{1}{(1+x)} \Bigg]
        \Big\{
            5 \zeta(3) \! 
          + \zeta(2) \bigl[ 8 H(-1;x)  \! 
          - H(0;x)   \nn \\
 & & 
          +  8 H(1;x) \bigr]
          + H(0,0,0;x) 
          + 2 \bigl[ H(-1,0,0;x) 
          + 2 H(-1,0,1;x)  \nn \\
& & 
          + H(1,0,0;x) 
          + H(0,1,0;x) 
          + H(0,0,1;x)  
          + 2 H(1,0,1;x)  \nn \\
& & 
          + 2 H(0,1,1;x) \bigr]
        \Big\} , \\
G_{2}^{(2)}(x) & = &  \frac{1}{32} \biggl[ \frac{1}{(1-x)} -
\frac{1}{(1+x)} \biggr] \bigg\{ 
            \frac{59}{10} \zeta^{2}(2)
          - \zeta(3) \bigl[
     2 H(0;x)
          - 10 H(1;x) \nn\\
& & 
          - 10 H(-1;x) \bigr]
          - \zeta(2) \bigl[
            H(0,0;x)
          + 2 H(0,1;x)
   + 2 H(1,0;x) \nn\\
& & 
   - 16 H(1,1;x)
   + 2 H(-1,0;x)
   - 16 H(-1,1;x)
   - 16 H(1,-1;x) \nn\\
& & 
   - 16 H(-1,-1;x) \bigr]
   + H(0,0,0,0;x)
   + 4 H(-1,-1,0,0;x) \nn\\
& & 
   + 8 H(-1,-1,0,1;x)
   + 2 H(-1,0,0,0;x)
   + 4 H(-1,0,0,1;x) \nn\\
& & 
   + 4 H(-1,0,1,0;x)
   + 8 H(-1,0,1,1;x)
   + 4 H(-1,1,0,0;x) \nn\\
& & 
   + 8 H(-1,1,0,1;x)
   + 2 H(0,0,0,1;x)
   + 2 H(0,0,1,0;x) \nn\\
& & 
   + 4 H(0,0,1,1;x)
   + 2 H(0,1,0,0;x)
   + 4 H(0,1,0,1;x) \nn\\
& & 
   + 4 H(0,1,1,0;x)
   + 8 H(0,1,1,1;x)
   + 4 H(1,-1,0,0;x) \nn\\
& & 
   + 8 H(1,-1,0,1;x)
   + 2 H(1,0,0,0;x)
   + 4 H(1,0,0,1;x) \nn\\
& & 
   + 4 H(1,0,1,0;x)
   + 8 H(1,0,1,1;x)
   + 4 H(1,1,0,0;x) \nn\\
& & 
   + 8 H(1,1,0,1;x) \biggl\} \, .
\eea
%%%%%%%%%%%%%%%%%%%%%%%%%%%%%%%%%%%%%%%%%%%%%%%%%%%%%%%%%%%%%%%%%%%%%

\end{fmffile}


\begin{thebibliography}{99} 
\def    \np     #1#2#3{{\it Nucl. Phys.} {\bf #1} (19#2) #3}
\def    \nptwoth     #1#2#3{{\it Nucl. Phys.} {\bf #1} (20#2) #3}
\def    \prep   #1#2#3{{\it Phys. Rep.} {\bf #1}  (19#2) #3}
\def    \pl     #1#2#3{{\it Phys. Lett.} {\bf #1} (19#2) #3}
\def    \pltwoth     #1#2#3{{\it Phys. Lett.} {\bf #1} (20#2) #3}
\def    \plold  #1#2#3{{\it Phys. Lett.} {\bf #1B} (19#2) #3}
\def    \prl    #1#2#3{{\it Phys. Rev. Lett.} {\bf #1}  (19#2) #3} 
\def    \pr     #1#2#3{{\it Phys. Rev.} {\bf #1}  (19#2) #3}
\def    \prd    #1#2#3{{\it Phys. Rev.} {\bf D#1}  (19#2) #3} 
\def    \zeit   #1#2#3{{\it Z. Phys.} {\bf C#1}  (19#2) #3}
\def    \cmp    #1#2#3{{\it Comm. Math. Phys.} {\bf #1}  (19#2) #3}
\def    \ibid   #1#2#3{{\it ibid.} {\bf #1} (19#2) #3}
\def    \nc     #1#2#3{{\it Nuovo Cim.} {\bf #1} (19#2) #3}
\def    \acta   #1#2#3{{\it Acta Phys. Polon.} {\bf #1} (19#2) #3}
\def    \tmp    #1#2#3{{\it Theor. Math. Phys.} {\bf #1} (19#2) #3}
\def    \comp    #1#2#3{{\it Comput. Phys. Commun.} {\bf #1} (20#2) #3}
\def    \hepph  #1 {{\tt hep-ph/#1}}
\def    \hepex  #1 {{\tt hep-ex/#1}}
\def    \mathph  #1 {{\tt math-ph/#1}}
\parskip 0pt
\itemsep=0pt


%%%%%%%%%%%%%% FB-Asym %%%%%%%%%%%%%%%%%%%
%
\bibitem{Yellow}
 Z Physics at LEP, vol.1, edited by G. Altarelli, CERN 89-08, 1989.

\bibitem{EWWorkingGroup}
 The LEP collaborations ALEPH, DELPHI, L3, OPAL, the LEP Electroweak
 Working Group, and the SLD Heavy Flavour Group, {\it A Combination of
 Preliminary Electroweak Measurements and Constraints on the Standard
 Model}, report LEPEWWG/2003-01, April 2003.

\bibitem{TOPexp}
 L. Salmi (\hepex{0301021}).
%%CITATION = HEP-EX 0301021;%%

\bibitem{Altarelli}
 G. Altarelli and B. Lampe, \np{B391}{93}{3}.
%%CITATION = NUPHA,B391,3;%%

\bibitem{vanNeerven}
 V. Ravindran and W. L. van Neerven, \pl{B445}{98}{214}.
(\hepph{9809411}).
%%CITATION = HEP-PH 9809411;%%


\bibitem{Catani}
 S. Catani and M.H. Seymour,
 {\it JHEP} {\bf 07} (1999) ({\tt hep-ph/9905424}).
%%CITATION = HEP-PH 9905424;%%

%%%%%%%%%%%%%% 2-loop %%%%%%%%%%%%%%%%%%%

\bibitem{RoPieRem1}
  R. Bonciani, P. Mastrolia and E. Remiddi, \nptwoth{B661}{03}{399}
  (\hepph{0301170}).
%%CITATION = HEP-PH 0301170;%%

\bibitem{Passa}
  G. Passarino, ``Loops to the Deepest''. Talk given at the XXVIIth
  International Conference of Theoretical Physics, September 2003, 
  Ustron, Poland.

\bibitem{RoPieRem2}
  R. Bonciani, P. Mastrolia and E. Remiddi, \nptwoth{B676}{04}{289}
  (\hepph{0307295}).
%%CITATION = HEP-PH 0307295;%%

%%%%%%%%%%%%%%%%%% Dim Reg %%%%%%%%%%%%%%%%%%%%%%%%%%%%%

\bibitem{DimReg}
  G. 't Hooft and M. Veltman, \np{B44}{72}{189}.\\
%%CITATION = NUPHA,B44,189;%%
  C. G. Bollini and J. J. Giambiagi, \plold{40}{72}{566}; 
  \nc{12B}{72}{20}. \\
%%CITATION = NUCIA,B12,20;%%
  J. Ashmore, {\it Lett. Nuovo Cimento} {\bf 4} (1972) 289.\\
%%CITATION = NCLTA,4,289;%%
  G. M. Cicuta and E. Montaldi, {\it Lett. Nuovo Cimento} {\bf 4} 
  (1972) 289. \\
%%CITATION = NCLTA,4,329;%%
  R. Gastmans and R. Meuldermans, \np{B63}{73}{277}.
%%CITATION = NUPHA,B63,277;%%


%%%%%%%%%%%%%%%%%% IBP %%%%%%%%%%%%%%%%%%%%%%%%%%%%%%%%% 

\bibitem{Chet}
  F.V. Tkachov, \pl{B100}{81}{65}.\\
%%CITATION = PHLTA,B100,65;%%
  K.G. Chetyrkin and F.V. Tkachov, \np{B192}{81}{159}.
%%CITATION = NUPHA,B192,159;%%


%%%%%%%%%%%%%%%%%% LI %%%%%%%%%%%%%%%%%%%%%%%%%%%%%%%%%

\bibitem{Rem3}
  T. Gehrmann and E. Remiddi, \nptwoth{B580}{00}{485} 
  ({\tt hep-ph/9912329}). 
%%CITATION = HEP-PH 9912329;%%
  

%%%%%%%%%%%%%%%%%% Diff. Eqs. Method %%%%%%%%%%%%%%

\bibitem{Kot}
  A. V. Kotikov, \pl{B254}{91}{158}.
%%CITATION = PHLTA,B254,158;%%

\bibitem{Rem1} 
  E. Remiddi, \nc{110A}{97}{1435} ({\tt hep-th/9711188}).
%%CITATION = HEP-TH 9711188;%%

\bibitem{Rem2}
  M. Caffo, H. Czy\.{z}, S. Laporta and E. Remiddi, 
  \acta{B29}{98}{2627} ({\tt hep-ph/9807119}).\\
%%CITATION = HEP-TH 9807119;%%
  M. Caffo, H. Czy\.{z}, S. Laporta and E. Remiddi, 
  \nc{A111}{98}{365} ({\tt hep-ph/9805118}).
%%CITATION = HEP-TH 9805118;%%

%%%%%%%%%%%%%%% Harm-Polylog %%%%%%%%%%%%%%%%%%%%%%%%%%%%%%%%%%%%

\bibitem{Polylog}
  E. Remiddi and J. A. M. Vermaseren, {\it Int. J. Mod. Phys.} 
  {\bf A15} (2000) 725 ({\tt hep-ph/9905237}). 
%%CITATION = HEP-PH 9905237;%%

\bibitem{Polylog3}
  T. Gehrmann and E. Remiddi, \comp{141}{01}{296} 
  ({\tt hep-ph/0107173}).
%%CITATION = HEP-PH 0107173;%%

%%%%%%%%%%%%%%%%%%%%%%%%%%%%%%%%%%%%%%%%%%%%%%%%%%%%%%%%%%%%%%%%%

\bibitem{webpage}
The results can be downloaded from: \\
{\tt http://pheno.physik.uni-freiburg.de/$^\sim$bonciani/} \\
as an input file for FORM.

%%%%%%%%%%%%%%%%%%%%%%%%%%%%%%%%%%%%%%%%%%%%%%%%%%%%%%%%%%%%%%%%%

\bibitem{UgoRo}
 U. Aglietti and R. Bonciani, {\tt hep-ph/0401193}.
%%CITATION = HEP-PH 0401193;%%

%%%%%%%%%%%%%%% Confronto con Davydychev %%%%%%%%%%%%%%%%%%%%%%%%%%%

\bibitem{Fleischer}
  J. Fleischer, A. V. Kotikov and O. L. Veretin, 
   \np{B547}{99}{343}  ({\tt hep-ph/9808242}).
%%CITATION = HEP-PH 9808242;%%
 
\bibitem{Davydychev}
 A. I. Davydychev and M. Yu. Kalmykov, {\tt hep-th/0303162v2}.
%%CITATION = HEP-TH 0303162;%%

\bibitem{Kuehn}
  B.~A.~Kniehl and J.~H.~Kuhn,
   \np{B329}{90}{547}.
%%CITATION = NUPHA,B329,547;%%

\bibitem{Fleischer1}
 J.~Fleischer, V.~A.~Smirnov, A.~Frink, J.~G.~Korner, D.~Kreimer, 
 K.~Schilcher and J.~B.~Tausk,
 {\it Eur. Phys. J.} {\bf C2} (1998) 747 
(\hepph{9704353}).
%%CITATION = HEP-PH 9704353;%%

%%%%%%%%%%%%%%%%%%% Programs %%%%%%%%%%%%%%%%%%%%%%%%%%%%

\bibitem{FORM} J.A.M.\ Vermaseren, Symbolic Manipulation with
               {\tt FORM}, Version 2, CAN, Amsterdam, 1991; \\
               New features of {\tt FORM}, (\mathph{0010025}).
%%CITATION = MATH-TH 0010025;%%

%%%%%%%%%%%%%%%%%%%%%%%%%%%%%%%%%%%%%%%%%%%%%%%%%%%%%%%%%%%%%%



\end{thebibliography}
\end{document}